\documentclass{aa}	     
\usepackage{epsfig}
\usepackage{graphicx}
\usepackage{epstopdf}
\usepackage{txfonts}
\usepackage{natbib}
\usepackage{color}
\usepackage{textcomp}
\usepackage{amsmath}
\usepackage{amssymb}
\usepackage{gensymb}
\usepackage{tikz}
\usepackage{longtable}
\usepackage{array,multirow}
\usepackage{mdwtab}
\usepackage[colorlinks=true,linkcolor=blue,allcolors=blue]{hyperref}

\newcommand{\lya}{Ly$\alpha\ $}

\newcommand{\angst}{\mbox{\normalfont\AA}}

\begin{document}
    \title{J-PLUS: Unveiling the brightest-end of the $\rm Ly\alpha$ luminosity function at $\rm 2.0 \!<\! z \!<\! 3.3$ over $\rm1000\,deg^2$}
   \author{Daniele Spinoso$\,^{1}$, Alvaro Orsi$\,^{1,2}$, Carlos L\'opez-Sanjuan$\,^1$, Silvia Bonoli$\,^{3,4}$, Kerttu Viironen$\,^{1}$, David Izquierdo-Villalba$\,^{1}$, David Sobral$\,^5$, Siddhartha Gurung-L\'opez$\,^{6,1}$, Antonio Hern\'an-Caballero$\,^{1}$, Alessandro Ederoclite$\,^{7,\,8}$, Jes\'us Varela$\,^1$, Roderik Overzier$\,^9$, Jordi Miralda-Escud\'e$^{10,11,12}$, David J. Muniesa$\,^{1}$, Jailson Alcaniz$\,^9$, Raul E. Angulo$\,^3$, A. Javier Cenarro$\,^1$, David Crist\'obal-Hornillos$\,^1$, Renato A. Dupke$\,^{9,13}$,  Carlos Hern\'andez-Monteagudo$\,^1$, Antonio Mar\'in-Franch$\,^1$, Mariano Moles$\,^1$, Laerte Sodr\'e Jr$\,^7$, H\'ector V\'azquez-Rami\'o$\,^1$
   }
   \institute{1 - Centro de Estudios de F\'isica del Cosmos de Arag\'on. Plaza San Juan 1, planta 2, 44001 Teruel, Spain\\
   2 - PlantTech Research Institute Limited. South British House, 4th Floor, 35 Grey Street, Tauranga 3110, New Zealand\\
   3 - Donostia International Physics Center. Paseo Manuel de Lardizabal, 4, 20018 Donostia-San Sebasti\'an (Gipuzkoa), Spain\\
   4 - IKERBASQUE, Basque Foundation for Science, E-48013, Bilbao, Spain\\
   5 - Department of Physics, Lancaster University, Lancaster, LA1 4YB, UK\\
   6 - Institute for Multi-messenger Astrophysics and Cosmology, Department of Physics, Missouri University of Science and Technology. 1315 N. Pine St., Rolla MO 65409, USA\\
   7 - Universidade de S\~ao Paulo, Instituto de Astronomia, Geof\'isica e Ci\^encias Atmosf\'ericas. 05508090 S\~ao Paulo, SP, Brazil\\
   8 - Asociaci\'on Astrof\'isica para la Promoci\'on de la Investigaci\'on, Instrumentaci\'on y su Desarrollo. 38205 La Laguna, Tenerife, Spain\\
   9 - Observat\'orio Nacional/MCTIC. Rua Jos\'e Cristino, 77, CEP 20921-400, S\~ao Crist\'ov\~ao, Rio de Janeiro (RJ), Brazil\\
   10 - Institut de Ci\`encies del Cosmos, Universitat de Barcelona, Mart\'i i Franqu\`es 1,08028 Barcelona\\
   11 - Instituci\'o Catalana de Recerca i Estudis Avan\c cats, 08034 Barcelona\\
   12 - Institute for Advanced Study, Princeton NJ 08544\\
   13 - Department of Astronomy, University of Michigan, Ann Arbor, MI 48109-1107, USA
   }
   \date{}
   

  \abstract
  {We present the photometric determination of the bright-end of the \lya luminosity function (at $\rm L_{Ly\alpha}\,{\gtrsim}\,10^{\,43.5}\, erg\,s^{-1}$) within four redshifts windows ($\rm\Delta\,z\,{<}\,0.16$) in the interval $\rm2.2\!\lesssim\!z\!\lesssim\!3.3$. Our work is based on the Javalambre Photometric Local Universe Survey (J-PLUS) first data-release, which provides multiple narrow-band measurements over $\rm{\sim}1000\,deg^2$, with limiting magnitude $r\!\sim\!22$. The analysis of high-z \lya$\!\!$-emitting sources over such a wide area is unprecedented, and allows to select a total of $\sim\!14,500$ hyper-bright ($\rm L_{Ly\alpha}\,{>}\,10^{43.3}\,erg\,s^{-1}$) \lya$\!\!$-emitting candidates. We test our selection with two spectroscopic follow-up programs at the GTC telescope, which confirm as line-emitting sources $\sim\!89\%$ of the targets, with $\sim\!64\%$ being genuine $\rm z\!\sim\!2.2$ QSOs. We extend the $\rm2.2\!\lesssim\!z\!\lesssim\!3.3$ \lya luminosity function for the first time above $\rm L_{Ly\alpha}\,{\sim}\,10^{44}\,erg\,s^{-1}$ and down to densities of $\rm\sim\!10^{-8}\,Mpc^{-3}$. Our results unveil with high detail the Schechter exponential-decay of the brightest-end of the \lya LF, complementing the power-law component of previous LF determinations at $\rm 43.3\!\lesssim\!Log_{10}(L_{Ly\alpha}/erg\,s^{-1})\!\lesssim\!44$. We measure $\rm\Phi^*=(3.33\pm0.19)\times10^{-6}$, $\rm Log(L^*)=44.65\pm0.65$ and $\rm\alpha=-1.35\pm0.84$ as an average over the redshifts we probe. These values are significantly different than the typical Schechter parameters measured for the \lya LF of high-z star-forming LAEs. This suggests that $\rm z\!>\!2$ AGN/QSOs (likely dominant in our samples) are described by a structurally different LF than $\rm z\!>\!2$ star-forming LAEs, namely with $\rm L^*_{QSOs}\sim100\,L^*_{LAEs}$ and $\rm\Phi^*_{QSOs}\sim10^{-3}\,\Phi^*_{LAEs}$. Finally, our method identifies very efficiently as high-z line-emitters sources without previous spectroscopic confirmation, currently classified as stars ($\sim\!2000$ objects in each redshift bin, on average). Assuming a large predominance of \lya$\!\!$-emitting AGN/QSOs in our samples, this supports the scenario by which these are the most abundant class of $\rm z\!\gtrsim\!2$ \lya emitters at $\rm L_{Ly\alpha}\,{\gtrsim}\,10^{\,43.3}\,erg\,s^{-1}$.}
  \titlerunning{J-PLUS. Unveiling the brightest-end of the $\rm Ly\alpha$ luminosity function at $\rm 2.0 \!<\! z \!<\! 3.3$}
  \authorrunning{Spinoso et al.}
  \maketitle

\section{Introduction}
An increasing number of recent works has been focusing on the study of   high-redshift Lyman-$\alpha$ emitters (LAEs), objects showing prominent rest-frame \lya emission within a spectrum (usually) devoided of other line features \citep[e.g.,][]{cassata2011,nakajima2018b}. The spectral properties of LAEs are usually interpreted as to be coming from young ($\rm\lesssim50\,Myr$) and low-mass ($\rm M_*\!<\!10^{10}\,M_\odot$) galaxies \citep[e.g.,][]{wilkins2011b,amorin2017,hao2018,santos2020} with small rest-frame UV half-light radii \citep[ $\rm R\!\lesssim\!1-2\,Kpc$, as in e.g.,][]{moller_warren1998,lai2008,bond2012,guaita2015,kobayashi2016,ribeiro2016,bouwens2017a, paulino-afonso2018} which are actively star-forming ($\rm SFR\!\sim\!1-100\,M_\odot/yr$) and dust-poor \citep[dust attenuation $\rm A_V\!<\!0.2$, see e.g.,][]{gawiser2006,gawiser2007,guaita2011,nilsson2011,bouwens2017b,arrabal-haro2020}. When observed at high redshift, isolated and grouped LAEs would represent the progenitors of present-day galaxies and clusters, respectively, hence providing extremely valuable insights about structure formation \citep[e.g.,][]{matsuda2004,matsuda2005,venemans2005b,gawiser2007,overzier2008,guaita2010,mei2015,bouwens2017b,khostovan2019}. A basic statistical tool to study the population of high-z LAEs is the description of their number density, at a given redshift, as a function of line luminosity ($\rm L_{Ly\alpha}$), 
namely the \lya luminosity function \citep[LF, see e.g.,][for a theoretical approach]{gronke2015}. 
Several recent works have focused on the construction of the \lya LF at $\rm z\!\geq\!2$ \citep[][]{gronwall2007,ouchi2008,blanc2011,clement2012,konno2016,sobral2017,sobral2018} 
by making use of deep observations carried over narrow sky regions \citep[up to few squared degrees, as in e.g.,][]{matthee2014,cassata2015,matthee2017b,ono2018}. Their findings describe a \lya LF which follows a Schechter function \citep{schechter1976} at relatively faint line luminosity \citep[i.e.
$\rm L_{Ly\alpha}\lesssim10^{42.5}$, see e.g.,][]{ouchi2008,konno2016,sobral2016a,matthee2017a}, a regime mostly occupied by low-mass star-forming galaxies \citep[e.g.,][]{hu1998,kudritzki2000,stiavelli2001,santos2004b,van-breukelen_jarvis_venemans2005,gawiser2007,rauch2008,guaita2011}.

\begin{figure}[t]
\centering
\includegraphics[width=0.49\textwidth]{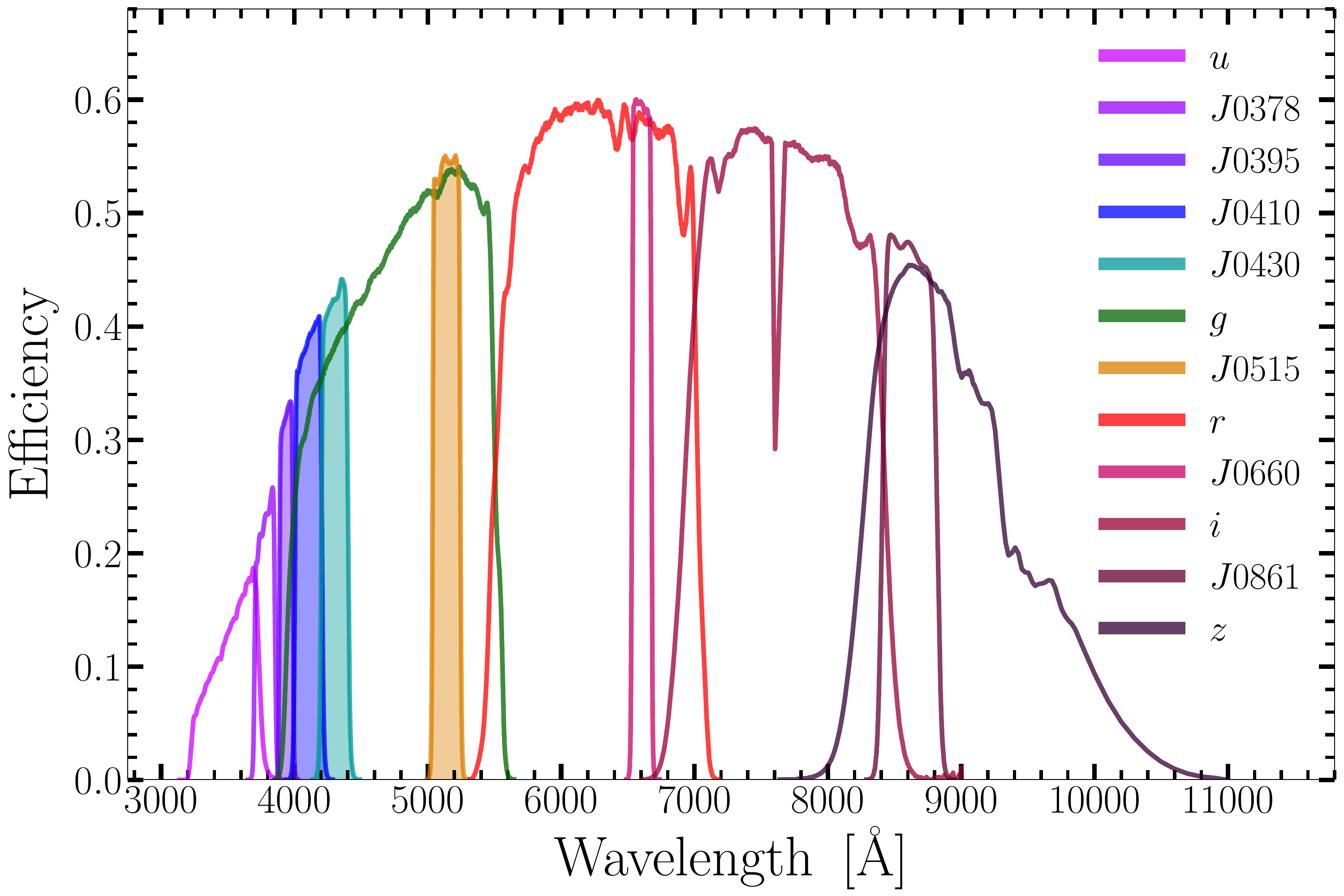}
\caption{\footnotesize The measured transmission curves for the J-PLUS filter set, after accounting for sky absorption, CCD quantum efficiency and the total effect of the JAST/T80 telescope optical system. The four NB we exploit to look for bright \lya emitters at $\rm z\!>\!2$ (namely, the $J$0395, $J$0410, $J$0430 and $J$0515 NBs) share their wavelength coverage with the $g$ band and are shown here as filled-area curves.}
\label{fig:fset}
\end{figure}

On the other hand, the bright-end of the \lya LF is populated by AGN/QSOs \citep{calhau2020} and rare, bright and SF-bursty \lya$\!\!$-emitting systems \citep[e.g.,][]{matsuda2011,bridge2013,cai2017b,cai2018}. Current constraints at high \lya luminosity are somewhat poor, given the relatively small cosmological volumes probed by past works focused specifically on detecting high-z \lya$\!\!$-emitting sources \citep[e.g.,][]{fujita2003,blanc2011,herenz2019}. In particular, recent measurements show hints about a number-density excess with respect to an  exponential (Schechter) decay, at $\rm L_{Ly\alpha}\gtrsim10^{\,43}$ \citep[e.g.,][]{konno2016}. This might be explained by means of a population of faint AGN contributing to the global LAE balance \citep[see e.g.,][]{matthee2017b, sobral2018}. Further support to this picture is provided by the tomographic analysis of the high-z \lya LF in the \texttt{COSMOS} field performed by \cite{sobral2018} by using a combination of optical, infrared and X-Ray data. In their work, the major contribution to the LF at $\rm L_{Ly\alpha}\gtrsim10^{43}$ is provided by sources showing X-Ray loud counterparts, thus likely to be AGN \citep[see also e.g.,][]{matthee2017b, calhau2020}. Their work shows how this contribution completely vanishes at $\rm z\!\gtrsim\!3.5$, thus paralleling the peak of AGN activity usually observed at $\rm z\!\sim\!2-3$ \citep[e.g.,][]{hasinger_miyaji_schmidt2005,miyaji2015}. Finally, the constraints on the bright-end of the \lya LF are prone to significant contamination by lower-redshift interlopers. For example, \cite{sobral2017} and \cite{stroe2017a} showed that a consistent fraction of bright LAE candidates at $\rm z\!>\!2$ are actually AGN at $\rm z\!\gtrsim\!1.5$ emitting \texttt{CIV}.

This work exploits the first data-release (DR1 hereafter) of the Javalambre Photometric Local Universe Survey \citep[J-PLUS,][]{cenarro2019}, which provides imaging of the Northern hemisphere in both narrow- and broad-bands (NB and BB, see Fig. \ref{fig:fset} and Table \ref{tab:filts}). The DR1 covers an area of $\rm>\!1000\,deg^2$, which is unprecedented for NB-surveys of $\rm z\!>\!2$ luminous line-emitters. Our goal is to exploit these characteristics for obtaining large samples of photometrically-selected bright \lya emitting sources, and probe the bright-end of their LF at four different redshifts (see Table \ref{tab:QSOcontamin}). The combination of large survey area and multi-NB data provides the means to assess the nature of bright \lya$\!\!$-emitting sources \citep{nilsson2011,shibuya2014} and sample their distribution over a luminosity regime which is yet poorly constrained \citep{gronwall2007,guaita2010,blanc2011,konno2016,sobral2018}. We complement our study by presenting the results of two follow-up spectroscopic programs aimed at assessing the performance and contamination of our methodology.

This paper is organised as follows: Sect. \ref{sec:jplus_survey} details the main features of the J-PLUS survey and the classes of sources we target. Our method for detecting NB excesses, our selection function and our sample of LAEs candidates are described in Sect. \ref{sec:selection}, along with our spectroscopic follow-up programs. Section \ref{sec:lumifunc} is focused on the computation of the four $\rm 2\!<\!z\!<\!3.3$ \lya LFs. Finally, we discuss our results in Sect. \ref{sec:scientific_results} and present our conclusions in Sect. \ref{sec:discussion_and_conclusions}. Throughout this paper, magnitudes are given in the AB system \citep{oke1974,oke_gunn1983}, and we assumed a flat $\Lambda$CDM cosmology described by \texttt{PLANCK15} parameters \citep{adam2016, ade2016}, namely: $\rm H_0=67.3\,Km\,s^{-1}\,Mpc^{-1}$, $\rm\Omega_{m;0}=0.315$, $\rm\Omega_{\Lambda;0}=0.685$.
\begin{table}[t]
    \centering
    \resizebox{9cm}{!}{
    \begin{tabular}{c|c|c|c}
    \hline 
    \hlx{v}
    Filter  &  FWHM [\AA] & $\rm mag_{AB}^{min}\,(3\sigma)$ & $\langle \mathit{f}_{\lambda}\rangle\ \rm(3\sigma)\ [erg\,cm^{-2}\,s^{-1}\,\text{\AA}^{-1}]$\\
    \hlx{v}
    \hline
    \hline
    \hlx{v}
    $u$       &  363.91 & 21.17 & $2.99\times10^{-17}$ \\
    \hlx{v}
    $J$0378     &  152.74 & 21.18 & $2.56\times10^{-17}$ \\
    \hlx{v}
    $J$0395     &  101.39 & 21.06 & $2.63\times10^{-17}$ \\
    \hlx{v}
    $J$0410     &  201.76 & 21.28 & $1.99\times10^{-17}$ \\
    \hlx{v}
    $J$0430     &  200.80 & 21.30 & $1.78\times10^{-17}$ \\
    \hlx{v}
    $g$       &  1481.92 & 22.09 & $7.05\times10^{-18}$ \\
    \hlx{v}
    $J$0515     &  207.19 & 21.35 & $1.18\times10^{-17}$ \\
    \hlx{v}
    $r$       &  1500.20 & 22.02 & $4.36\times10^{-18}$ \\
    \hlx{v}
    $J$0660     &  146.13 & 21.34 & $7.27\times10^{-18}$ \\
    \hlx{v}
    $i$       &  1483.59 & 21.54 & $4.47\times10^{-18}$ \\
    \hlx{v}
    $J$0861     &  410.50 & 20.67 & $7.94\times10^{-18}$ \\
    \hlx{v}
    $z$       &  1055.93 & 20.80 & $6.54\times10^{-18}$ \\
    \hlx{v}
    \hline
    \end{tabular}}\vspace{1mm}
    \caption{ Tabulated FWHMs and $3\sigma$ detection limits of J-PLUS filters. Additional details and information about J-PLUS DR1 can be found at: \texttt{https://archive.cefca.es/catalogues/jplus-dr1}. For the sake of simplicity, we generally refer to the filters $J$0378, $J$0395, $J$0410, $J$0430, $J$0515, $J$0660 and $J$0861 as J-PLUS NBs, even though some of these filters (e.g. $J$0861) could be defined as \textit{medium bands}.}
    \label{tab:filts}
\end{table}

\section{\lya emitters in the J-PLUS photometric survey}
\label{sec:jplus_survey}
J-PLUS is an ongoing wide-area photometric survey performed at the Observatorio Astrof\'isico de Javalambre \citep[OAJ,][]{cenarro2014} in Arcos de las Salinas (Teruel, Spain). Here we summarize its technical features \citep[detailed in][]{cenarro2019} and we define the class of \lya$\!\!$-emitting sources we target.

\begin{table*}[t]
    \centering
    \resizebox{18.3cm}{!}{
    \begin{tabular}{c|c|c|c|c|c|c|c|c||c|c|c|c}
    \hline
Narrow & \multicolumn{8}{ c|| }{ \texttt{$\rm Ly{\alpha}$}-related properties }  & \multicolumn{4}{ c }{ \tiny{$\ \langle\rm z\rangle$ of contaminant QSO lines} }\\
Band   & $\ \langle\rm z\rangle$  & $\rm z_{\,p}$ & $[\rm z_{min}\ ;\ z_{max}]$  & $\rm A_{eff}\ [deg^2]$ & $\rm Vol\ [cGpc^3]$ & $\rm F_{Ly\alpha}^{\,min}\,[erg\,cm^{-2}\,s^{-1}]$  &   $\rm L_{Ly\alpha}^{min}\,[erg\,s^{-1}]$ &  $\rm Log\,\left(L_{Ly\alpha}^{min}\right)$  & \tiny{\texttt{SiIV}}  & \tiny{\texttt{CIV}}  & \tiny{\texttt{CIII]}}  & \tiny{\texttt{MgII}} \\
    \hlx{vv}
    \hline
    \hline
    \hlx{v}
    $J$0395  & 2.24 & 2.25 & $2.20 - 2.28$ & 897.44 & 0.961 & $5.23\times10^{-16}$ &  $2.16\times10^{43}$ &  43.33 &  \tiny{1.82} &  \tiny{1.54} &  \tiny{1.06} & \tiny{0.41} \\
    \hlx{v}
    $J$0410  & 2.38 & 2.37 & $2.29 - 2.46$ & 897.46 & 1.917 & $4.46\times10^{-16}$ &  $2.21\times10^{43}$ &  43.34 &  \tiny{1.94} &  \tiny{1.65} &  \tiny{1.15} & \tiny{0.47} \\
    \hlx{v}
    $J$0430  & 2.54 & 2.53 & $2.46 - 2.62$ & 897.41 & 1.907 & $3.99\times10^{-16}$ &  $2.25\times10^{43}$ &  43.35 &  \tiny{2.08} &  \tiny{1.78} &  \tiny{1.25} & \tiny{0.54} \\
    \hlx{v}
    $J$0515  & 3.23 & 3.24 & $3.14 - 3.31$ & 965.99 & 2.044 & $2.65\times10^{-16}$ &  $2.68\times10^{43}$ &  43.43 &  \tiny{2.68} &  \tiny{2.32} &  \tiny{1.69} & \tiny{0.84} \\
    \hlx{v}
    \hline
    \end{tabular}
    }
    \vspace{1.5mm}
    \caption{Second to ninth columns from the left: properties of the filters related to the \lya line. From left to right: median redshift in the filter bandwidth, redshift associated to the filter \textit{pivot} wavelength \citep[see][]{tokunaga_vacca2005}, redshift interval covered by the NB FWHM, effective DR1 area ad volume (see Sect. \ref{sec:lyalum_and_volume_computation}), minimum detectable line flux and luminosity (both in linear and logarithmic units, see Sect. \ref{sec:lyaLimits}). Last four columns to the right: redshift associated to strong QSOs lines \citep[e.g.,][]{vandenberk2001} which can act as contaminants in our selection.}
    \label{tab:QSOcontamin}
\end{table*}

\subsection{Survey description and source catalogs}
\label{sec:jplus_definition_and_catalogs}
J-PLUS observations are being carried out by the T80Cam instrument on the JAST/T80 83cm telescope \citep{marin-franch2015}. The JAST/T80 optical system provides a wide field of view ($\rm FoV\!\sim\!1.96\,deg^2$) while ensuring a high spatial resolution \citep[$0.55$ arcsec/pixel, see][for technical details]{cenarro2019}. J-PLUS nominal depth is shallower than that of comparably-wide optical surveys, i.e. $r=22$ at signal-to-noise ratio $\rm SNR=3$ \citep[as compared to e.g. $\rm r'=23.1$ at $\rm SNR=5$ for SDSS, see][]{york2000}. Nevertheless, it offers NB measurements over an unprecedented sky-area, making it suitable for extensive searches of bright emission-line galaxies (ELGs). The J-PLUS filter set is composed by 12 photometric pass-band filters (see Fig. \ref{fig:fset}) which can be divided into 5 broad-bands (BBs) and 7 narrow-bands (NBs) of width $\sim\!800\!-\!2000\,\angst$ and $\sim\!150\!-\!500\,\angst$, respectively (table \ref{tab:filts}). Their measured transmission curves (i.e. accounting for optical elements, CCD quantum efficiency and sky transparency) are shown in Fig. \ref{fig:fset}.

J-PLUS images are automatically reduced in order to obtain public catalogs of sources\footnote{J-PLUS catalogs can be found at: http://archive.cefca.es/catalogues}. This work is based on the recent DR1, obtained with stable pipelines for data reduction and source-extraction, specifically calibrated and tested on J-PLUS data \citep[as detailed in e.g.,][]{cenarro2019,lopezsanjuan2019a}. We use the standard J-PLUS dual-mode objects lists, constructed with $r$ as the band for source detection
and for defining their associated sky position and photometric apertures. The latter are then used to extract sources' photometry in the remaining filters. We note that relying on dual-mode catalogs has non-trivial implications on the completeness of our final LAEs samples, which we address in Sect. \ref{sec:bivariate_completeness}. Finally, this work is based on the DR1 \texttt{auto}-aperture\footnote{For details about J-PLUS aperture-photometry definitions see:\\ \texttt{\tiny http://archive.cefca.es/catalogues/jplus-dr1/help\_adql.html}\normalsize} photometry. We ensure that this choice allows to recovery the total \lya line flux of point-like sources
(see Sect. \ref{sec:line_flux_retrieval}) and exploit the measurement of detection completeness in each survey pointing provided in the DR1, which was tested on \texttt{auto}-aperture photometry (see Sect. \ref{sec:completeness}).

\subsection{Detection of \lya emission with J-PLUS}
\label{sec:jplus_NB_and_emission_lines_contamination}
The design of the J-PLUS filters potentially allows to detect \lya emission within seven redshift windows, one per NB, respectively centered at $\rm z\!\sim\!2.11,\,2.24,\,2.38,\,2.54,\,3.23,\,4.43$ and $6.09$. In particular, we employ the $J$0395, $J$0410, $J$0430 and $J$0515 filters (see Fig. \ref{fig:fset}) for targeting $\rm z\!\sim\!2.24,\,2.38,\,2.54$ and $3.23$, as shown in Table \ref{tab:QSOcontamin}. 
Our selection is based on measuring NB excesses with respect to the continuum traced by BB photometry (see Sect. \ref{sec:3filters}). Consequently, it is prone to contamination by prominent emission lines. In particular, we expect our samples to be significantly contaminated by both nebular emission due to star-formation (e.g. \texttt{H}$\,_\beta$, \texttt{[OIII]}$\!_{\,4959+5007}$ and \texttt{[OII]}$\!_{\,3727}$ lines) and AGN/QSOs ionizing radiation \citep[e.g. \texttt{CIV}$\!_{\,1549}$, \texttt{CIII]}$\!_{\,1908}$, \texttt{MgII}$\!_{\,2799}$ and \texttt{SiIV}$\!_{\,1397}$ lines, see also][]{stroe2017a,stroe2017b}. The latter ones and their associated redshift intervals in J-PLUS are listed in Table \ref{tab:QSOcontamin}. We note that \texttt{SiIV} and  \texttt{MgII} are minor sources of contamination since: i) they are significantly fainter than \lya \citep[e.g.,][]{telfer2002,selsing2016}, ii) J-PLUS probes relatively small cosmological volumes at $0.35\!<\rm\!z\!<\!0.85$ and iii) the number density of AGN/QSOs at $\rm z\!<\!1$ is lower than at $\rm z\!>\!2$ \citep[e.g.,][]{palanque-delabrouille2016,paris2018}. We exclude the $J$0378 NB after checking that our method does not reliably detect photometric excess in this NB (see Sect. \ref{sec:line_flux_retrieval}). We also exclude the $J$0660 and $J$0861 NBs since they provide very scarce samples of candidates ($\lesssim100$ sources) whose contamination cannot be reliably estimated, due to the absence of cross-matches with SDSS spectroscopic data (see Sect. \ref{sec:sample_cleaning}). We note that this is in agreement with the work of \cite{sobral2018}, which shows no significant detection of bright ($\rm L_{Ly\alpha}\gtrsim10^{\,43}\,erg\,s^{-1}$) \lya$\!\!$-emitting sources at $\rm z\!\gtrsim\!3.5$, i.e. at the redshift probed by the $J$0660 and $J$0861 NBs.

\subsubsection{$\rm L_{Ly\alpha}$ and $\rm EW_{Ly\alpha}$ detection limits}
\label{sec:lyaLimits}
The minimum luminosity of an emission line measurable with a NB filter ($\rm L_{Ly\alpha}^{min}$) can be computed by knowing the relative contribution of line and continuum to the total flux in the band, and the source redshift. In other words, by knowing the line equivalent width ($\rm EW$ hereafter, see appendix \ref{sec:append:3FM_equations}) and the wavelength position of the line-peak in the NB. Unfortunately, these are not provided by a single NB measurement without further hypothesis. To compute $\rm L_{Ly\alpha}^{min}$ for each J-PLUS NB, we first assume that faint sources are detected with higher probability at the wavelength of the transmission curve peak. Consequently their line would be redshifted to the observed $\rm\lambda_{\,obs}\!=\!\lambda_{T^{\,max}}$. The choice of $\rm EW$, on the other hand, as a higher degree of arbitrariness. Despite $\rm EW$ as low as $5\,\text{\AA}$ have been explored in the past \citep[e.g.,][]{sobral2017}, high-z \lya$\!\!$-emitting sources typically exhibit $\rm EW\!>\!15-20\,\text{\AA}$ \citep[as in e.g.,][]{gronwall2007,guaita2010,santos2020}. We hence select $\rm EW\!=\!25\text{\AA}$ as our lower limit to estimate $\rm L_{Ly\alpha}^{min}$ \citep[see e.g.,][]{ouchi2008, santos2016, konno2018}. In detail, we use the detection limits of J-PLUS bands (table \ref{tab:filts}) to compute the minimum line-flux measurable with each NB ($\rm F^{\,min}_{Ly\alpha}$, see Sect. \ref{sec:3filters} and appendix \ref{sec:append:3FM_equations} for details). We then link the latter to $\rm L_{Ly\alpha}^{min}$ using our assumptions on $\rm\lambda_{\,obs}$ and $\rm EW$.
\begin{figure*}[t]
\centering
\includegraphics[width=0.85\textwidth]{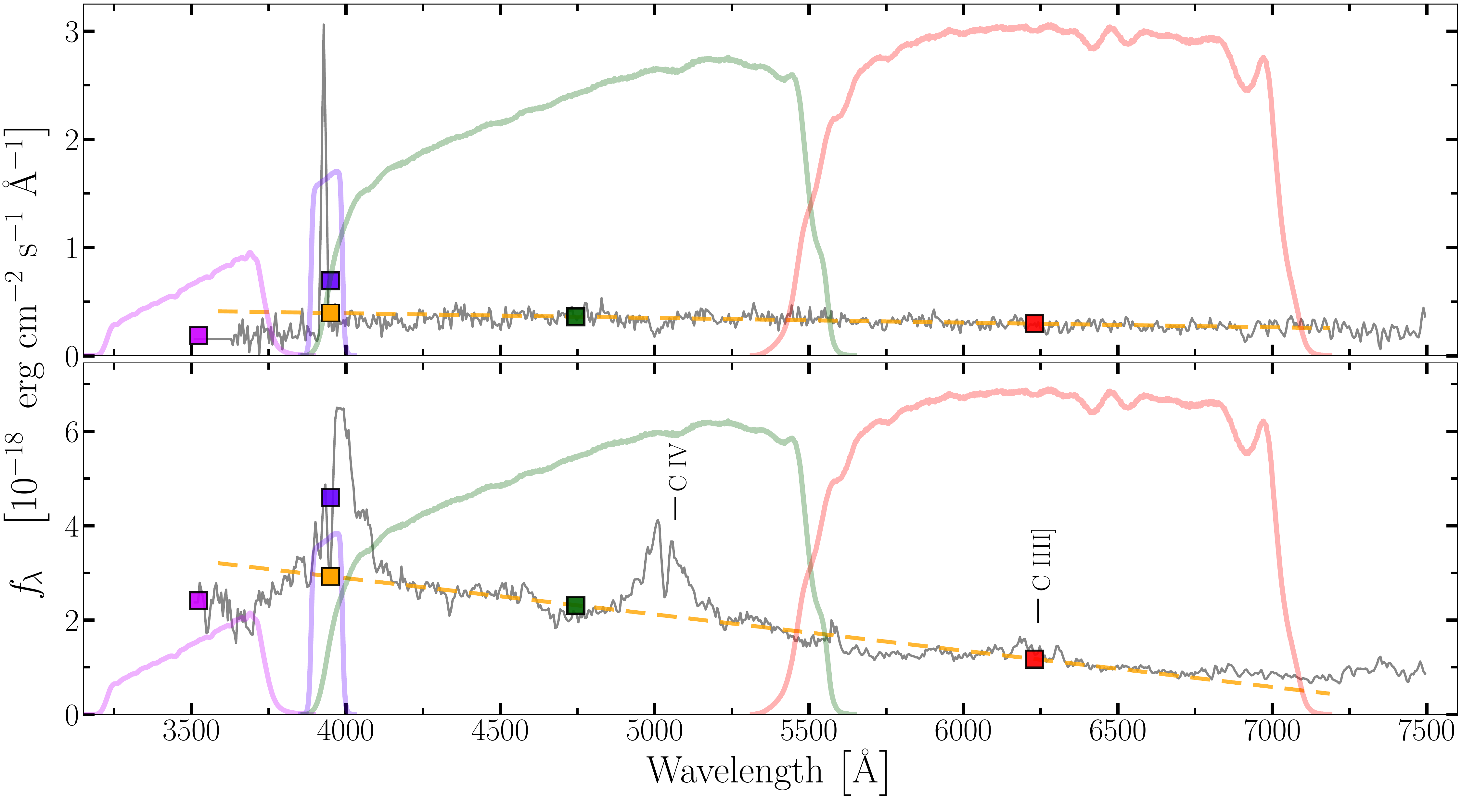}
\caption{\footnotesize Representation of our NB excess detection method. Grey lines in both panels show the observed spectra of typical $\rm z\!\sim\!2$ \lya$\!\!$-emitting sources \citep[from the publicly available VUDS DR1 spectroscopic dataset, see e.g.,][]{lefevre2015, tasca2017}. Upper panel: a SF LAE spectrum showing a single, prominent \lya line (here redshifted at $\rm\lambda_{obs}\!\sim\!3900\,\angst$) and no other significant features. Bottom panel: a QSO spectrum with evident \texttt{CIV} and \texttt{CIII]} lines in addition to \lya (at $\rm\lambda_{obs}\!\sim\!4000\,\angst$). We show the transmission curves and associated synthetic photometry of four J-PLUS bands as colored lines and squares. From left to right, $u$ (purple), $J$0395 NB (violet), $g$ (green) and $r$ (red). In brief: our method uses $g$ and $r$ photometry for estimating a linear continuum (yellow dashed line in the plots) which is then evaluated at the NB pivot wavelength (yellow square). Finally, the ratio between the latter and the NB measurement (violet square) is used as a proxy for the \lya line flux (see Eq. \ref{eq:deltaNB}). By using $u$ and $g$ instead of $g$ and $r$ this method would provide a poorer handle of the non-linear continuum in the region affected by the \lya line profile.}
\label{fig:LAEspectra_3FM}
\end{figure*}

The characteristics of J-PLUS filters and its observing strategy make its data sensitive to very bright \lya emission ($\rm L_{Ly\alpha}\!>\!10^{\,43.3}\,erg\,s^{-1}$, see Table \ref{tab:QSOcontamin}). We note that few studies have explored this range of $\rm L_{Ly\alpha}$, mostly due to the limited sky areas of their associated deep photometric surveys \citep[see e.g.,][]{blanc2011,konno2016,matthee2017b,sobral2018}. On the contrary, J-PLUS DR1 provides multi-band imaging over $\rm\sim\!\!1000\,deg^2$, which is unprecedented for studies targeting high-z \lya$\!\!$-emitting sources. The effective survey area after masking artifacts and bright stars sums up to $\rm\sim\!900\,deg^2$, which correspond to $\rm\gtrsim\!1\,Gpc^3$ (comoving) in each z window we sample (see Table \ref{tab:QSOcontamin}).
This allows to measure with high precision the \lya luminosity function at $\rm2.2\!\lesssim\!z\!\lesssim\!3.3$ and $\rm L_{Ly\alpha}\gtrsim2\times10^{\,43}\,erg\,s^{-1}$.

\subsubsection{AGN/QSOs or Star-Forming galaxies}
\label{sec:LAEs_classes_definition}
Recent compelling hints point towards identifying the majority of high-z \lya$\!\!$-emitting sources at $\rm L_{Ly\alpha}\!>\!2\times10^{43}\,erg\,s^{-1}$ as AGN/QSOs \citep[see e.g.,][]{nilsson2011,konno2016,matthee2017b,sobral2018b,sobral2018, calhau2020}. The work of \cite{sobral2018b}, in particular, pointed out the co-existence of two different classes of luminous $\rm z\!\sim\!2-3$ LAEs at roughly $\rm3\,L^*$, namely dust-free, highly star-forming galaxies and AGN. In addition, a significant fraction (at least $\gtrsim20\%$) of bright LAEs selected by \cite{matthee2017b} and \cite{sobral2018}, respectively on the Bo\"otes and COSMOS fields (with areas of $\rm\sim\!0.7\,deg^2$ and $\rm\sim\!2\,deg^2$) show X-Ray counterparts, which strongly points towards confirming them as AGN/QSOs. Finally, \cite{calhau2020} shows how the fraction of AGN/QSOs within a sample of $\rm z\!>\!2$ \lya$\!\!$-emitting candidates approaches $\sim\!100\%$ at $\rm L_{Ly\alpha}\gtrsim10^{\,43.5}\,erg\,s^{-1}$.

We broadly expect the above findings to hold valid over the much wider area of DR1 (bigger by a factor of $\sim\!500$), hence to select a mixture of extremely \lya$\!\!$-bright, rare star-forming galaxies \citep[e.g.,][]{sobral2016a,hartwig2016b,cai2017a,shibuya2018b,cai2018,marques-chaves2019} and luminous AGN/QSOs, numerically dominated by the latter source class. Indeed, our work selects objects showing strong and reliable NB excess, without employing any further criterion to disentangle its nature. Figure \ref{fig:LAEspectra_3FM} shows typical spectra of high-z SF galaxies and QSOs, pointing out their significant diversity \citep[see e.g.,][for a comparison with narrow-line AGN spectra]{hainline2011}. Ideally, this difference should be mirrored by bi-modalities in the photometric properties of our selected samples, assuming that i) both the \lya emitting source classes are significantly present in our selection results and ii) J-PLUS filters can effectively capture their spectral difference. For generality, we conduct our analysis by considering all the sources in our selected samples as \lya$\!\!$-emitting candidates (LAE candidates, in brief). We then look for eventual bi-modalities in their photometric properties as hints for the presence of two distinct classes of objects. Where needed, we explicitly refer to the two categories of \lya$\!\!$-emitting sources as either QSOs or SF LAEs to clearly state this distinction.

\subsubsection{Morphology of \lya$\!\!$-emitting sources in J-PLUS data}
\label{sec:morpho}
Due to resonant scatter of \lya photons by neutral hydrogen, SF LAEs can be surrounded by faint \lya$\!\!$-emitting halos and then appear more extended at \lya wavelengths than in their continuum \citep[e.g.,][but see also \citealt{bond2010,bond2012} and \citealt{feldmeier2013}]{moller_warren1998,fynbo_moller_thomsen2001,fynbo2003,nilsson2009a,finkelstein2011a,guaita2015,wisotzki2016,shibuya2019}. As shown in Sect. \ref{sec:jplus_definition_and_catalogs}, the DR1 dual-mode catalog is based on detection in $r$-band, which probes UV-continuum wavelengths in the rest-frame of $\rm z\!\gtrsim\!2.2$ sources. UV observations show typical rest-frame half-light radii of about $\rm r_{50}\!\lesssim\!2\,kpc$ for $\rm z\!\gtrsim\!2$ SF LAEs \citep[see e.g.,][]{venemans2005b, taniguchi2009, bond2009, bond2012, kobayashi2016, ribeiro2016, paulino-afonso2017, paulino-afonso2018}. This translates into apparent sizes comparable to the spatial resolution of T80cam ($\rm R=0.5$"$/pixel$) and to the typical J-PLUS seeing \citep[i.e. $s\lesssim1''$,][]{cenarro2019}. Since QSOs are point-like by definition, we then expect both SF LAEs and QSOs at $\rm 2.2\!\lesssim\!z\!\lesssim\!3.3$ to show compact morphology in the J-PLUS $r$ band. Section \ref{sec:morphology_cut} details how we exploit this assumption to look for potential low-z interlopers.

Furthermore, the extended \lya halos of SF LAEs are usually characterized by low surface brightness and hence observed by means of very deep NB imaging \citep[e.g. $\rm m_{NB}\gtrsim26-27$, see][]{leclerq2017,badescu2017,erb_steidel_chen2018} or IFU surveys \citep[e.g.,][]{bacon2015,drake2017}. This also applies to the peculiar class of high-z \lya$\!\!$-emitting systems showing rest-frame very extended ($\rm d\gtrsim20-30$ kpc) and bright ($\rm L_{Ly\alpha}\!>\!10^{\,43}\,erg\,s^{-1}$) \lya emission, namely Ly$\alpha$-nebulae or \textit{blobs} \citep[i.e. LABs, see e.g.,][]{matsuda2004,bridge2013,ao2015,cai2017b,cantalupo2019,lusso2019}. Despite extended \lya emission
being usually too faint for J-PLUS detection limits, extremely rare but sufficiently bright \lya$\!\!$-emitting extended sources might still be observed within the very large area of J-PLUS DR1. These should be targeted by not relying on dual-mode catalogs but instead on analysing the 511 continuum-subtracted NB images of J-PLUS DR1 and applying specific source extraction criteria \citep[as in e.g.,][]{sobral2018}. Nevertheless, we did not focus on these tasks since they deserve a separate and detailed analysis which lies outside the goals of this work.

\section{\lya$\!$-emitting candidates selection}
\label{sec:selection}
In order to select our candidates from the J-PLUS DR1, we first look for secure \textit{NB emitters} (i.e. objects showing a reliable NB excess) for each of the four NBs we use. We then exploit cross-matches with external databases and the remaining J-PLUS NBs to remove low-z interlopers. Our selection rules are detailed in Sect. \ref{sec:selection_rules} and \ref{sec:sample_cleaning}, while the following section explains how we target \lya emission with J-PLUS NBs.

\subsection{Detection of NB excess with a set of three filters}
\label{sec:3filters}
Our method to estimate the eventual NB excess for all DR1 sources and assess its significance is based on the works of \cite{vilellarojo2015} and \cite{logrono2019} which parallel well-established methodologies \citep[see e.g.,][]{venemans2005b, pascual2007, gronwall2007, guaita2010}. 
We employ sets of three filters composed as: [NB; $g$; $r$], where NB stands for either $J$0395, $J$0410, $J$0430 or $J$0515. By using spectroscopically identified $\rm z\!>\!2$ QSOs, we checked that filter-sets defined as [NB; $u$; $g$] provide less accurate \lya flux measurements than [NB; $g$; $r$]. As detailed in \cite{vilellarojo2015}, our method assumes that:
\begin{enumerate}
    \item the emission line profile can be approximated by a Dirac-delta centered at a given wavelength $\rm\lambda_{EL}$,
    \item  the source continuum is well traced by a linear function over the wavelength range covered by the three filters. \label{hypo:two}
\end{enumerate}

\noindent
Hypothesis \ref{hypo:two} implies that NB measurements affected by an emission line should exhibit a photometric excess with respect to the straight line graced by $g$ and $r$ photometry (see Fig. \ref{fig:LAEspectra_3FM}). The goal of our method is to measure this excess and relate it to the line flux which is producing it.

All the NBs we use share their probed wavelength ranges with the $g$ filter, hence the eventual emission-line flux would affect also the $g$ measurement and must be removed in order to estimate the source continuum. As detailed in appendix \ref{sec:append:3FM_equations}, we combine the NB, $g$ and $r$ fluxes (respectively $\rm f_{\lambda}^{\,NB}$, $\mathrm{f}_\lambda^{\,g}$ and $\mathrm{f}_\lambda^{\,r}$)\footnote{Throughout the paper, all the \textit{flux-density} measurements indicated by $\rm f_\lambda$ are expressed in $\rm f_\lambda$ units, i.e. $\rm erg\,cm^{-2}\,s^{-1}\,\text{\AA}^{-1}$. Capital $\rm F$, on the other hand, denotes \textit{integrated} flux in units of $\rm erg\,cm^{-2}\,s^{-1}$.} to estimate the line-removed continuum-flux in the $g$ and NB filters (respectively $\rm f_{\lambda\,;\,cont}^{\,\mathit{g}}$ and $\rm f_{\lambda\,;\,cont}^{\,NB}$). In this way, we can estimate the eventual NB excess due only to an emission-line as:
\begin{equation}
    \rm \Delta m^{NB} = m^{NB}_{cont} - m^{NB} = 2.5\,Log\,\left(\frac{ f_{\lambda}^{\,NB} }{ f_{\lambda\,;\,cont}^{\,NB} }\right)\ ,
    \label{eq:deltaNB}
\end{equation}
where the last equality follows by the definition of AB magnitudes
$\rm f^{\,NB}_\lambda$ ($\rm m^{NB}$) is the total NB flux (magnitude) including continuum \textit{and} line contributions, while $\rm f^{\,NB\,;\,cont}_\lambda$ ($\rm m^{NB}_{cont}$) is the continuum-only NB flux (magnitude), shown as a yellow square in Fig. \ref{fig:LAEspectra_3FM}. $\rm\Delta m^{NB}$ is an indirect probe of $\rm F_{Ly\alpha}$, i.e. the continuum-subtracted integrated line flux
emitted by a given source. As fully detailed in appendix \ref{sec:append:3FM_equations}, by introducing the coefficients
\begin{equation}
    \rm \alpha_x = \frac{\int\lambda^2\,T_\lambda^{\,x}\,d\lambda}{\int T_\lambda^{\,x}\,\lambda\,d\lambda} \quad;\quad\quad \beta_x = \frac{T_{\lambda}^{\,x}(\lambda_{EL})\,\,\lambda_{EL}}{\int T_\lambda^{\,x}\,\lambda\,d\lambda}\ ,
    \label{eq:alphabeta_params_3FM}
\end{equation}
which only depend on the transmission curve of a given filter ``x'' (i.e. $\rm T^{\,x}_\lambda$) and on $\rm\lambda_{\,EL}$ (i.e. the wavelength position of the line-peak in the NB), our methodology can directly estimate $\rm F_{Ly\alpha}$ via the quantity:
\begin{equation}
    \mathrm{ F^{\,3FM}_{Ly\alpha} } = \frac{ \left(\mathrm{f}^{\,g}_\lambda - \mathrm{f}^{\,r}_\lambda\right) + \frac{\alpha_r\, -\  \alpha_g}{\alpha_{\mathrm{NB}}\, -\  \alpha_{r}}\cdot\left(\mathrm{f}_{\mathrm{NB}} - \mathrm{f}^{\,r}_\lambda\right)}{ \beta_{g} + \frac{\alpha_{r}\, -\ \alpha_{g}}{\alpha_{\mathrm{NB}}\, -\ \alpha_{r}}\cdot\beta_{\mathrm{NB}} }\ .
    \label{eq:line_flux}
\end{equation}
We use $\rm\Delta m^{NB}$ for selecting reliable NB excesses (section \ref{sec:selection_rules}), while $\rm F^{3FM}_{Ly\alpha}$ for computing the luminosity of our candidates (section \ref{sec:lumifunc}). In Eq. \ref{eq:line_flux}, the superscript 3FM (as in three-filters method) points out that our method provides a photometric \textit{estimate} of $\rm F_{Ly\alpha}$. The biases affecting $\rm F^{\,3FM}_{Ly\alpha}$ are addressed in Sect. \ref{sec:line_flux_retrieval}.

Figure \ref{fig:LAEspectra_3FM} graphically explains our method, when applied to both a SF LAE and a QSO spectrum\footnote{From VUDS public data \citep[see][]{lefevre2015,tasca2017}}. In general, SF LAEs show narrow \lya$\!\!$-line profiles as opposed to QSOs, whose emission can easily cover (observed) intervals of few $\sim\!100\text{\AA}$. This implies that part of QSOs' \lya flux can lie outside the NB wavelength coverage, hence might be undetected by J-PLUS NBs. The importance of this bias on $\rm F^{\,3FM}_{Ly\alpha}$ depends on e.g. line profile details and the position of its peak in the NB. In turn, these are determined by a number of complex aspects, such as the QSOs accretion status \citep[e.g.,][]{calhau2020}, the transfer of \lya photons in the hydrogen-rich ISM and IGM \citep[e.g.,][]{dijkstra2017, gurung2018} or the sources' metals and dust content \citep[e.g.,][]{christensen2012}. These details can be extracted by high-resolution spectroscopic data but not from J-PLUS photometry. For this reason, we apply Eq. \eqref{eq:line_flux} to all our selected candidates and then statistically correct $\rm F^{3FM}_{Ly\alpha}$ to account for the line-flux loss, as detailed in Sect. \ref{sec:filter_width_corr}.

\subsection{Selection Function}
\label{sec:selection_rules}
We extract our $\rm2.2\!\lesssim z\lesssim\!3.3$ LAE candidates from a parent sample of $\rm N\!\sim\!1.1\times10^7$ sources, obtained from the J-PLUS DR1 $r$-band selected, dual-mode catalog \citep[see Sect. \ref{sec:jplus_definition_and_catalogs} and][]{cenarro2019}. Our selection targets strong NB excesses with respect to the BB-estimated continuum and removes secure contaminants (see Sect. \ref{sec:sample_cleaning}). Its overall performance was significantly improved thanks to the spectroscopic follow-up programs described in Sect. \ref{sec:gtcProgram}. The selection results are presented in Sect. \ref{sec:LAE_samples}, while the implications of using dual-mode catalogs are addressed in Sect. \ref{sec:line_flux_retrieval} and \ref{sec:completeness}.

\begin{description}
\item[\textit{Magnitude cut in $g$ and $r$ bands}]{\ \\The photometry of too-bright or too-faint objects is likely to be either saturated or severely affected by noise. Hence we apply a very broad cut on $g$, $r$ magnitudes and their associated errors ($\rm\sigma_\mathit{g}$ and $\rm\sigma_\mathit{r}$), namely:\vspace{1mm}
\begin{itemize}
    \item[]{$14\leqslant$ $g$ $\leqslant24\ \bigwedge\ \rm\sigma_\mathit{g}<1$}\hspace{3mm};\hspace{3mm}$14\leqslant$ $r$ $\leqslant24\ \bigwedge\ \rm\sigma_\mathit{r}<1$.\vspace{1mm}
\end{itemize}
We check that these conditions do not significantly affect the final number of our candidates. Nevertheless, we account for eventual losses of continuum-faint sources with relatively bright \lya emission (see Sect. \ref{sec:completeness}). Spurious detections eventually included in these $g$ and $r$ intervals are removed by adequate SNR cuts (see below). \vspace{1.0mm}}

\item[\textit{Detection confirmation in the three-filters set}]{\ \\We additionally require single-mode detection in each of the NB, $g$ and $r$ bands, since all are necessary for our excess-detection method. For this, we exploit the detection flags provided by the DR1 database\footnote{For details see the information provided at:\\ \texttt{\tiny http://archive.cefca.es/catalogues/jplus-dr1/help\_adql.html}\normalsize}. This condition implies that we are only sensitive to low EW at faint \lya flux; we account for this in our completeness estimates (section \ref{sec:completeness}).\vspace{1.0mm}}

\item[\textit{Effective exposure time cut}]{\ \\The normalized effective exposure time $\rm t^{\,eff}_{\,exp}$ (provided in the DR1) can be used as a proxy for the number of exposures contributing to the photometry of each source. The limit $\rm t^{\,eff}_{\,exp}>0.5$ excludes objects whose detection is affected by the dithering pattern of J-PLUS pointings, which might compromise the removal of cosmic rays or their extraction process.\vspace{1.0mm}}

\item[\texttt{MANGLE} \textit{mask}]{\ \\Sources' photometry can be affected by optical artifacts or bright stars. J-PLUS makes use of the \texttt{MANGLE} software \citep[][]{swanson2008} in order to mask-out areas affected by these defects. For each of our selection, we apply the cumulative \texttt{MANGLE} mask associated to the three-filters [NB; $g$; $r$]. This reduces the total sky-coverage of our data to an effective area of $\rm A_{eff}\sim900\,deg^2$ (see Table \ref{tab:QSOcontamin} for details).}
\end{description}

\subsubsection{Pointing-by-pointing selection}
\label{sec:NBexcess_selection}
The combined action of the previous cuts produce four different lists (one per NB) of $\rm N\gtrsim2\times10^6$ sources each (see Table \ref{tab:selection_cuts_effect}). To proceed, we take into account that J-PLUS DR1 is composed by 511 different pointings (or \textit{tiles}) which exhibit e.g. varying depths, source counts and colors. Consequently, we apply the following conditions on each tile separately build a selection function as uniform as possible.

\begin{description}
\item[\textit{NB excess significance}]{\ \\
In order to select line-emitters candidates, we look for outliers in the $\rm \Delta m^{NB}$ vs. $\rm m^{NB}$ distribution of each tile, after considering photometric uncertainties \citep[as in e.g.,][and Fig. \ref{fig:colmag}]{bunker1995,fujita2003,sobral2009a,bayliss2011,matthee2017b}. In particular, using Eq. \ref{eq:deltaNB} we compute the error:
\begin{equation}
  \rm\sigma_{\Delta m^{NB}}(m^{NB}) = \sqrt{\rm\sigma_{m^{NB}_{cont}}^2 + \sigma_{m^{NB}}^2}\ ,
\end{equation}
and identify reliable NB-emitters as the objects satisfying:
\begin{equation}
    \rm \Delta m^{NB} > \Sigma\cdot\sigma_{\Delta m^{NB}}+\langle\Delta m^{NB}\rangle\ ,
\end{equation}
with $\Sigma=3$. We account for pointing variations by anchoring our cut to the average color $\rm \langle\Delta m^{NB}\rangle$ of each tile, which acts as a rigid offset. Figure \ref{fig:colmag} shows the results of this procedure on a J-PLUS tile with $\rm \langle\Delta m^{NB}\rangle=-0.27$. As expected, only $\lesssim\!10-15\%$ of our parent-sample pass this cut (see Table \ref{tab:selection_cuts_effect}).\vspace{1.0mm}}

\item[\textit{NB signal-to-noise}]{\ \\We explicitly exclude objects with low-SNR NB measurements by imposing $\rm m_{\,NB} \!>\! m^{NB}_{cut}$, where $\rm m^{NB}_{cut}$ is the NB magnitude at which the average NB SNR of each pointing is equal to $5$. This threshold is relatively impacting on the whole DR1, since only $\sim35\%$ of sources is able to pass it. We checked that imposing $\left\langle\rm SNR\right\rangle=3$ would lead to significantly higher contamination of our selected samples.\vspace{1.0mm}
\begin{figure}[t]
\centering
\includegraphics[width=0.49\textwidth]{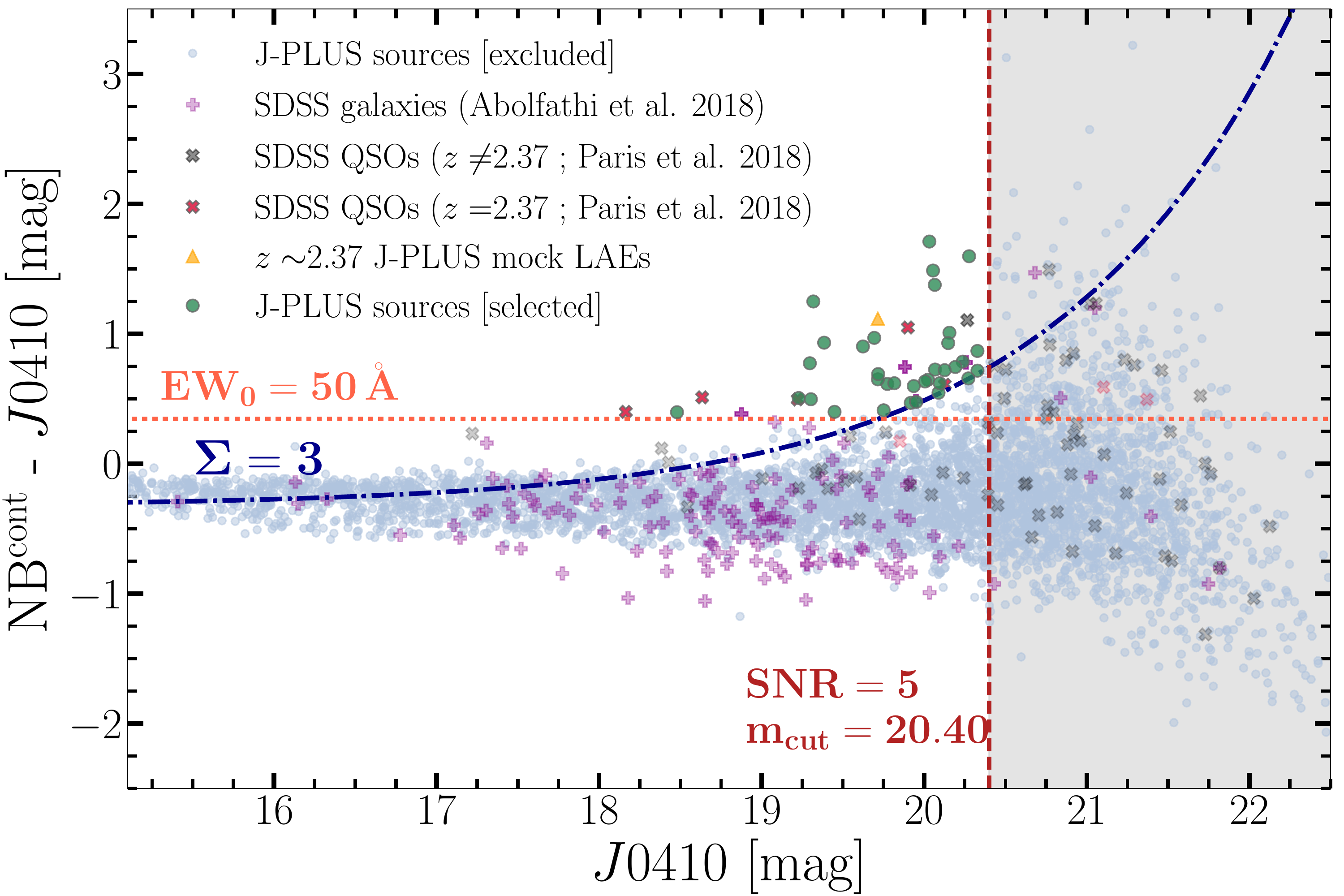}
\caption{\footnotesize Example of a color-magnitude diagram obtained for the NB filter $J$0410 on a DR1 pointing (out of 511). Our photometric cuts are summarized as follows: the blue dashed-dotted line shows the $\rm\Delta m^{NB}$-significance threshold, while the vertical red line marks the NB SNR limit. We exclude sources below the blue dashed-dotted line and inside the grey shaded area. The orange horizontal dotted line shows $\rm\Delta m^{NB}$ associated to $\rm EW = 50\text{\AA}$ (see Eq. \ref{eq:mincolor}). Grey-blue dots mark all the J-PLUS detections in the pointing, while red and purple crosses show $\rm z\!\sim\!2.4$ QSOs and low-z galaxies from \texttt{SDSS DR14}. Yellow triangles show J-PLUS mock data of $\rm z\!\sim\!2.4$ SF LAEs \citep[][]{izquierdo-villalba2019}. Finally, our \lya$\!\!$-emitting candidates are shown as green dots.}
\label{fig:colmag}
\end{figure}}

\item[\textit{BB signal-to-noise}]{\ \\Clean BB photometry is a key requirement to estimate the sources NB excess. We exclude objects with $g$ $\rm\!>\!g_{cut}$ and $r$ $\rm\!>\!r_{cut}$, where $\rm g_{cut}$ and $\rm r_{cut}$ are defined as the magnitudes at which $\rm\langle SNR\rangle=5$ in each BB and pointing. Despite its effect on the parent samples being small (see Table \ref{tab:selection_cuts_effect}), this cut might exclude genuine continuum-faint candidates with bright \lya. We account for this as described in Sect. \ref{sec:completeness}.\vspace{1.0mm}}

\item[\textit{Minimum NB-color cut}]{\ \\
In principle, \lya can be distinguished from e.g. \texttt{CIV} and \texttt{CIII]} of AGN/QSOs spectra \citep[e.g.,][]{stroe2017a, stroe2017b} or nebular \texttt{H}$\,_\beta$, \texttt{[OIII]}$\!_{\,4959+5007}$ and \texttt{[OII]}$\!_{\,3727}$ by exploiting its generally higher intrinsic strength and EW \citep[e.g.,][]{vandenberk2001,hainline2011,selsing2016,nakajima2018b}. Therefore, we impose a NB-color cut defined by assuming a minimum \textit{rest-frame} EW for our candidates \citep[as in, e.g. ][]{fujita2003,gawiser2006,gronwall2007,hayes_schaerer_ostlin2010,adams2011,clement2012,santos2016}. Observed- and rest-frame EWs (respectively $\rm EW_{obs}$ and $\rm EW_0$) are related via:
\begin{equation}
    \rm EW_{obs} = EW_0\ (1+z)\ .
    \label{eq:ew}
\end{equation}
We set $\rm EW_0^{\,min}=50\,\text{\AA}$ and obtain the corresponding $\rm EW_{obs}^{\,min}$ from Eq. \eqref{eq:ew}. We then link $\rm EW_{obs}$ and $\rm\Delta m^{NB}$ (defined in Eq. \ref{eq:deltaNB}) with the analytic expression:
\begin{equation}
    \rm \Delta m^{NB}_{min} = 2.5\,Log\,\left(1 + \beta_{NB}\cdot EW_{obs}^{min}\right)+\langle m^{NB}\rangle\ ,
    \label{eq:mincolor}
\end{equation}
\citep[see][and appendix \ref{sec:append:3FM_equations}]{guaita2010}, where $\rm\beta_{NB}$ is defined in Eq. \ref{eq:alphabeta_params_3FM} and $\rm\langle m^{NB}\rangle$ is the average color in the tile. By requiring $\rm\Delta m^{NB}\!>\!\Delta m^{NB}_{min}$ (orange horizontal dotted line in Fig. \ref{fig:colmag}) we exclude $\gtrsim96\%$ of our DR1 parent sample, since most sources do not show line-emission. We note that the choice of $\rm EW_0^{\,min}$ has a certain degree of arbitrariness indeed past works have explored a wide range of limiting values \citep[see e.g.,][]{gronwall2007,ouchi2008,bond2009,nilsson2009a,guaita2010,konno2016,matthee2016,badescu2017,sobral2017}. We fix $\rm EW_0=50\text{\AA}$ after checking our EW estimates on publicly-available spectroscopic catalogs of $\rm z\!\gtrsim\!2$ SF LAEs and QSOs \citep[namely \texttt{DR14}, VUDS and VVDS][]{cassata2011,lefevre2015,paris2018} and on the confirmed $\rm z\!\sim\!2$ QSOs in our follow-up data (see Table \ref{tab:GTC_obs_summary} in Sect. \ref{sec:gtcProgram}). In particular, $50\text{\AA}$ provides a good compromise between the retrieval of $\rm z\!\gtrsim\!2$ sources and the exclusion of $\rm z\!<\!2$ interlopers. We note that this relatively high $\rm EW_0^{\,min}$ is still close to the lower limits of EW distributions usually measured for high-z \lya$\!$-emitting sources \citep[e.g.,][]{nilsson2009b,bond2012,amorin2017,hashimoto2017,santos2020}. Besides, low EWs can be accessed with very-narrow bands ($\rm FWHM\!\lesssim\!50\text{\AA}$) and deep observations \citep[$r\!>\!22$, e.g.,][]{sobral2017}, which both act as limiting factors in our case. Finally, we stress that this condition is not directly applied on $\rm EW_0$, hence it does not pose a strict limit on the measured $\rm EW$ of our candidates \citep[see][for a similar discussion]{ouchi2008}.}
\end{description}
\noindent
These cuts select respectively 12251, 19905, 24813 and 15213 objects for $J$0395, $J$0410, $J$0430 and $J$0515 NBs (i.e. $<\!1\%$ of the parent catalog, see Table \ref{tab:selection_cuts_effect}). These samples are still likely to be contaminated by interlopers, such as lower-z QSOs, ELGs and faint blue stars, which are usually targeted with BB-based color cuts \citep[e.g.,][]{ross2012, ivezic2014, peters2015, richards2015}. We checked that, in our case, these methods significantly affect also the number of selected $\rm z\!\gtrsim\!2$ QSOs from SDSS \texttt{DR14}. We hence decided to drop any color cut because of its non-trivial effect on our selection.
\begin{table*}[t]
    \centering
    \resizebox{18.4cm}{!}{
    \begin{tabular}{ c|c|c|c|c|c||c }
    \hline
	\hlx{v}
	Filters  &  \texttt{DR1} parent sample  &  $\rm\Delta m^{\,NB}$ significance  &  NB SNR  &  BB SNR  &  $\rm EW_{obs}^{\,min}$  &  First selection \\
	\hlx{v}
	\hline
	\hline
	\hlx{v}
	$J$0395 & 2,036,657 & 348,613 (17.1\%) & 1,324,373 (65.0\%) & 2,017,720 (99.1\%) & 57,800 (2.8\%) & 12,251 (0.6\%)\\
	\hlx{v}
	$J$0410 & 2,730,135 & 232,753 (8.5\%) & 1,846,144 (67.6\%) & 2,679,515 (98.2\%) & 150,321 (5.5\%) & 19,905 (0.7\%)\\
	\hlx{v}
	$J$0430 & 3,015,684 & 235,685 (7.8\%) & 2,024,629 (67.2\%) & 2,930,026 (97.2\%) & 173,388 (5.8\%) & 24,813 (0.8\%)\\
	\hlx{v}
	$J$0515 & 4,520,911 & 244,550 (5.4\%) & 2,956,154 (65.4\%) & 3,797,178 (84.0\%) & 143,662 (3.2\%) & 15,213 (0.3\%)\\
	\hlx{v}
	\hline
    \end{tabular}
    }
    \caption{\footnotesize{Number counts of sources passing each cut of our selection, for the four J-PLUS NBs we use. Here we report the effects of each cut separately to highlight its effect, hence the fractions reported in the Table do not add to 100\%. The most impacting cuts are those on $\rm EW_{obs}^{\,min}$ cut and on NB excess significance. The number of sources passing all these conditions, for each NB, produce our final samples of LAE candidates and is shown in the last column to the right. These \textit{partial} samples are likely to be highly contaminated by interlopers showing reliable NB excess. Table \ref{tab:interlopers_and_final_candidates} shows a summary of the samples contamination and the final number of selected sources.}}
    \label{tab:selection_cuts_effect}
\end{table*}

\subsection{Removal of residual contaminants}
\label{sec:sample_cleaning}
Despite efficiently identifying NB-emitters, the conditions in Sect. \ref{sec:selection_rules} might also select line-emitting interlopers (see Sect. \ref{sec:jplus_NB_and_emission_lines_contamination}). Previous works based on similar methods have usually explored limited sky regions already surveyed by deep multi-wavelength data, which supported the identification of contaminants \citep[e.g. COSMOS, UDS, SXDS, SA22 and Bo\"otes fields, see][]{warren2007,scoville2007,furusawa2008,geach2008,kim2011,bian2012,stroe_sobral2015}. Unfortunately, few previous surveys uniformly cover the very wide area of J-PLUS DR1, hence limiting our ability to identify contaminants.

\subsubsection{Cross-matches with public external databases}
Interlopers with a secure identification (either spectroscopic, astrometric or photometric) can be removed via cross-matches with public catalogs. We employ a radius of $\rm r_{max}^{\,match} = 3.5$" after checking that this provides a high matching completeness while keeping low the number of multiple matches, for all the matched databases. More in detail, we recover the 80\% (95\%) of all QSOs from SDSS \texttt{DR14} (within the DR1 footprint) respectively at $r\!\sim\!21.25$ ($r\!\sim\!20.80$) and $\rm Log\,(L_{Ly\alpha})\!\sim\!44.25$ ($\sim\!44.70$).

\begin{description}
\item[\textit{SDSS DR14}]{\ \\We exploit the lists of spectroscopically-identified galaxies \citep[][]{bundy2015, hutchinson2016}, stars \citep[][]{majewski2017} and QSOs \citep[][]{paris2018} provided by the recent \texttt{SDSS-IV DR14} \citep[\texttt{DR14} hereafter,][]{blanton2017, abolfathi2018}.
Given the wide overlap with J-PLUS DR1 and the higher depth of \texttt{DR14} \citep[][]{cenarro2019}, this cross-match ensures the removal of secure contaminants from our selection. As discussed in Sect. \ref{sec:jplus_NB_and_emission_lines_contamination}, QSOs can act as both interlopers and genuine candidates depending on their z, hence we need to rely on a list of \textit{securely} identified QSOs. The \cite{paris2018} catalog includes $\rm N\gtrsim5.3\times10^5$ sources observed by \texttt{BOSS} and \texttt{eBOSS} surveys \citep{dawson2013, dawson2016} and confirmed as QSOs by careful inspection. We keep genuine \lya$\!\!$-emitting sources at the z sampled by each NB, while the rest are identified as contaminants and removed. The cross-match with \texttt{DR14} shows a generally low contamination (table \ref{tab:interlopers_and_final_candidates}), with low-z galaxies accounting respectively for 5.1\%, 4.3\%, 5.3\% and 3.1\% of our $J$0395, $J$0410, $J$0430 and $J$0515 NB samples. On the other hand, the $\rm z\!\lesssim\!2$ QSOs fraction drops from 11.1\% to 0.3\%, paralleling the drop of \texttt{DR14} $\rm z\!\gtrsim\!2.2$ QSOs. Finally, SDSS stars account for $\lesssim\!2\%$ of our samples. These fractions are likely to be underestimated, given the different depth of the two surveys and eventual mis-matches between \texttt{DR14} and DR1 catalogs. Nevertheless, being measured on spectroscopically confirmed sources, these are secure contamination estimates.\vspace{1.5mm}}

\item[\textit{Gaia DR2}]{\ \\Our spectroscopic follow-up program 2018A (see Sect. \ref{sec:gtcProgram}) showed a non-negligible contamination from stars in our samples. To limit this issue, we built a specific criterion for excluding stars, based on the very accurate measurements offered by \texttt{Gaia DR2} data \citep[][]{brown2018}. Since the latter do not include source classification, we define \textit{secure stars} by using the significance of their proper-motion assessments. More in detail, we exclude the J-PLUS sources with a counterpart in \texttt{Gaia DR2}, showing significant measurements ($\sigma\!>\!3$) in each proper motion component, i.e.:
\begin{equation}
\rm\sigma_{pm}=\sqrt{\rm \sigma_{pmra}^2+\sigma_{pmdec}^2+\sigma_{\mu}^2}>\sqrt{27}\sim5.2\ ,
\end{equation}
where $\rm\sigma_{pmra}$, $\rm\sigma_{pmdec}$ and $\rm\sigma_{\mu}$ are respectively the errors on proper motion (\texttt{ra} and \texttt{dec}) and parallax.
With this cut, we explicitly remove objects showing significant apparent motion from our list of LAE candidates. The good performance of this criterion was confirmed by the results of our second follow-up program, whose targets were selected from the results of our updated pipeline (see  Sect. \ref{sec:gtcProgram} for details). The contamination from \texttt{Gaia DR2} is presented in Table \ref{tab:interlopers_and_final_candidates}.\vspace{1.5mm}}

\item[\textit{GALEX-UV}]{\ \\ \lya$\!\!$-emitting sources at $\rm z\!>\!2$ are generally expected to appear faint at (observed) UV wavelengths due to the dimming action of the \lya$\!$-break and Lyman-break \citep[e.g.,][]{steidel1992, steidel1996a, steidel1999a, shapley2003}. On the contrary, $\rm z\!<\!2$ AGN/QSOs, blue stars and low-z star-forming galaxies can show significant UV emission. We exploit this property for removing $\rm z\!<\!2$ interlopers by cross-matching our catalogues with \texttt{GALEX} all-sky UV observations \citep[][]{gildepaz2009}. In particular, we remove sources with a $\rm SNR\!>\!3$ detection in either of the two FUV and NUV \texttt{GALEX} bands \citep[see e.g.,][]{ciardullo2012}. Table \ref{tab:interlopers_and_final_candidates} shows the fraction of interlopers identified with this cross-match in each NB. In order to check our assumption according to which only $\rm z\!<\!2$ sources are expected to be significantly observed in UV, we additionally matched the J-PLUS sources with counterparts in \texttt{GALEX} to the spectroscopic sample of \texttt{DR14}. This analysis confirmed that $>\!99.5\%$ of sources with UV-bright \texttt{GALEX} detection show a spectroscopic $\rm z\!<\!2$, hence act as contaminant in our selection.\vspace{1.5mm}}

\item[\textit{LQAC-3}]{\ \\The third release of the Large Quasar Astrometric Catalog \citep[][]{souchay2015a, souchay2015catalog} is a complete archive of spectroscopically identified QSOs. By combining data from available catalogs, it provides the largest complement to the \texttt{DR14} list \citep{paris2018}. We exclude sources included in LQAC-3 with spectroscopic z lying outside the range probed by each NB. As expected, this step identifies only few additional interlopers (see Table \ref{tab:interlopers_and_final_candidates}).}
\end{description}

\subsubsection{Multiple NB excesses}
\label{sec:multiNB_excess}
We target additional interlopers by exploiting the whole set of J-PLUS NBs. In particular, we look for LAE candidates showing significant excesses (with respect to adjacent BBs) in the six NBs not used for their selection. Indeed, we expect SF LAEs to not show any additional NB feature \citep[e.g.,][]{shapley2003, nakajima2018b}, while QSOs at the targeted z can exhibit only particular combinations of NB excesses.
\begin{figure}[t]
\centering
\includegraphics[width=0.47\textwidth]{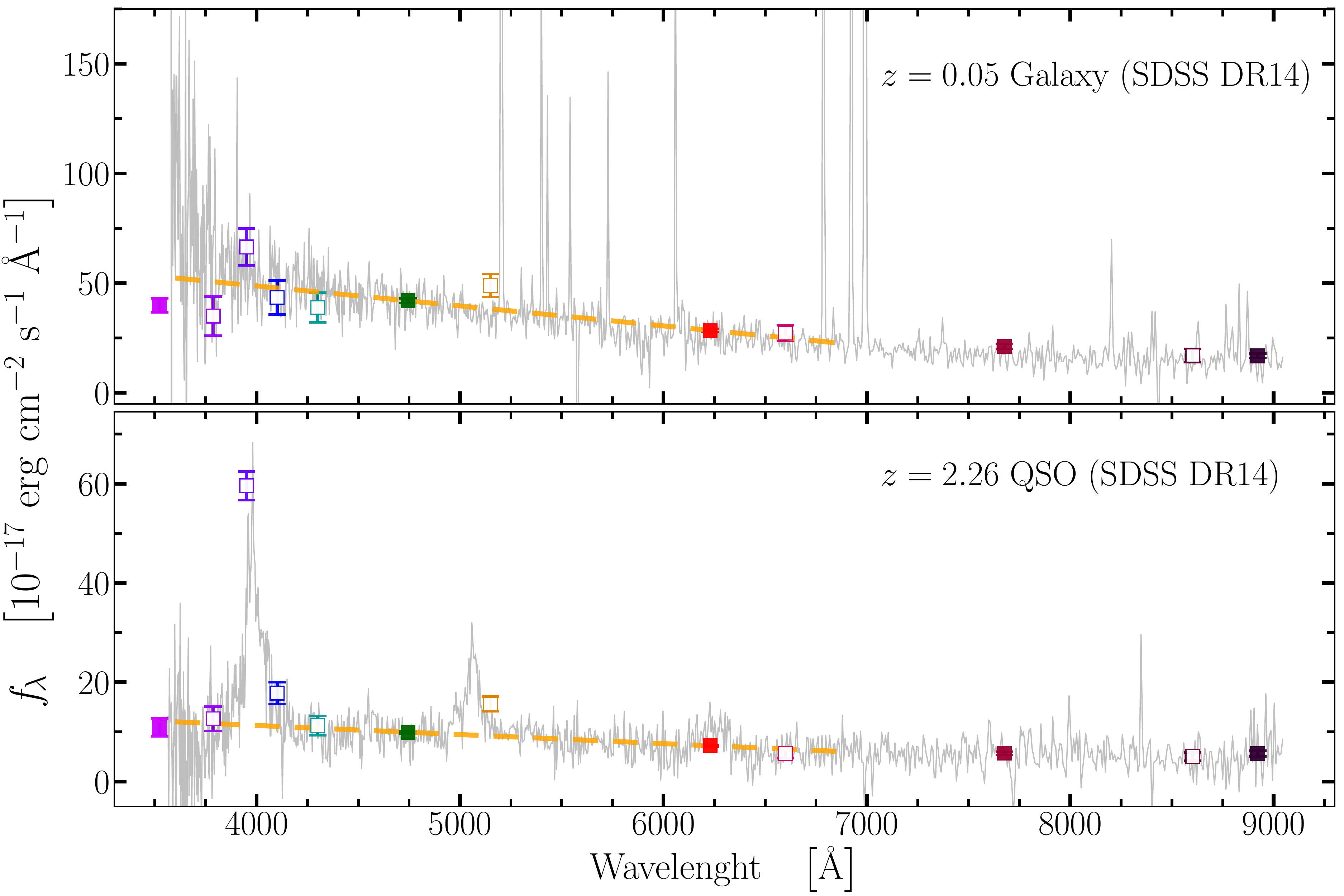}
\caption{\footnotesize Examples of multiple NB excess in J-PLUS photo-spectra. Empty and filled squares respectively mark NB and BB photometry, while the dashed yellow line shows the linear continuum we estimate through $g$ and $r$ BBs (in green and red, respectively). This comparison shows that both a $\rm z\!\sim\!0.05$ galaxy (upper panel) and a $\rm z\!\sim\!2.25$ QSO (lower panel) exhibit significant excesses in $J$0395 and $J$0515 NBs (respectively second and fifth  empty squares from the left). The $J$0515 excess is produced respectively by \texttt{H}$\,_\beta$ at $\rm z\!\sim\!0.05$ and \texttt{CIV} line at $\rm z\!\sim\!2.25$, but its nature is hardly distinguishable by J-PLUS photometry.}
\label{fig:multiNB}
\end{figure}
Consequently, we remove the sources showing multiple excesses not compatible with $\rm z\!>\!2$ spectral features \citep[e.g.,][]{matthee2017b}. On the other hand, sources showing multiple excesses compatible with $\rm z\!>\!2$ sources can hardly be separated into different classes by J-PLUS data. As an example, Fig. \ref{fig:multiNB} shows the photo-spectra of a $\rm z\!\sim\!0.05$ galaxy (upper panel) and a $\rm z\!\sim\!2.25$ QSO (bottom panel) from the \texttt{DR14} spectroscopic samples. Both sources show simultaneous excesses in $J$0395 and $J$0515 filters (respectively purple and yellow empty squares) with respect to the linear continuum traced by $g$ and $r$ BBs (yellow dashed line). On top of this, both photo-spectra exhibit comparable BB colors and might hence be confused by our selection. Since we are not able to directly measure this source of contamination, we estimate a statistical correction as explained in Sect. \ref{sec:purity_of_samples}.
\begin{table*}[t]
    \centering
    \resizebox{18.0cm}{!}{
    \begin{tabular}{ c|c|c|c|c|c|c|c||c }
    \hline
    \hlx{v}
	Filters  &  First selection  &  SDSS spectra  &  \texttt{GALEX}  &  \texttt{Gaia DR2} stars  &  LQAC QSOs  &  Multiple NB & Extended &  Final [N; $\rm deg^{-2}$]\\
	\hline
	\hline
	\hlx{v}
	$J$0395 & 12,251 & 2,192 (17.9\%) & 2,003 (16.4\%) & 857 (7.0\%) & 87 (0.7\%) & 1,312 (10.7\%) & 6,307 (51.5\%) & 2,547 ; 2.8\\
	\hlx{v}
	$J$0410 & 19,905 & 1,983 (9.9\%) & 2,003 (10.1\%) & 2,738 (13.8\%) & 56 (0.3\%) & 16,48 (8.3\%) & 9,557 (48.0\%) & 5,556 ; 6.2 \\
	\hlx{v}
	$J$0430 & 24,813 & 2,083 (8.4\%) & 2,597 (10.5\%) & 2,441 (9.8\%) & 40 (0.2\%) & 3,313 (13.4\%) & 15,468 (62.3\%) & 4,994 ; 5.6 \\
	\hlx{v}
	$J$0515 & 15,213 & 523 (3.4\%) & 1,249 (8.2\%) & 531 (3.5\%) & 7 (0.05\%) & 1,282 (8.4\%) & 12,992 (85.4\%) & 1,467 ; 1.5 \\
	\hlx{v}
	\hline
    \end{tabular}
    }
    \caption{Number counts (and fractions) of secure interlopers among the sources passing our photometric selection, for each J-PLUS NB we use (see discussion in Sect. \ref{sec:sample_cleaning}). We note that the extended fraction of our samples is particularly high for the $J$0515 NBs, suggesting that this filter is affected by high level of contamination from extended low-z interlopers. Indeed, this is the only NB among the four which is susceptible to contamination from the strong \texttt{[OIII]}$\,_{4959+5007}$ doublet and \texttt{H}$\,_\beta$ line, in addition to \texttt{[NeIII]} and \texttt{[OII]}. Sources with at least one identification as secure interloper are excluded; the final number counts of \lya$\!\!$-emitting candidates are shown in the last column to the right. The average sky density of these objects shows significant variation among the four lists, with an average of $\rm\sim4\,deg^{-2}$ sources, per filter. The complete catalogs of genuine candidates (i.e. after excluding securely identified interlopers) can be found at: \texttt{https://www.j-plus.es/ancillarydata/dr1\_lya\_emitting\_candidates} }
    \label{tab:interlopers_and_final_candidates}
\end{table*}

\subsubsection{Morphological cut}
\label{sec:morphology_cut}
We expect our $\rm z\!\gtrsim\!2.2$ candidates to appear compact in J-PLUS data (see Sect. \ref{sec:morpho}), hence the candidates showing extended morphology are likely to be low-z interlopers. The DR1 catalog provides a morphological parameter $\mathcal{C}$ which allows to discriminate between compact ( $\mathcal{C}\!\sim\!1$) and extended objects \citep[$\mathcal{C}\!\sim\!0$, see][for details]{lopezsanjuan2019b}. By cross-matching the whole DR1 sample to SDSS spectroscopic catalogs of galaxies and QSOs, we checked that more than $\gtrsim90\%$ of galaxies in SDSS \citep[$\rm z\!\lesssim\!1$, see][]{hutchinson2016} and only $\lesssim5\%$ of \texttt{DR14} QSOs (at any z) are found at $\rm\mathcal{C}\leqslant0.1$. We then remove objects with $\rm\mathcal{C}\leqslant0.1$ from our selection. Table \ref{tab:interlopers_and_final_candidates} (previous-to-last column to the right) shows the abundance of extended sources in each of the four lists.

\subsection{Spectroscopic follow-up at the GTC telescope}
\label{sec:gtcProgram}
This section presents two spectroscopic follow-up programs executed at the Gran Telescopio Canarias (GTC) telescope\footnote{Observatorio del Roque de los Muchachos, La Palma, Canary Islands} in the semesters 2018A and 2019A. The spectroscopic confirmation of a sub-sample of our candidates allowed to assess the performance of our selection, to refine our methodology and to estimate its residual contamination. Overall, these programs confirmed 45 sources selected among our $J$0395 NB-emitters.

\subsubsection{Programs description}
To ensure uniform observations and comparable results, we performed the same target selection and required identical observing conditions for both programs (namely \texttt{GTC2018A} and \texttt{GTC2019A}). In particular, we randomly selected a sample of 24 (21) \lya$\!$-bright candidates ($\rm L_{Ly\alpha}\!>\!10^{\,43.5}\,erg\,s^{-1}$) for program \texttt{GTC2018A} (\texttt{GTC2019A}), spanning the entire luminosity range covered by our candidates. We stress that targets for \texttt{GTC2019A} were selected after refining our selection with the help of \texttt{GTC2018A} results. We requested to use the \texttt{OSIRIS} spectrograph and the R500B grism, in order to exploit its good spectral resolution ($\rm R\!\sim\!500$, which translates to $\rm\Delta\lambda\,pixel^{-1}\!\sim\!3.65\text{\AA}$ for the 0.8" slit width we requested). The exposure times for our targets were computed by assuming the observing conditions summarized in the header of Table \ref{tab:GTC_obs_summary} (appendix \ref{sec:append:spectraMeasurement}). These were calibrated to achieve $\rm SNR\!\geqslant\!3$ (in each $\lambda$ bin) over the whole \texttt{OSIRIS} spectral range, in order to identify eventual emission lines and measure their integrated flux. 

We limited our programs length to $<\!20$ hours, to ensure their completion. Due to the high observing times required by our targets, we followed-up only candidates selected by $J$0395 NB. The target selection balanced the total observing time and the uniform sampling of our candidates $\rm L_{Ly\alpha}$ distribution. Finally, we excluded objects with previous spectroscopic identifications (at any z). Our proposals were respectively awarded with 11.56 and 18.95 hours of observations and were both fully executed.

\subsubsection{Spectroscopic results}
\label{sec:GTC_program_results}
\begin{figure}[t]
\centering
\includegraphics[width=0.45\textwidth]{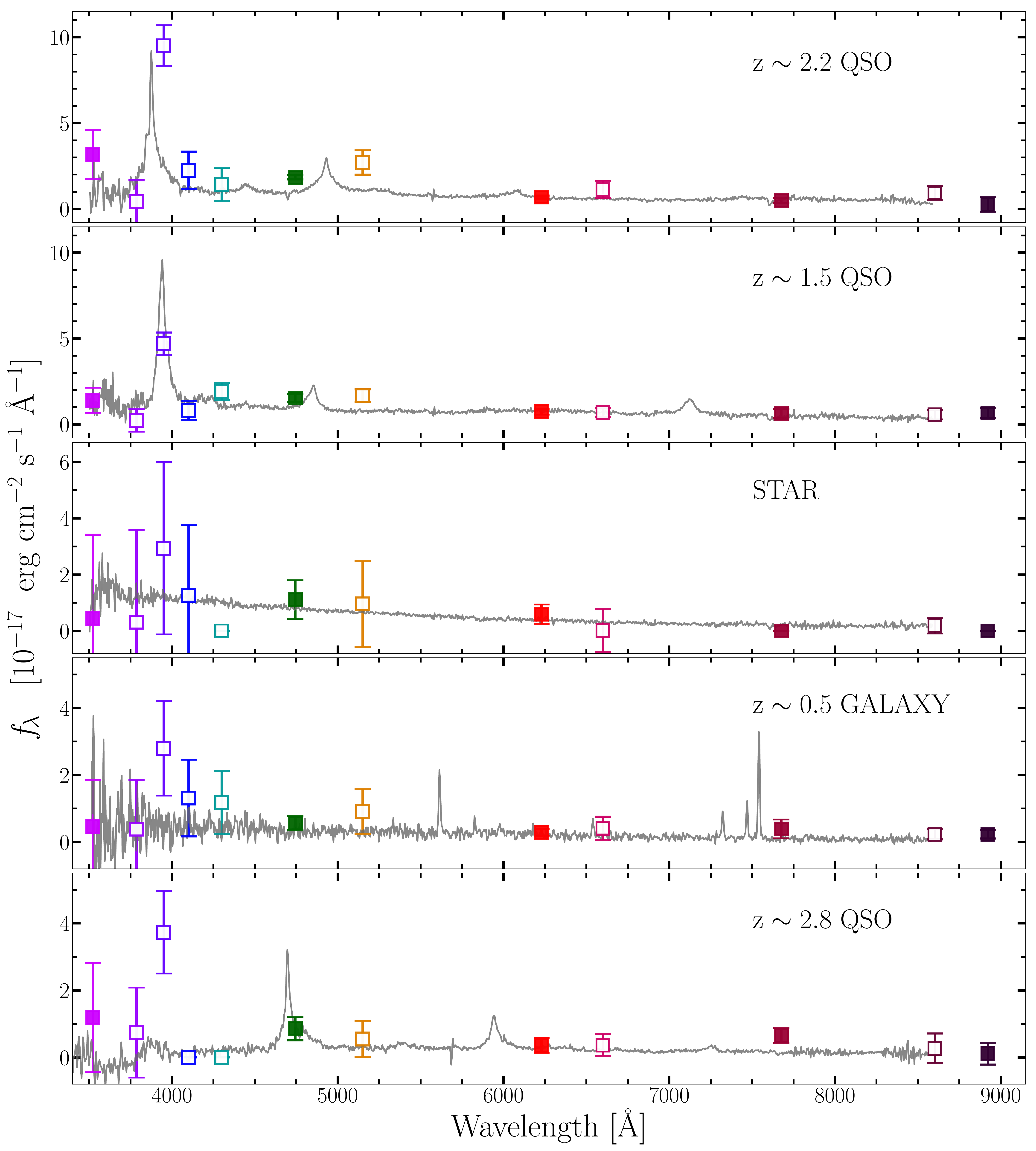}
\caption{\footnotesize Summary of our spectroscopic results, showing one spectrum (grey line in each panel) for each source classes identified in our target lists. From top to bottom: $\rm z\!\sim\!2.2$ QSO, $\rm z\!\sim\!1.5$ QSO, star, $\rm z\!\sim\!0.5$ galaxy and $\rm Ly\beta\!\!$-emitting QSO. The corresponding J-PLUS photometry is shown as coloured squares. The star and galaxy targets show low-significance excesses in $J$0395 NB (third square from the left). Indeed, these interlopers were selected as targets by the first version of our methodology, i.e. before applying the improvements due to \texttt{CTG2018A} results and the re-calibration of J-PLUS data \citep{lopezsanjuan2019a}.}
\label{fig:GTC_spectral_classes}
\end{figure}

The results of both programs are shown in Table \ref{tab:GTC_obs_summary}. Overall, we identified 29/45 targets ($64.4\%$) as genuine $\rm z\!\sim\!2.2$ \lya$\!\!$-emitting sources, 8/45 ($17.7\%$) as $\rm z\!\sim\!1.5$ QSOs emitting \texttt{CIV} at $\rm\lambda_{obs}\sim4000\,\texttt{\AA}$, 1 ($2.2\%$) $\rm Ly\beta$-emitting QSO at $\rm z\!\sim\!2.76$, 5 ($11.3\%$) blue stars and 2 ($4.4\%$) low-z galaxies selected because of their narrow emission lines. As an example, Fig. \ref{fig:GTC_spectral_classes} shows a spectra for each different source class together with its associated J-PLUS photometry.
Both $\rm z\!\sim\!2.2$ and $\rm z\!\sim\!1.5$ QSOs show prominent line emission at $\rm\lambda_{obs}\sim3950$ and are consequently selected as genuine $J$0395 NB-emitters. The same applies to the $\rm z\!\sim\!2.8$ QSO emitting $\rm Ly\beta$ at $\rm\lambda_{obs}\!\sim\!3950$. On the contrary, the remaining sources do not show significant spectral features, indeed their selection is due to strong blue colors combined to a barely-significant NB-excess (see e.g. third panel from above). In particular, the star and galaxy interlopers (i.e. third and fourth panels from the top) were picked as targets before we refined our selection rules and the J-PLUS DR1 was re-calibrated \citep{lopezsanjuan2019a}. With the current J-PLUS photometry and our updated selection these objects are not re-selected (right column of Table \ref{tab:GTCresults2018A}). Given the absence of emission lines at $\rm\lambda_{obs}\!\!\sim\!\!3900\text{\AA}$ for these objects, their low-significance NB-excess is likely due to imperfections in their photometry. In the case of the $\rm z\!\!\sim\!\!0.5$ galaxy (fourth panel from the top in Fig. \ref{fig:GTC_spectral_classes}), we additionally observe a discrepancy between the spectrum and J-PLUS data. A number of possible explanation can account for this, such as errors in the spectrum extraction and calibration, too-low spectroscopic SNR at $\rm\lambda_{obs}\lesssim4500\,\text{\AA}$ or artifacts biasing only the $J$0395 photometry. On the contrary, the excess of the $\rm z\!\sim\!2.8$ QSO (bottom panel in Fig. \ref{fig:GTC_spectral_classes}) is due to the $\rm Ly\beta$ line redshifted at $\rm\lambda_{obs}\sim3950$ in the observed spectrum, although in tension with J-PLUS photometry. In this case, QSO variability might play a role \citep[e.g.,][]{hook1994,kozlowski2016} as well as photometric imperfections.


Overall, $40/45$ targets ($88.9\%$) are genuine line emitters, hence confirming the efficiency of our selection. Moreover, the stars contamination is reduced from $\sim\!17\%$ to $\lesssim\!5\%$ between the two programs (see Tables \ref{tab:GTCresults2018A} and \ref{tab:GTCresults2019A}).
Indeed, guided by the \texttt{GTC2018A} results, we i) excluded sources with significant apparent motion according to \texttt{Gaia DR2} and ii) selected $\rm EW_0=50\text{\AA}$ as our limiting value for defining the $\rm\Delta m^{NB}$ cut (see Sect. \ref{sec:selection_rules}).
\begin{table}[t]
    \centering
    \resizebox{9cm}{!}{
    \begin{tabular}{c|c|c|c}
    \hline
    \hlx{v}
    Object  & Fraction   & Percentage & Retrieved after      \\
     class  &   \#       &    (\%)    & improvement \\
    \hline
    \hline
    \hlx{v}
     $\rm z\sim2.2$ QSOs  & 15/24 & 62.5\% & 11/15  \\
    \hlx{v}
     $\rm z\sim1.5$ QSOs  & 4/24  & 16.7\%  & 4/15  \\
    \hlx{v}
     Stars                & 4/24  & 16.7\%  & 0/15  \\
    \hlx{v}
     Low-z Galaxies     & 1/24  & 4.1\%   & 0/15  \\
    \hlx{v}
    \hline
    \end{tabular}}\vspace{2mm}
    \caption{Number counts resulting from the \texttt{GTC2018A} program. Targets are divided in: $\rm z\!\sim\!2.2$ QSOs, whose $J$0395 NB-excess is due to \lya emission, and 3 contaminant species. Among these, $\rm z\!\sim\!1.5$ QSOs are also genuine NB-emitters due to their \texttt{CIV} line.}
    \label{tab:GTCresults2018A}
\end{table}
Our improved methodology retrieves 15/24 original \texttt{GTC2018A} targets, with 11/15 ($\sim\!74\%$) being $\rm z\!\sim\!2.2$ QSOs and 4/15 ($\sim\!26\%$) being $\rm z\!\sim\!1.5$ QSOs (i.e. no star is re-selected). Nevertheless, 4 out of 15 $\rm z\!\sim\!2.2$ QSOs from the original sample are not re-identified as line emitters. The new calibration of the entire J-PLUS survey occurred after \texttt{GTC2018A} \citep{lopezsanjuan2019a} plays a role in this since 2 out of the 4 non-reselected $\rm z\!\sim\!2.2$ QSOs do not pass the NB SNR criterion due to their recomputed NB photometry.
\begin{table}[t]
    \centering
    \resizebox{7.2cm}{!}{
    \begin{tabular}{c|c|c}
    \hline
    \hlx{v}
    Object class  & Fraction \#  & Percentage (\%) \\
    \hline
    \hline
    \hlx{v}
     $\rm z\sim2.2$ QSOs  & 14/21 & 66.6\% \\
    \hlx{v}
     $\rm z\sim1.5$ QSOs  & 4/21  & 19.0\% \\
    \hlx{v}
     $\rm z\sim2.8$ QSOs  & 1/21  & 4.8\%  \\
    \hlx{v}
     Low-z Galaxies & 1/21  & 4.8\%  \\
    \hlx{v}
     Stars  & 1/21  & 4.8\% \\
    \hlx{v}
    \hline
    \end{tabular}}\vspace{2mm}
    \caption{Number counts for \texttt{GTC2019A} program, including five source classes: $\rm z\!\sim\!2.2$, $\rm z\!\sim\!1.5$ and $\rm z\!\sim\!2.76$ QSOs, low-z galaxy and star. Except for the star, all targets are genuine $J$0395 NB-emitters due to, respectively: \lya, \texttt{CIV}, $\rm Ly\beta$ and \texttt{[OIII]} emission lines. The contamination from blue stars significantly dropped to $\lesssim5\%$ (from $\sim17\%$ in \texttt{GTC2018A} results), mainly due to the cross-match with \texttt{Gaia DR2} data.}
    \label{tab:GTCresults2019A}
\end{table}
Finally, the fraction of genuine NB-emitters significantly improved from $\sim\!74\%$ for \texttt{GTC2018A} to over $95\%$ for \texttt{GTC2019A} thanks to the improved methodology.

\subsection{Selected samples of \lya$\!$-emitting candidates}
\label{sec:LAE_samples}
\begin{figure*}[t]
\centering
\includegraphics[width=0.478\textwidth]{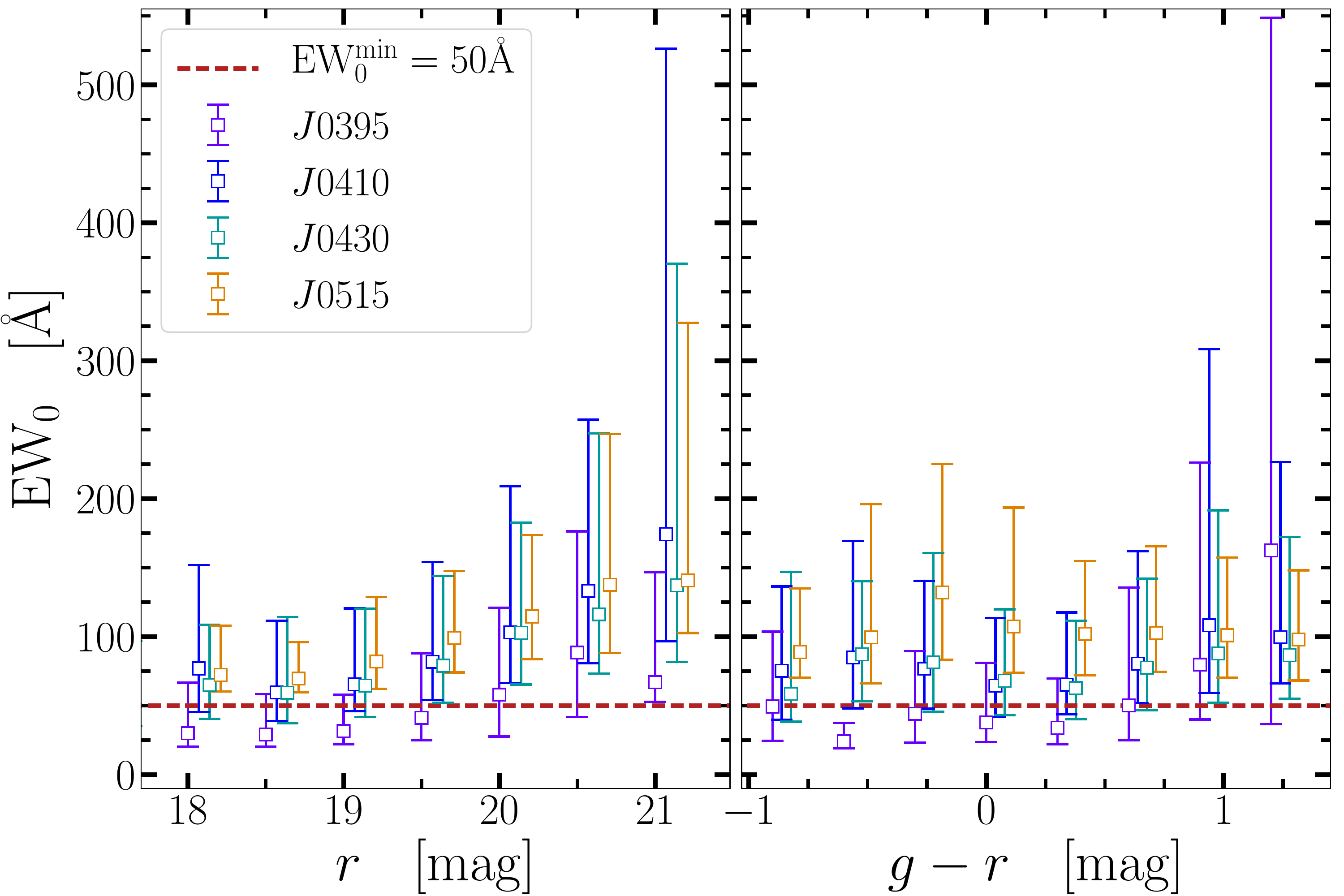}\hspace{5mm}
\includegraphics[width=0.478\textwidth]{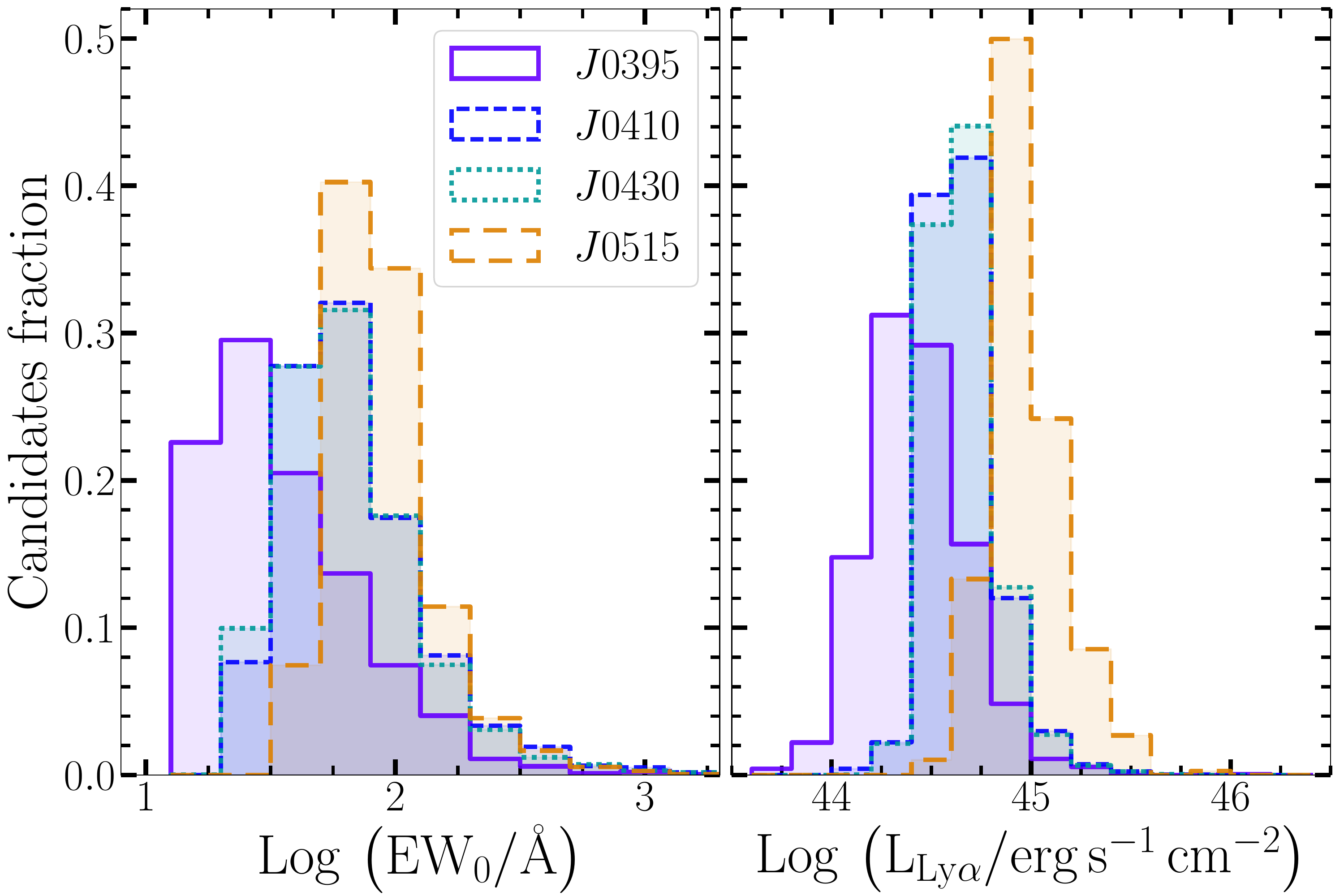}
\caption{\footnotesize \textit{Left figure:} $\rm EW_0$ distribution of our selected candidates as a function of $r$ and $g-r$ color (left and right panels, respectively). Squared points and error bars show respectively the distribution median and 16th-84th percentiles, in each magnitude and color bin. Points have been artificially shifted for a better visualization. The values for $J$0395 filter at $r\!<\!20$ ($g-r\!<\!0.75$) are systematically below the theoretical $\rm EW_0^{min}=50\text{\AA}$ cut we apply (section \ref{sec:selection_rules}). This is due to the little overlap beteween this NB and the $g$ BB, which is reflected into a poor extrapolation of the linear-continuum at the NB filter pivot wavelength \citep[see e.g.,][and the discussion in Sect. \ref{sec:selection_rules}]{ouchi2008}. \textit{Right figure:} Normalized distributions of our candidates in EW and \lya luminosity, for each filter. This result clearly shows that filters sampling higher redshifts also sample brighter \lya luminosity. This is a direct effect of J-PLUS detection limits which only allow to observe brighter and rare objects at higher redshifts. We address this issue by applying the completeness corrections described in Sect. \ref{sec:completeness}.}
\label{fig:ew_and_lyalum_distributions}
\end{figure*}
The final results of our selection procedure are four samples\footnote{Our candidates catalogs can be found on the J-PLUS website:\\ \tiny{\texttt{https://www.j-plus.es/ancillarydata/dr1\_lya\_emitting\_candidates}}.} of $\rm z\!>\!2$ \lya$\!\!$-emitting candidates which meet all the following requirements: i) reliable excess in the NB used for their selection, ii) secure detection and photometry in the filter triplet [NB; $g$; $r$], iii) no spectroscopic counterparts in \texttt{DR14} with redshift outside the ranges probed by each NB, iv) no apparent motion according to \texttt{Gaia DR2} data, v) no significant observed-frame UV detection in \texttt{GALEX}, vi) compact morphology and, eventually, vii) multiple NB excesses compatible with being $\rm z\!\gtrsim\!2$ sources. These lists account for 2547, 5556, 4994, 1467 sources respectively for $J$0395, $J$0410, $J$0430 and $J$0515 NBs (see Table \ref{tab:interlopers_and_final_candidates}), which translates into approximately $2.8$, $6.2$, $5.6$ and $1.5$ objects per squared degree, respectively. We underline that these samples are the largest-to-date collections of \lya$\!\!$-emitting candidates within the narrow redshift bins we can access to \citep[see e.g.,][]{guaita2010,cassata2015,konno2016,matthee2017b,sobral2018}.

The drop of number counts for $J$0515 NB can be ascribed to the combination of J-PLUS data depth and the cosmological decrease of bright SF LAEs and AGN/QSOs number densities at $\rm z\!\gtrsim\!2.5$ \citep[e.g.,][]{nilsson2009a,ciardullo2012,sobral2018}. Indeed, the right panels of Fig. \ref{fig:ew_and_lyalum_distributions} show that $J$0515 NB can only access to ranges of $\rm Log\,(L_{Ly\alpha})$ and \lya $\rm  Log\,(EW_0)$ which are significantly higher than the other NBs. In general, filters sampling smaller wavelengths can access to fainter \lya luminosity and smaller $\rm EW_0$, as a result of the combination between J-PLUS depth and the probed z interval.

\subsubsection{$\rm EW_0$ and $\rm L_{Ly\alpha}$ distributions}
\label{sec:EW_LLya_distributions}
The left panels in Fig. \ref{fig:ew_and_lyalum_distributions} show the distribution of $\rm EW_0$ measured on our samples, as a function of both $r$ magnitude and $g\!-\!r$ color. As commented in Sect. \ref{sec:selection_rules}, our cut on $\rm\Delta m^{NB}$ derives from a theoretical expected limit of $\rm EW_0=50\,\text{\AA}$. Nevertheless, not all selected sources display $\rm EW_0\!>\!50\,\text{\AA}$ \citep[see also][for a similar discussion]{ouchi2008}. This is evident for the $J$0395 NB candidates, whose $\rm EW_0$ distribution is systematically below $\rm50\,\text{\AA}$ for $r\lesssim20$ and $g-r\lesssim0.75$. Indeed, the little overlap between $J$0395 and $g$ transmission curves ultimately provides a relatively poor extrapolation of the linear continuum up to the pivot wavelength of $J$0395, which translates into an under-estimation of $\rm EW_0$. This induces a bias on our \lya luminosity measurement, which we account for as described in Sect. \ref{sec:filter_width_corr}. On the other hand, no significant nor systematic bias affects $\rm EW_0$ with respect to color, as shown by the flat $g\!-\!r$ distribution in Fig. \ref{fig:ew_and_lyalum_distributions}. We confirmed this by using the spectra of $\rm z\!\sim\!2$ \texttt{DR14} QSOs, but we do not show the results for the sake of brevity.

Overall, our distributions are broadly consistent with previous determinations of the rest-frame EW of $\rm z\!\sim\!2-3$ LAEs \citep[either SF LAEs and AGN/QSOs, see e.g.,][]{gronwall2007,guaita2010,hainline2011,ciardullo2012,shibuya2014,hashimoto2017,santos2020}. Interestingly, our samples include a moderate fraction of sources ($\lesssim7\%$, on average) showing $\rm EW_0\!>\!240\,\text{\AA}$ \citep[e.g.,][]{ouchi2008,santos2020}. High-EW LAEs have been studied with particular interest \citep[e.g.,][]{cantalupo2012,kashikawa2012,shibuya2014} since nebular emission of Pop-II stellar populations can only account for $\rm EW_0^{Ly\alpha}\!\lesssim\!500\text{\AA}$ \citep[e.g.,][]{charlot_fall1993, hernan-caballero2017}. At the same time, high \lya EWs can be easily produced by AGN/QSOs which are likely to dominate our selected samples. Since analyzing high-EW LAEs would require a careful separate analysis, we refrain to comment further on this topic. Nevertheless, we underline that our lists of selected candidates can provide catalogs of high-EW LAE targets for upcoming studies.

\subsubsection{Relative abundance of QSOs and SF LAEs}
\label{sec:LAE_QSOfractions}
\begin{figure}[t]
\centering
\includegraphics[width=0.47\textwidth]{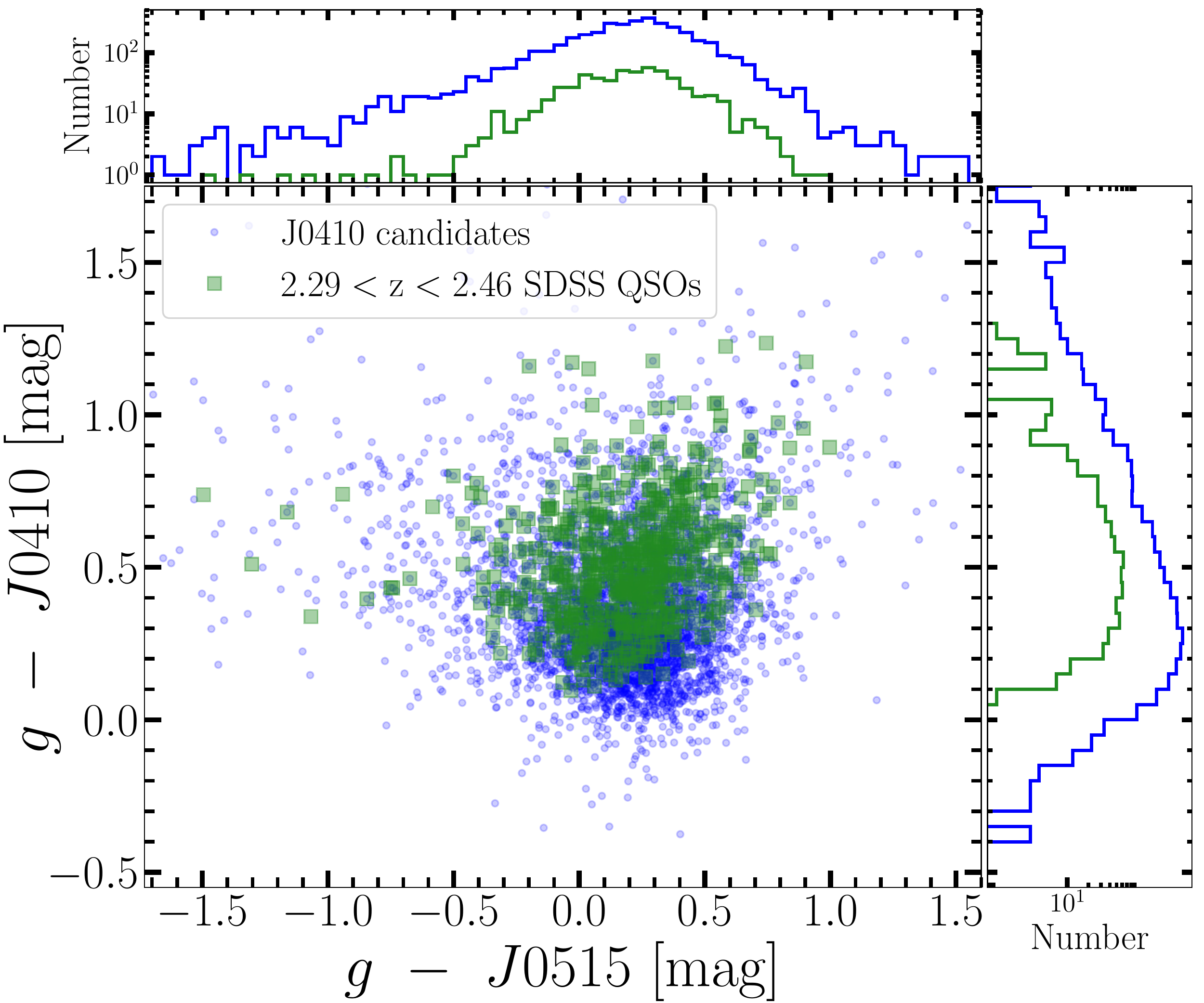}
\caption{\footnotesize Comparison between the color-color distributions of \texttt{DR14} QSOs (green squares) within our $J$0410 sample and of the remaining $J$0410 genuine candidates (blue dots, after removing known interlopers; Sect. \ref{sec:sample_cleaning}). The two source classes occupy comparable color regions, suggesting that our selection results might be effectively dominated by $\rm z\!\sim\!2.3$ AGN/QSOs. This scenario is also supported by the results of our spectroscopic program (section \ref{sec:gtcProgram}).}
\label{fig:color-color_candidates_and_QSOs}
\end{figure}
The design of J-PLUS filters potentially allows to capture at the same time peculiar combination of high-z lines with different NBs (see Sect. \ref{sec:multiNB_excess}). For instance, QSOs emitting \lya at $\rm z\!\sim\!2.3$ could show simultaneous NB excesses in $J$0410 and $J$0515 NBs (the latter being due to \texttt{CIV} emission). This offers the possibility of investigating the relative fraction of AGN/QSOs and SF LAEs in our samples, since the latter should not exhibit such double-NB emission. We hence separate the \texttt{DR14} QSOs selected with $J$0410 from the rest of $J$0410 candidates and plot the color distribution of these two source classes.
\begin{figure}[t]
\centering
\includegraphics[width=0.47\textwidth]{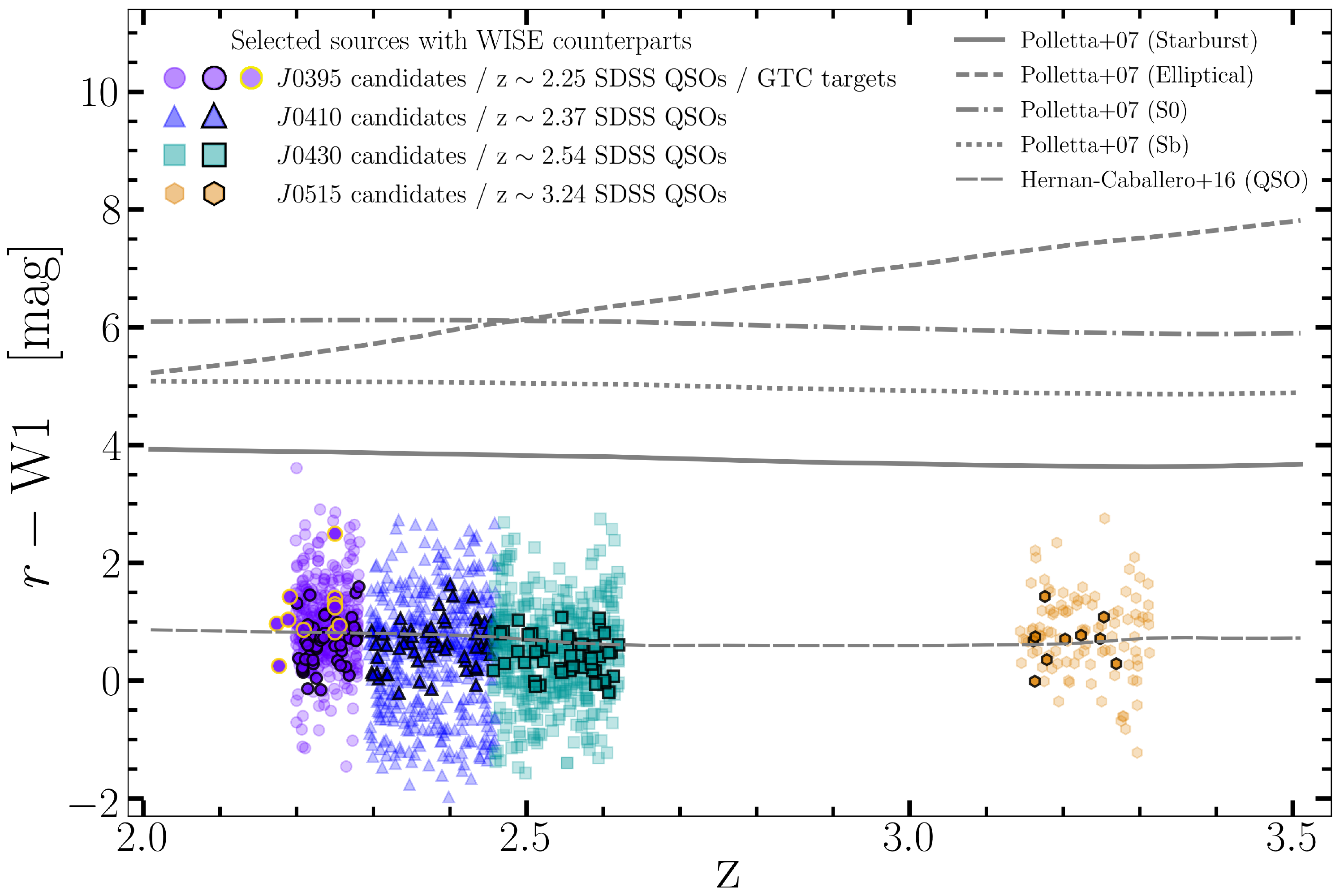}
\caption{\footnotesize \textit{Colored points}: $r-\mathrm{W1}$ color versus redshift of all our candidates with a counterpart in WISE. Our sources are uniformly spread within the z interval sampled by each NB. \textit{Grey lines}: tracks for different galaxy templates \citep[from the SWIRE library,][]{polletta2007} and the QSO template of \cite{hernan-caballero2016}. Black-contoured and yellow-contoured points respectively mark the SDSS QSOs in our selection and the GTC targets (only for $J$0395 NB). Our candidates are all compatible with the high-z QSO template, hence suggesting that the fraction of SF LAEs in our sample is very low. In addition, this suggests that we identify as high-z QSOs a large number of sources without previous spectroscopic identification.}
\label{fig:wise-color_vs_redshift}
\end{figure}

Figure \ref{fig:color-color_candidates_and_QSOs} shows the color space defined by $J$0410 and $J$0515 NBs with respect to $g$ BB. In this plane, both SF LAEs and QSOs should exhibit $g\!-\!J0410\!>\!0$ due to \lya emission, but the \texttt{CIV} line \citep[usually much stronger in QSOs than in SF LAEs, see e.g.,][]{hainline2011,stroe2017a,stroe2017b, nakajima2018b} should displace QSOs at $g\!-\!J0515\!>\!0$. Nevertheless, our color distribution (blue solid histogram) does not show any evident bimodality and no significant overdensity at $g\!-\!J0515\!\sim\!0$, as expected for SF LAEs. This suggest that either i) our $J$0410 candidates are mostly dominated by $\rm z\!\sim\!2$ QSOs or ii) the J-PLUS filter set does not reliably disentangle the different spectral features of high-z SF LAEs and AGN/QSOs.

We further analyze the nature of our candidates by exploiting the cross-match with the all-sky WISE data \citep{wright2010} provided by the J-PLUS DR1 database. In particular, we compare the $r-\mathrm{W1}$ color of our candidates with WISE counterparts to the synthetic-photometry color-tracks of galaxy and QSOs templates \citep[respectively from][and \citealt{hernan-caballero2016}]{polletta2007}. Figure \ref{fig:wise-color_vs_redshift} shows how the color of our candidates are clearly compatible with the ones of QSOs, while being significantly different from the galaxy ones. We also highlight the SDSS QSOs (black-contoured dots) and the confirmed QSOs of our GTC programs (yellow-contoured dots, only for $J0395$ NB) to underline the comparability of our candidates properties with those of spectroscopically-confirmed QSOs.

Interestingly, by joining these evidences with the results of our spectroscopic follow-up programs, we expect our samples of candidates to be dominated by $\rm z\!\sim\!2.3$ QSOs which yet lack a spectroscopic identification. Indeed, by considering the number of our genuine candidates without SDSS identification (namely 2057, 4959, 4494 and 1377 respectively for $J$0395, $J$0410, $J$0430 and $J$0515 NBs) and conservatively applying a residual contamination of $\gtrsim\!35\%$ (as suggested by our GTC follow-up) our method identifies for the first time as $\rm z\!>\!2$ QSOs respectively $\gtrsim\!1300, 3200, 2900$ and $\sim\!900$ $\rm z\!>\!2$ sources in each NB. This is shown in both Fig. \ref{fig:color-color_candidates_and_QSOs} and \ref{fig:wise-color_vs_redshift} by the wide difference between the number counts of \texttt{DR14} QSOs within our selection and our remaining genuine candidates. We interpret this as an effect of the NB-based selection we perform, which efficiently targets the line-emission features of these objects, eventually missed by previous target-selections based on BB-colors \citep[e.g.,][]{richards2009, ross2012, ivezic2014}. Nevertheless, a systematic and uniform spectroscopic confirmation of our samples is needed to validate these findings.

\section{Construction of the \lya luminosity function}
\label{sec:lumifunc}
The luminosity function $\rm\Phi(L)$ of a given class of sources is usually defined as their comoving number density per unit luminosity \citep[see e.g.,][]{schmidt1968}. Following a common convention in literature, we express our LFs in logarithmic units of luminosity and hence use the following definition:
\begin{equation}
    \rm \Phi[\,Log\,(L_{Ly\alpha})]\,=\,\frac{\sum_i\ (P_i\,/\,C_i)}{V\ \cdot\,\Delta\,Log(L_{Ly\alpha})}\ ,
    \label{eq:LF_definition}
\end{equation}
where the sum at the numerator is extended to all the objects in a given bin of (logarithmic) luminosity $\rm\Delta\,Log(L_{Ly\alpha})$, while the coefficients $P_i$ and $C_i$ are statistical weights that account respectively for the sample purity and completeness (as detailed below). We exploit our lists of candidates selected with $J$0395, $J$0410, $J$0430 and $J$0515 NBs to build four determinations of the \lya LF at the redshifts given by Table \ref{tab:QSOcontamin}. The next sections detail the steps we perform for assessing the reliability of our \lya flux measurements (section \ref{sec:line_flux_retrieval}), measuring the \lya luminosity of our candidates and the cosmological volume probed by J-PLUS NBs (section \ref{sec:lyalum_and_volume_computation}) and estimating the purity (section \ref{sec:purity_of_samples}) and completeness (section \ref{sec:completeness}) of our selection.

\subsection{Retrieval of the total \lya flux}
\label{sec:line_flux_retrieval}
Our \lya flux measurements ($\rm F^{\,3FM}_{Ly\alpha}$ hereafter) can be affected by systematic uncertainties due to both the J-PLUS aperture photometry and our measuring method (see Eq. \ref{eq:line_flux} and Eq. \ref{eq:app:fline_3fm}. In order to build our \lya LFs, we first study the differences between $\rm F^{\,3FM}_{Ly\alpha}$ and a corresponding spectroscopic measurement (i.e. $\rm F_{Ly\alpha}$), assuming that the latter provides a reliable estimate of the sources' total emitted \lya flux. We then compute statistical corrections which account for the bias between $\rm F^{\,3FM}_{Ly\alpha}$ and its spectroscopic analog, using spectroscopically identified QSOs \citep{paris2018} at the redshift sampled by each NB and their counterparts in the DR1 catalog. We obtain the \lya flux from QSOs spectra with the methodology shown in appendix \ref{sec:append:spectraMeasurement}.

\subsubsection{Aperture correction}
\label{sec:aper_corr}
To make sure that \texttt{auto}-aperture photometry (see Sect. \ref{sec:jplus_definition_and_catalogs}) do not introduce any bias on the photometry of our candidates, we compare the synthetic flux $\rm \langle \mathit{f}^{\,\mathit{r}}_{\lambda}\,\rangle^{synth}$ of SDSS QSOs to the analogous measurements obtained from J-PLUS DR1. For the sake of brevity, the details of this check are presented in the appendix \ref{sec:app:lya_flux_definitions}, while here we summarize our findings. In general, we no significant bias ($\lesssim0.2\,\sigma_r$) affects the \texttt{auto}-aperture flux of point-like sources for each NB. Consequently, we do not apply aperture corrections to $\rm F^{\,3FM}_{Ly\alpha}$. On the other hand, the flux comparison points out the need for an additional statistic uncertainty on top of the J-PLUS photometric errors for $\rm F^{\,3FM}_{Ly\alpha}$ (see \ref{sec:app:lya_flux_definitions} for details). We then re-scale the $r$ band uncertainties and propagate them on $\rm F^{\,3FM}_{Ly\alpha}$. Finally, we account for these on our LF determinations, as discussed in Sect. \ref{sec:errors_on_the_LF}.

\subsubsection{Filter width correction}
\label{sec:filter_width_corr}
A fraction of the flux of broad lines (i.e. broader than the FWHM of the measuring NB) can be systematically lost by photometric measurements, especially if the line-peak is displaced at the edge of the NB transmission curve. SF LAEs usually show a narrow \lya as opposed to the usually broad line profile of QSOs \citep[e.g.,][and Fig. \ref{fig:LAEspectra_3FM}]{vandenberk2001, telfer2002, selsing2016}. For these reasons, we expect this bias to significantly affect the $\rm F^{\,3FM}_{Ly\alpha}$ measurements of QSOs, while not influencing those of SF LAEs. At the same time, no SF LAEs were observed among our followed-up targets (section \ref{sec:GTC_program_results}), in line with previous results suggesting that AGN/QSOs dominate the samples of photometrically-selected LAEs at $\rm L_{Ly\alpha}\gtrsim2\times10^{43}\,erg\,s^{-1}$ \citep[see e.g.,][]{santos2004b,konno2016,matthee2017b,sobral2018,calhau2020}. Furthermore, the (expected) low fraction of SF LAEs in our final selection cannot be reliably disentangled from QSOs by J-PLUS photometry (section \ref{sec:LAE_QSOfractions}). This hinders the possibility of applying a flux correction exclusively to a sub-class of our candidates. Consequently, we consider valid our method for measuring $\rm F^{\,3FM}_{Ly\alpha}$ and then apply a statistical correction to all our candidates. In particular, we obtain the corrected \lya flux as follows (see appendix \ref{sec:app:lya_flux_definitions}):
\begin{equation}
    \rm F^{\,3FM\,;\,corr}_{Ly\alpha} = (1-\Delta F)\cdot F^{\,3FM}_{Ly\alpha}\ .
    \label{eq:lya_flux_correction}
\end{equation}
The quantity $\rm\Delta F$ is a rigid offset obtained from the normalized distribution of flux difference: $\rm (F^{\,3FM}_{Ly\alpha}-F^{\,spec}_{Ly\alpha})/F^{\,3FM}_{Ly\alpha}$, where $\rm F^{\,3FM}_{Ly\alpha}$ is our \lya flux estimate and $\rm F^{\,spec}_{Ly\alpha}$ is its spectroscopic analog measured on SDSS QSOs (see appendix \ref{sec:app:lya_flux_definitions} for details). We obtain a $\rm\Delta F$ for each NB and then use the corrected values $\rm F^{\,3FM\,;\,corr}_{Ly\alpha}$ for our luminosity function computation. With this a analysis we also obtain a correction for the error on $\rm F^{\,3FM\,;\,corr}_{Ly\alpha}$, which we propagate on our \lya LF determination (see Sect. \ref{sec:errors_on_the_LF}).

Finally, the $\rm F^{\,3FM}_{Ly\alpha}$ obtained with $J$0378 NB are affected by a significant bias ($\rm\Delta F=1.75\pm0.35$). This can be ascribed to wavelength separation between this NB and $g$, which reflects into a poor extrapolation of the linear continuum approximation. Consequently, we exclude $J$0378 from the list of NBs we use.

\subsection{Computation of $\rm\, L_{Ly\alpha}$ and cosmological volume}
\label{sec:lyalum_and_volume_computation}
We compute the \lya luminosity as:
\begin{equation}
    \rm L_{Ly\alpha}^{3FM} = 4\,\pi\,d_L^2(z)\ F_{Ly\alpha}^{\,3FM\,;\,corr} = 4\,\pi\,[d_c\,(1+z)]^2\,F_{Ly\alpha}^{\,3FM\,;\,corr}\ ,
    \label{eq:lya_luminosity}
\end{equation}
where $\rm d_L(z)\!=\!d_c(z)\!\cdot\!(1+z)$ and $\rm d_c(z)$ are the luminosity and comoving distances of our sources, computed by assuming \texttt{PLANCK2015} cosmology \citep{adam2016, ade2016}.

In order to compute the $\rm d_L(z)$ of our candidates without spectroscopic determination it is necessary to assume a value of z. Being blind towards their nature, we use the $\rm z_{\,p}$ obtained by shifting the \lya rest-frame wavelength to the pivot wavelength \citep[][]{tokunaga_vacca2005}
of the NB used for selection (see Table \ref{tab:QSOcontamin}). Consequently, the uncertainty $\rm\sigma_z$ is obtained from the half-width of each NB (see Table \ref{tab:QSOcontamin}). This does not apply to the candidates with a spectroscopic counterpart, as in these cases we use the \texttt{DR14} z and $\sigma_z$. Finally, we propagate the redshift errors on the total $\rm L_{Ly\alpha}$ uncertainty and on our LF determinations (see section  \ref{sec:errors_on_the_LF}).

The cosmological volume sampled by our data depends on the z windows associated to \lya detection and the DR1 area not affected by masking, for each NB. In our case, the redshift intervals are given by the FWHM of each NB (see Table \ref{tab:QSOcontamin}) and converted to cosmological volumes by assuming the \texttt{PLANCK2015} cosmology \citep{adam2016, ade2016}. On the other hand, the effective area observed in a given band can be obtained with the \texttt{MANGLE} software \citep{hamilton_tegmark2004,swanson2008}. Since we require single-detection in [NB; $g$ and $r$], we computed the intersection between the three associated \texttt{MANGLE} masks (see Table \ref{tab:QSOcontamin}). We assume negligible errors on volume estimates for our LF computation.

\subsection{Estimate of the samples contamination}
\label{sec:purity_of_samples}
The steps detailed in Sect. \ref{sec:sample_cleaning} do not ensure to identify \textit{all} the contaminants, as confirmed by our follow-up results (section \ref{sec:GTC_program_results}). For this reason, we estimate the residual contamination of our samples by computing a statistical \textit{purity weight} for our candidates as a function of their $r$-magnitude:
\begin{figure}[t]
    \centering
    \includegraphics[width=0.47\textwidth]{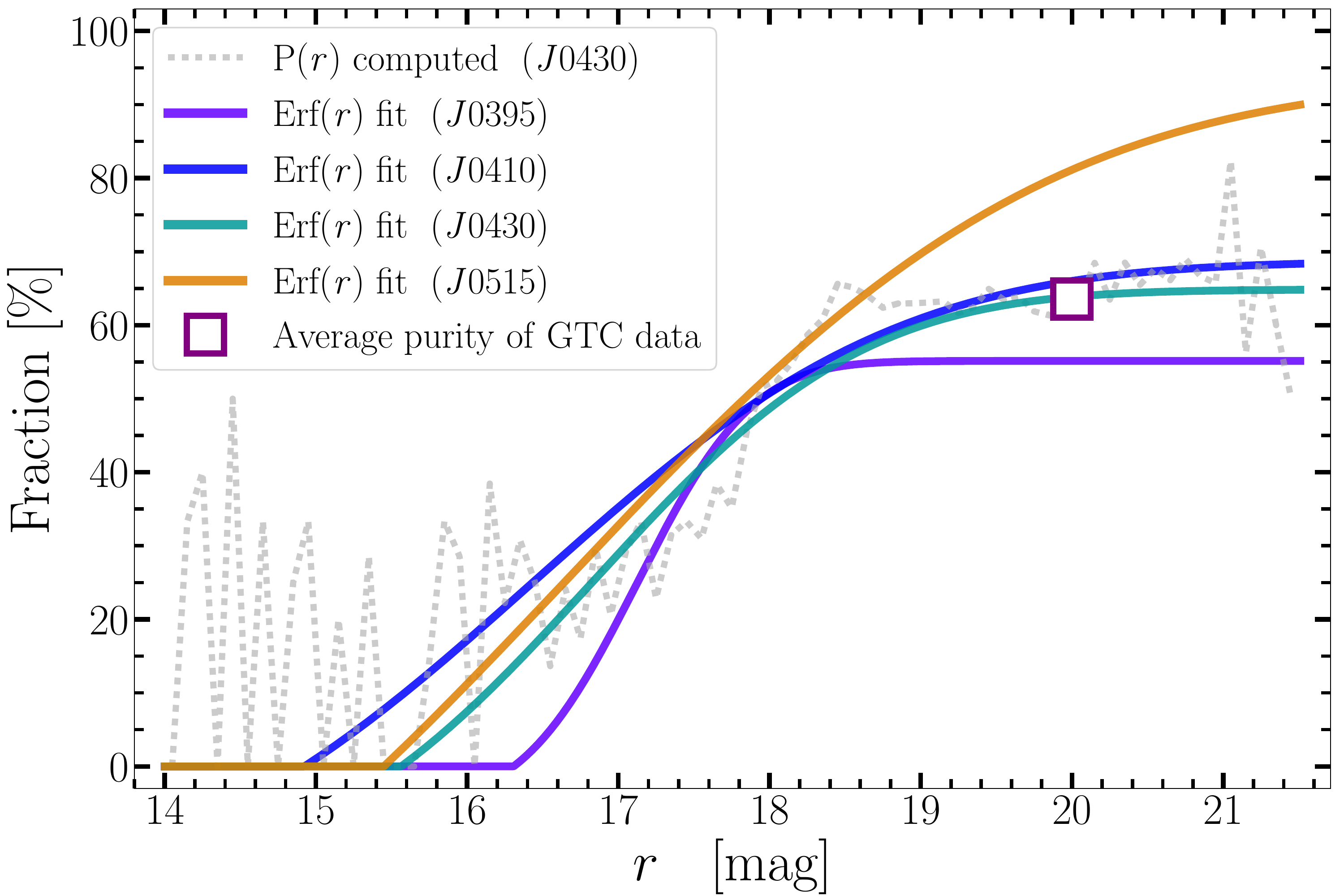}
    \caption{Statistical purity weight for each NB (coloured solid lines), as estimated by fitting an error-function to the computed purity. The grey-dotted line shows the computed purity of $J$0430 NB as an example. All filters show similar purity weights, rising to $\gtrsim\!60\%$ at $r\gtrsim18.5$. This is in agreement with the average purity of our 45 spectroscopic targets (section \ref{sec:GTC_program_results}), shown as a purple empty square. }
    \label{fig:purity}
\end{figure}
\begin{equation}
    \rm P(\mathit{r}) = 1-\frac{N_{interlopers}(\mathit{r})}{N_{total}(\mathit{r})}\ .
    \label{eq:purity}
\end{equation}
$\rm N_{interlopers}(\mathit{r})$ and $\rm N_{total}(\mathit{r})$ are respectively the number of \textit{secure} interlopers (see Sect. \ref{sec:sample_cleaning}) and the total number of candidates at a given magnitude. We then fit $\rm P(\mathit{r})$ with an error-function and use the latter to obtain the statistical weight of each genuine candidate to the final LF. Figure \ref{fig:purity} shows the error-function fits for each NB (solid colored lines) and the computed $\rm P(\mathit{r})$ values for the $J$0430 filter, as an example (dotted grey line). The purple empty square shows the average purity measured on the complete sample of both our spectroscopic follow-up programs. This is in good agreement with the statistical weights of each NB (i.e. $\rm P(\mathit{r})\!\gtrsim\!60\%$ at $r\gtrsim18.5$). The high values reached by $J$0515 ($\sim80\%$) are driven by the drop of interlopers with spectroscopic identification at $\rm z\!\gtrsim\!3$.

\subsection{Estimate of the samples completeness}
\label{sec:completeness}
Genuine line-emitting candidates might be lost by our selection due to the J-PLUS detection limits and source extraction, the effect of photometric errors and the $r$-band pre-selection of our parent samples \citep[see e.g.,][]{geller2012, loveday2012, gunawardhana2013}. In order to correct for these known issues, we estimate the completeness\footnote{We define the completeness $C$ as the ratio between the number of \textit{genuine} targets effectively selected (\textit{true positives}, TP) and the \textit{total} number $\rm N_{tot}$ of genuine targets in the survey footprint, either detected or undetected. $\rm N_{tot}$ is generally unknown and can be thought as the sum of TP, \textit{false negatives} (i.e. genuine targets detected but lost by the selection) and \textit{undetected} candidates. In other words: $\rm C = N_{TP} / (N_{TP}+N_{FN}+N_{UD})$.} of our samples by considering three different components. In detail, we account for: i) the DR1 source-extraction process (i.e. \textit{detection weight} $\rm C^{\,d}$), ii) our selection methodology (i.e. \textit{selection weight} $\rm C^{\,s}$) and iii) $r$ band pre-selection of dual-mode catalogs (i.e. \textit{dual-mode weight} $\rm C^{\,dm}$). We obtain the total completeness weight of each candidate to the final LF as: $\rm C_i=C_i^{\,d}+C_i^{\,s}+C_i^{\,dm}$.
\begin{figure}[t]
    \centering
    \includegraphics[width=0.5\textwidth]{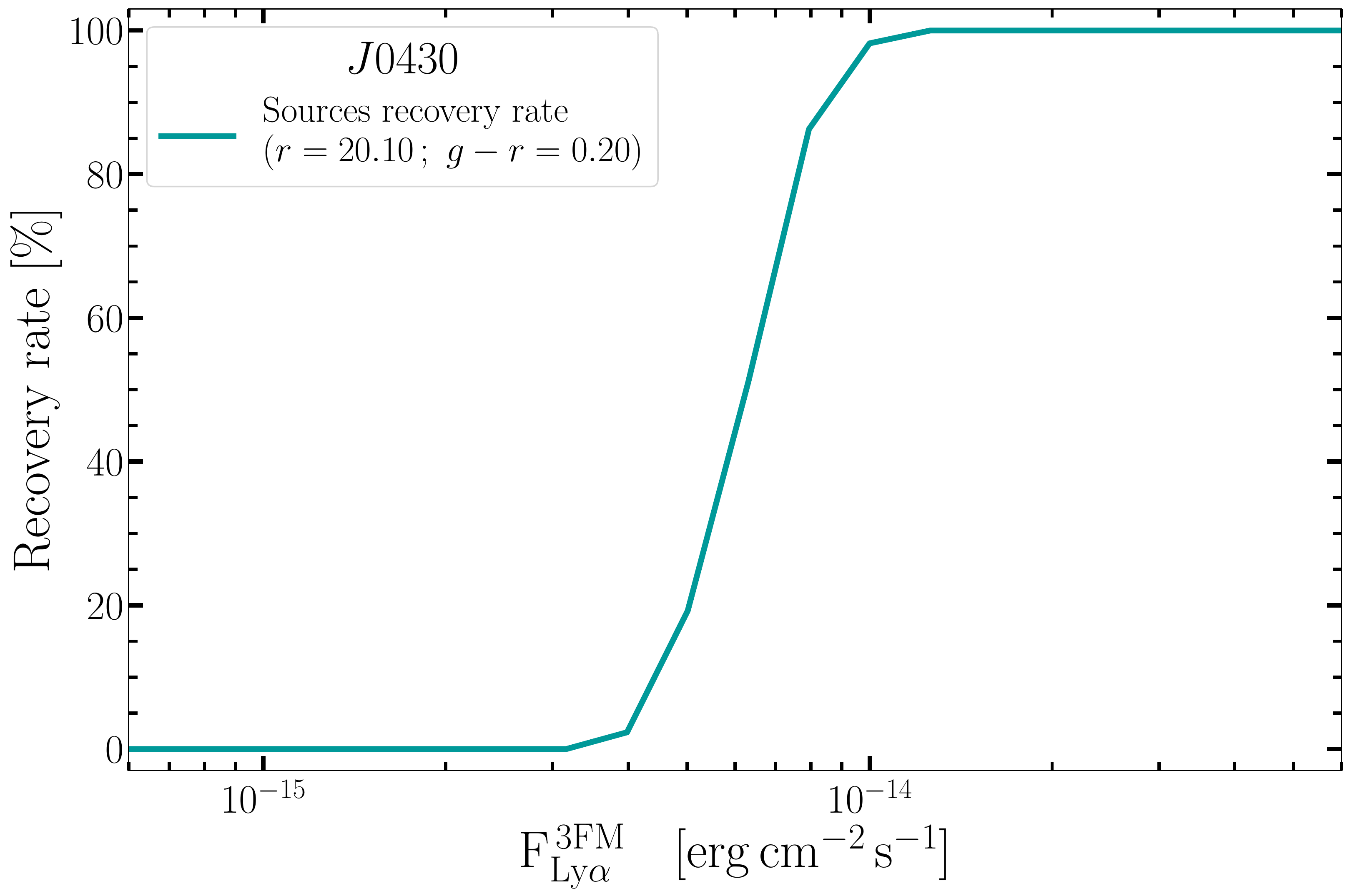}
    \caption{\footnotesize{Example of the recovery fraction of our selection as a function of \lya flux, computed for $J$0430 NB in a bin of $r$ magnitude and $g\!-\!r$ color (namely at $r=20.1$ and $g\!-\!r=0.2$). The full 3D grid is shown in appendix \ref{sec:app:completeness_computation} for the same NB. All filters show comparable values of recovery fractions, hence we just report the case of $J$0430 for brevity.}}
    \label{fig:completeness_example_bin}
\end{figure}

\subsubsection{Detection completeness}
\label{sec:detection_completeness}
The detection completeness of each J-PLUS pointing (for each filter) is automatically computed by the standard source-extraction pipeline as:
\begin{equation}
    \rm C_i^{\,d}(\textit{r}) = 1 - \frac{1}{e^{\,-k_s\,\cdot\,(\textit{r}\, -\, \mathit{r}_s)}\, +\, 1}\ ,
    \label{eq:completeness_model}
\end{equation}
where $\rm k_s$ and $\rm \mathit{r}_s$ are computed for each pointing. They are respectively the decay-rate of $\rm C_i^{\,d}(\mathit{r})$ and the magnitude at which $\rm C_i^{\,d}(\textit{r})$ reaches 50\%. All details of this computation are provided by the J-PLUS DR1 datababse. We obtain $\rm C^{\,d}_i$ from the $\rm[k_s;\mathit{r}_s]$ parameters and $r$ corresponding to each DR1 source.

\subsubsection{Selection completeness}
\label{sec:flux_completeness}
Starting from $r$-detected catalogs, our selection makes use of NB-excess significance and a linear estimate of the sources continuum slopes, related to their $g\!-\!r$ color (see Sect. \ref{sec:3filters}, \ref{sec:selection_rules} and figures \ref{fig:LAEspectra_3FM} and \ref{fig:colmag}). In order to capture its multiple dependencies, we test the retrieval efficiency of our selection as a function of $r$ magnitude, \lya flux and $g-r$ color. In particular, we compute the recovery rate of simulated candidates over wide ranges of these three quantities, by re-applying each of our selection rules. This accounts for source loss at different \lya flux, continuum and EW. We organize the measured recovery rates in a 3D-grid which we interpolate at the measured position of each genuine candidate to compute its selection weight $\rm C^{\,s}_i$. The details of this computation are given in appendix \ref{sec:app:completeness_computation}.

\subsubsection{$r$ - $\rm L_{Ly\alpha}$ bivariate completeness}
\label{sec:bivariate_completeness}
The use of $r$-band detected catalogs makes our selection prone to the loss of continuum-faint $\rm z\!\gtrsim\!2$ \lya$\!\!$-emitting sources, with non-trivial effects on the EW distribution of our selected samples. At low \lya flux, for instance, the $r$-detection requirement might favour the selection of high-EW \lya$\!\!$-emitting sources. This issue has been pointed out by previous works whose selection function was built on the convolution of $r$-band detection and NB-excess significance. In particular, \cite{gunawardhana2015} showed that accounting for this effect requires a multi-variate approach. In other words, the fraction of undetected continuum-faint line-emitters can be estimated by modelling the full-2D luminosity function of candidates in the $r$ vs. line-luminosity plane.

\begin{figure}[t]
    \centering
    \includegraphics[width=0.45\textwidth]{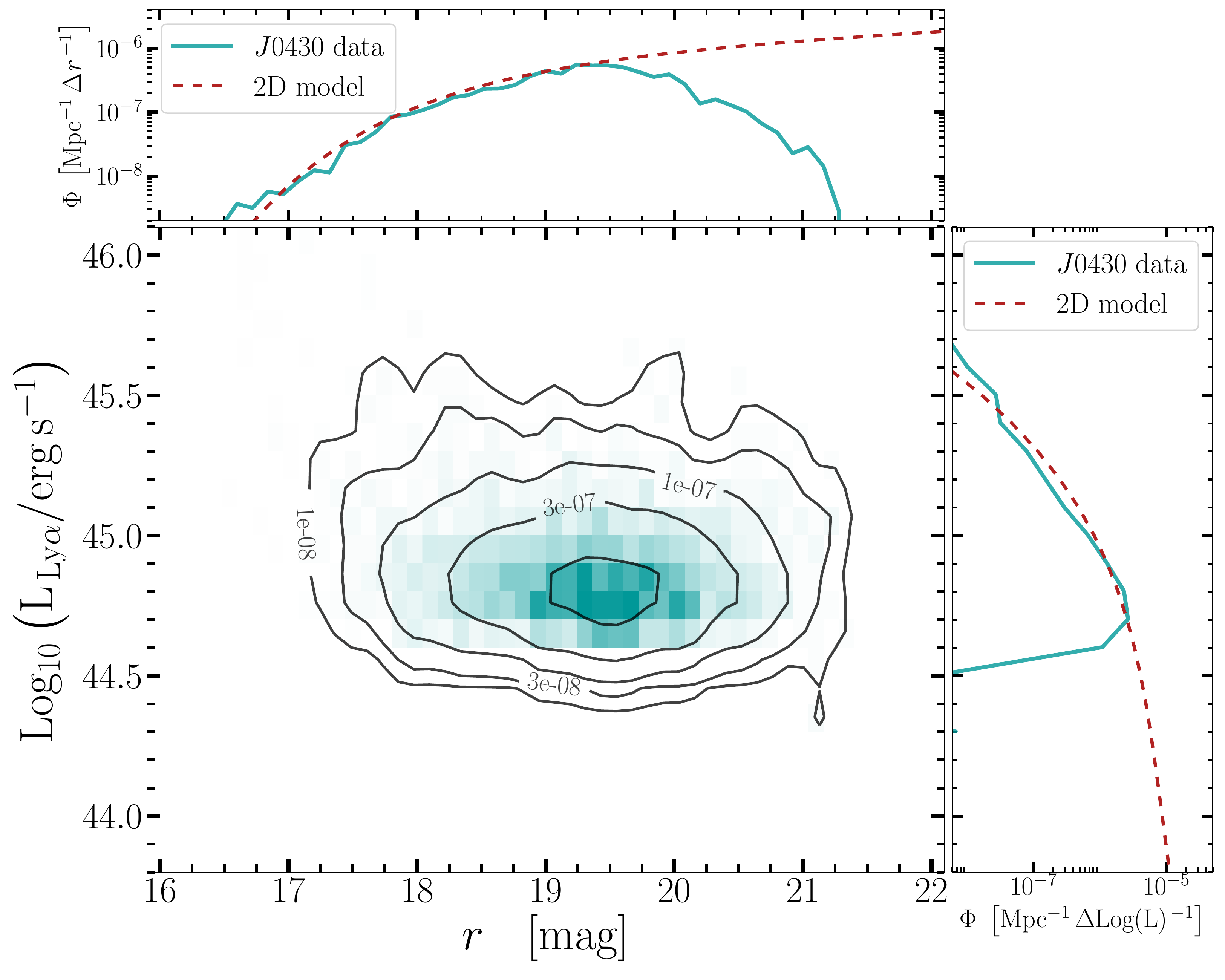}
    \caption{\footnotesize Central panel: full-2D luminosity function of our $J$0430 candidates, as a function of $r$ and $\rm Log(L_{Ly\alpha})$. Green solid lines in the top and right panels show the projections of the 2D LF respectively along the $\rm Log(L_{Ly\alpha})$ and $r$ axis. The red dashed lines show the projection of the 2D model along the same axis. This model was fitted on the 2D distribution shown in the central panel (see appendix \ref{sec:app:completeness_computation} for computational details) and it allows to extrapolate our data distribution at faint $r$ and \lya luminosity. We use the data-to-model ratio (in the $r$ vs.$\rm Log(L_{Ly\alpha})$ 2D plane) to compute the $\rm C^{\,dm}_i$ weight of each candidate.}
    \label{fig:bivariate_completeness}
\end{figure}

We closely follow the methods of \cite{gunawardhana2015} applying their computations to the $r$ vs $\rm L_{Ly\alpha}$ space. The details of this procedure and its main equations are presented in appendix \ref{sec:app:completeness_computation}. In brief, we assume that the 2D LF can be modelled by the product of two functions, describing respectively the $r$ and $\rm Log(L_{Ly\alpha})$ distributions \citep[see also][]{corbelli_salepeter_dickey1991}. We combine a Schechter (in logaritmic form) and a Gaussian (in $\rm Log\,L_{Ly\alpha}$) functions \citep[as in][see appendix \ref{sec:app:completeness_computation}]{gunawardhana2015}. By fitting this 2D model to our measured 2D LF, we can model the number density of sources in regions of the $r$ vs $\rm L_{Ly\alpha}$ plane affected by our incompleteness. Finally, the ratio of our data to the the 2D model (in the 2D space $r$ vs. $\rm L_{Ly\alpha}$) allows us to compute the statistical weight $\rm C^{\,dm}(\mathit{r},Ly\alpha)$ for each source, which accounts for the loss of $r$-faint \lya$\!\!$-emitting sources. Figure \ref{fig:bivariate_completeness} shows the results of our 2D modelling for the $J$0430 filter. In particular, the top and right panels show the projection of both our 2D LF (green solid lines) and 2D model (red dashed line) respectively along the  $\rm Log(L_{Ly\alpha})$ and $r$ axis. It is clear how the model extrapolates our measurements at $r\!>\!19.5$ and $\rm Log(Ly\alpha/erg\,s^{-1})\!<\!44.2$.

\subsubsection{Errors on the $\rm Ly\alpha$ luminosity function}
\label{sec:errors_on_the_LF}
The uncertainties on sources redshift and \lya flux, the binning in \lya luminosity and the internal variance of the samples (due to differences among each J-PLUS pointing) jointly contribute to the errors on our final LFs \citep[e.g.,][]{sobral2018b}. We measure separately each source of uncertainty and finally sum in quadrature their different contributions. To account for $\rm L_{Ly\alpha}$ uncertainties, we repeat the determination of our LF 1000 times by perturbing each time the sources flux according to its uncertainty. During this procedure, we keep the sources redshift fixed to $\rm z_{\,p}$ (see Sect. \ref{sec:lyalum_and_volume_computation}) in order to evaluate only the contribution of flux perturbations to the final errors on our LFs. We then compute the asymmetric errors from the percentiles of the LFs distribution as $\rm\sigma^-=50^{th}-16^{th}$ and $\rm\sigma^+=84^{th}-50^{th}$, where $\rm84^{th}$, $\rm50^{th}$ and $\rm16^{th}$ are the corresponding distribution percentiles. The contribution of redshift errors is accounted in the analogous way by fixing the flux measurements. To account for the internal variance of our LAE candidates sample due to field variations in J-PLUS DR1 we perform random realizations of the luminosity function by splitting our samples into 10 independent sub-samples and computing a LF for each sub-sample. We repeat this process 1000 times and ultimately extract the errors from the $\rm16^{\,th}$ and $\rm84^{\,th}$ percentiles of the LFs distribution (see above). Finally we also add the poissonian errors ($\sqrt{N}$) associated to the sources number counts in each bin to the total LF uncertainties.

\section{Results}
\label{sec:scientific_results}
In this section we present and discuss the four \lya luminosity functions we compute from our samples of candidates. In particular, we compare our measurements to previous results in the literature in Sect. \ref{sec:lya_LF_comparison_literature}, we describe the computation of its Schechter parameters in Sect. \ref{sec:schechter_fits} and finally we estimate the fraction of AGN/QSOs as a function of luminosity in Sect. \ref{sec:AGN_fractions_from_LFs}.

\subsection{The \lya luminosity functions at $\rm\,2\!<\!z\!<\!3.3$}
\label{sec:lya_LF_discussion}
Figure \ref{fig:lf} shows the four determinations of the \lya LF we compute at $\rm z\!\sim\!2.25,\ 2.37,\ 2.54$ and $\rm z\!\sim\!3.24$ (colored empty squares). For each NB, we only consider the candidates with a total completeness weight $\rm C=C^{\,d}\times C^{\,s}\times C^{\,dm}\!>\!0.85$ (see Sect. \ref{sec:completeness} and appendix \ref{sec:app:completeness_computation}). This excludes sources whose contribution is severely affected by the completeness correction, especially at $\rm Log(L_{Ly\alpha}/\,erg\,s^{-1})\!\!\lesssim\!\!44$. Overall, our results probe a luminosity interval of $\rm\sim\!\!1.5\,dex$, from $\rm Log(L_{Ly\alpha}/\,erg\,s^{-1})\!\sim\!44$ to $\rm Log(L_{Ly\alpha}/\,erg\,s^{-1})\!\sim\!45.5$. These regimes are expected to be significantly populated by \lya$\!\!$-emitting AGN/QSOs \citep[e.g.,][]{borisova2016,matthee2017b,sobral2018b,calhau2020}. Interestingly, our results extend by $\rm\sim\!\!1\,dex$ into a previously-unconstrained \lya luminosity range, allowing to probe it with high precision. In addition, our data extend 
down to $\rm\sim\!\!10^{-8}\,Mpc^{-3}$, a limit which is hardly reached by previous studies \citep[see e.g.,][]{sobral2018}. These remarkable features are ultimately attained because of the very wide area covered by J-PLUS NB imaging (unprecedented for \lya LF determinations), which balances the J-PLUS depth ($r\!<\!22$).

The shaded grey areas in each panel of Fig. \ref{fig:lf} mark the regions which are not accessible by our data, respectively due to the limiting $\rm L_{Ly\alpha}$ (vertical limit, see Table \ref{tab:QSOcontamin}) and the survey area (horizontal limit). In particular, the latter marks the comoving number density (per $\rm\Delta\,Log\,L_{Ly\alpha}$) obtained if only a single object were detected in the whole survey footprint. Errors on $\rm\Phi_{L_{Ly\alpha}}$ are computed as described in Sect. \ref{sec:errors_on_the_LF}, and show a clear prevalence of the completeness correction at the lowest luminosity bins. On the other hand, the bright-end of our LFs are dominated by the internal variance of our samples, as the number density of our candidates approaches the survey limit. To stress the impact of low-statistics on the bright tail, we marked with faded colors the data points at $\rm\leq\!1\,dex$ above the density limit.

\subsection{Comparison with previous determinations}
\label{sec:lya_LF_comparison_literature}
We compare our \lya LFs to a collection of previous determinations at similar z, after uniforming their underlain cosmology to the \texttt{PLANCK2015} one. This task is complicated by the significant differences between the technical features of J-PLUS and previous high-z \lya surveys \citep[as noted in e.g.,][]{blanc2011}. Indeed, these can reach up to $\sim\!5$ magnitudes in depth and a factor of $10^3$ on the surveyed area \citep[see e.g.,][]{ouchi2008,konno2016}. Nevertheless, the comparisons at $\rm z\!\sim\!2.25$ and $\rm z\!\sim\!2.37$ (respectively, $J$0395 and $J$0410 NBs) are remarkable, showing an overlap of our faint-end to the works of \cite{konno2016} and \cite{matthee2017b}.
\begin{figure*}[h]
\centering
\includegraphics[width=0.49\textwidth]{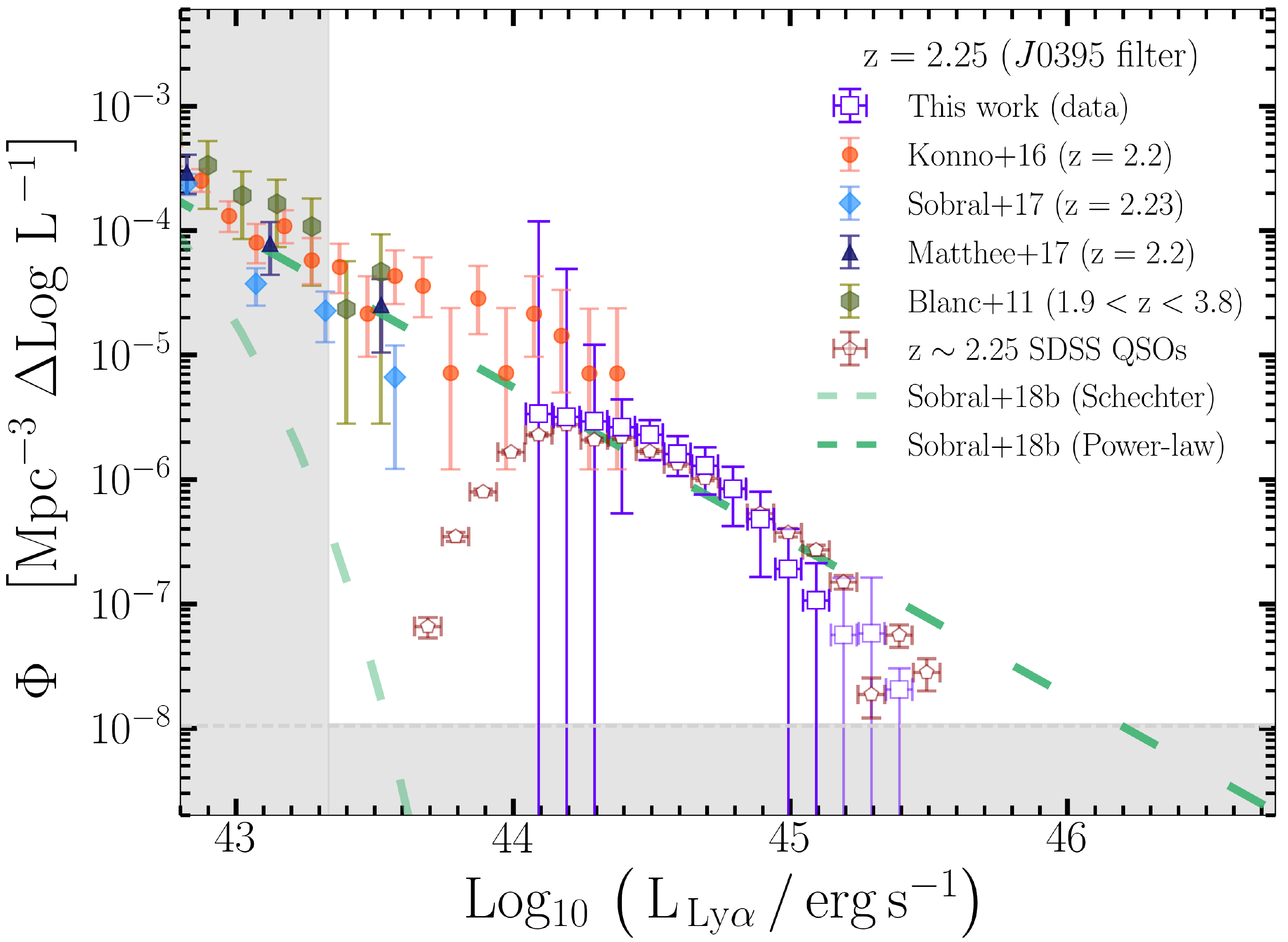}
\includegraphics[width=0.49\textwidth]{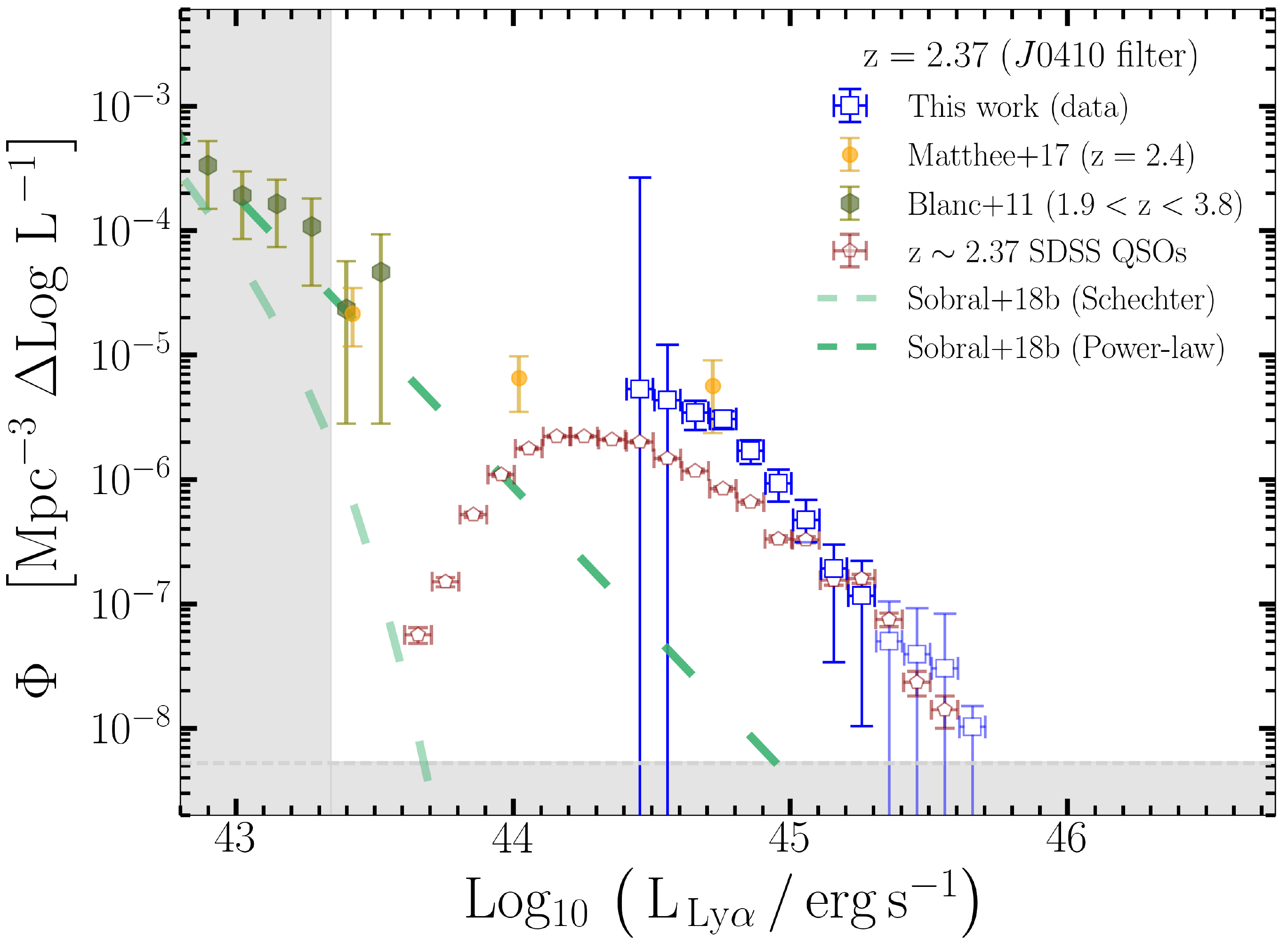}\\
\includegraphics[width=0.49\textwidth]{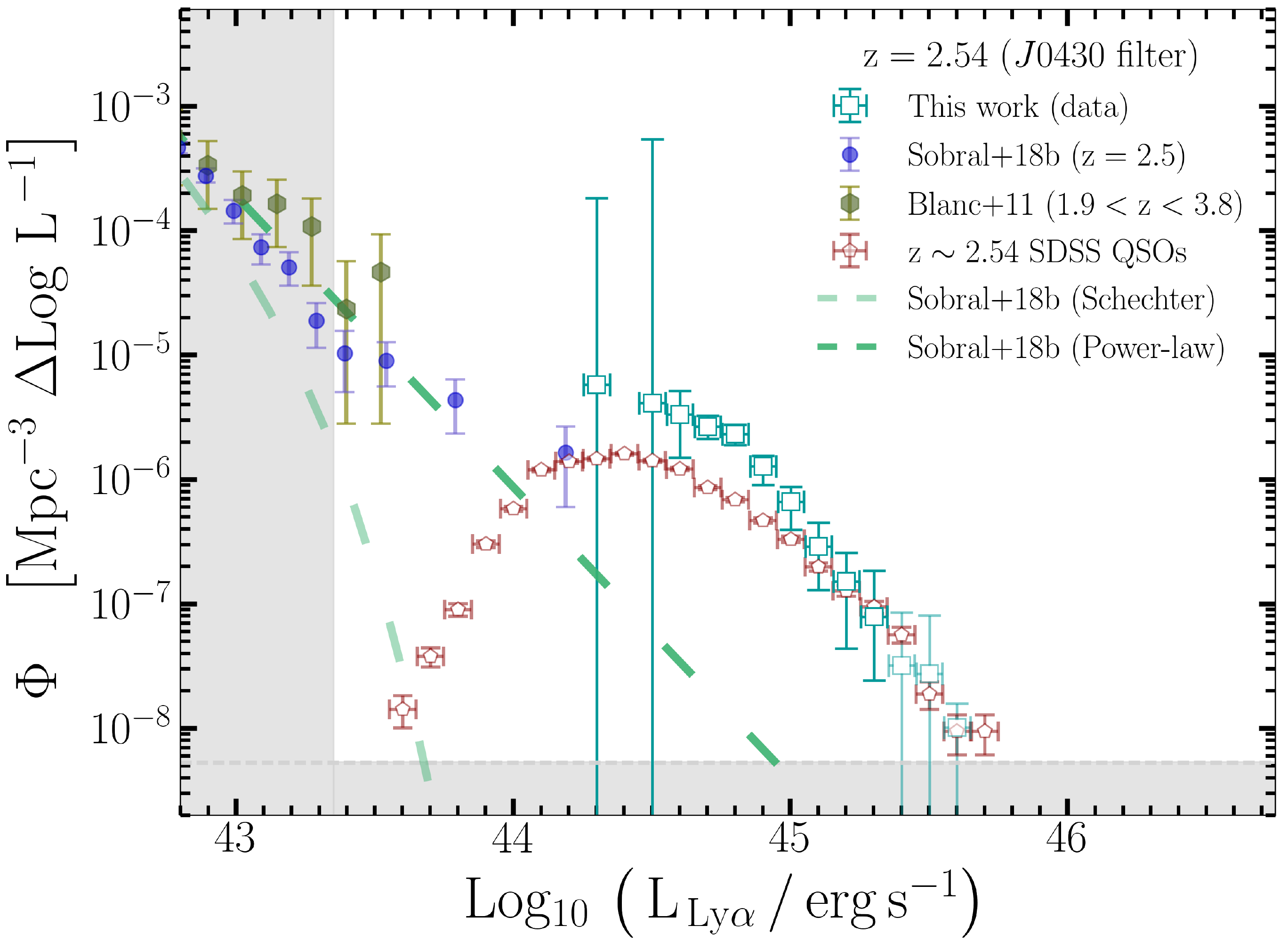}
\includegraphics[width=0.49\textwidth]{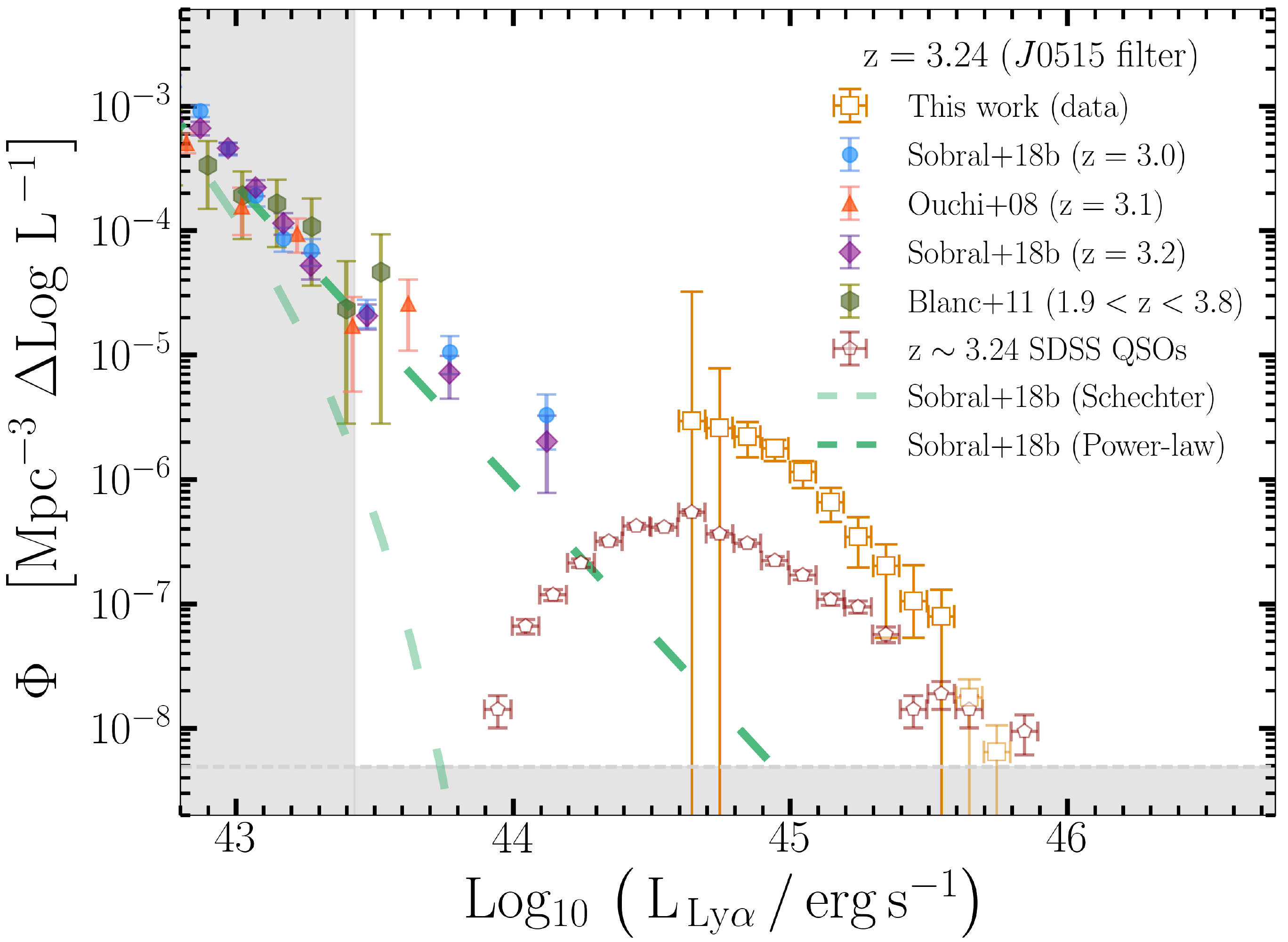}
\caption{\footnotesize \lya luminosity functions for each of the NB filter we used in our study (colored squares). The grey shaded areas show the \lya luminosity limit (vertical limit) and the limiting number density measurable by J-PLUS (horizontal limit). The wide area explored by the narrow-bands of J-PLUS survey allow to remarkably extend the range of luminosity sampled by previous studies (coloured circles, triangles hexagons and diamonds in each plot) and to explore previously-unconstrained $\rm L_{Ly\alpha}$ intervals. Dashed lines marks the best-fit determinations by \cite{sobral2018}, split respectively into a Schecther (light green) and a power-law contribution (dark green). Our results provide tight constraints at $\rm44.5\lesssim Log\,(L_{Ly\alpha}/erg\,s^{-1})\lesssim45.5$, a regime currently unexplored by previous \lya LFs determinations. Our errors are dominated by the completeness correction at low luminosity, while poor statistics due to low number counts (i.e. poissonian errors) dominate the bright tail of our distributions.}
\label{fig:lf}
\end{figure*}

We report the best fits from \cite{sobral2018} at each redshift since these highlight both the Schechter and power-law components of the LFs (respectively, light-green and dark-green dashed lines in Fig. \ref{fig:lf}).
These are obtained from a mixed Schechter/power-law model adapted to $\rm Log(L_{Ly\alpha}/\,erg\,s^{-1})\!\lesssim\!44.5$ data, showing a transition between the two regimes at $\rm Log(L_{Ly\alpha}/\,erg\,s^{-1})\!\sim\!43.5$. Despite the small overlap of luminosity regimes, our $\rm z\!\sim\!2.25$ LF shows a remarkably good agreement with the power-law of \cite{sobral2018}, as shown in the upper-left panel of Fig. \ref{fig:lf}. Interestingly, this component well accounts for the population of X-ray bright objects in their samples, suggesting that these sources might belong to a separate class described by a different luminosity distribution than SF LAEs at $\rm Log(L_{Ly\alpha}/\,erg\,s^{-1})\!\lesssim\!43.3$. On the contrary, a significant discrepancy between our data and the power-law components is evident at higher z. We ascribe this to the wider separation between the $\rm L_{Ly\alpha}$ ranges probed by our data and those on which the fits of \cite{sobral2018} are obtained at these z.

We note that our $\rm z\!\sim\!2.25$ data nicely complement also the bright-end determination of \cite{konno2016} (orange dots in the upper-left panel of Fig. \ref{fig:lf}). This work clearly showed an excess with respect to the exponential decay of a Schechter function at $\rm Log(L_{Ly\alpha}/\,erg\,s^{-1})\!\gtrsim\!43$. Their explanation relied on the contribution of a population of \lya$\!\!$-emitting AGN/QSOs, as in e.g. \cite{matthee2017b} and \cite{sobral2018}. By joining these hints to the results of our spectroscopic follow-up and our sample analysis (Sect. \ref{sec:gtcProgram} and \ref{sec:LAE_samples}), our work further supports the picture according to which \lya$\!\!$-emitting AGN/QSOs are responsible for the bright-end excess observed on the $\rm2\!\lesssim\!z\!\lesssim\!3$ \lya luminosity function at $\rm43.3\!\lesssim\! Log(L_{Ly\alpha}/\,erg\,s^{-1})\!\lesssim\!44.5$.

\subsubsection{Comparison with SDSS DR14 QSOs}
\label{sec:lya_LF_comparison_SDSS_QSOs}
Figure \ref{fig:lf} additionally shows the \lya LF of all the \texttt{DR14} QSOs in the J-PLUS footprint \citep[from][]{paris2018}, with spectroscopic redshift in the intervals sampled by each NB (red pentagons). We obtain this determination by performing synthetic photometry of SDSS QSOs with J-PLUS filters and applying the same flux corrections as those computed for our data (see Sect. \ref{sec:line_flux_retrieval}). For simplicity, we only associate poissonian errors to the SDSS LF.

Despite the comparison being only qualitative, the agreement between the SDSS QSOs distribution and our data is good, especially at low z. Interestingly, the fraction of our genuine candidates showing SDSS QSOs counterparts at the redshift probed by each NB
is $\lesssim30\%$, in each NB. Assuming that the \cite{paris2018} catalog represents a $\sim\!100\%$ complete sample of QSOS and considering the low fraction of SDSS QSOs in our data, the agreement between the two LFs could be explained in terms of a significant residual contamination of our samples ($\sim\!70\%$). Nevertheless, this is in contrast with both our purity estimates and our spectroscopic follow-up (Sect. \ref{sec:purity_of_samples} and \ref{sec:GTC_program_results}). A more interesting explanation is that our NB-based selection might actually be sensitive to high-z QSOs which lack spectroscopic determination in SDSS \citep[due e.g. to their BB colors, see][]{ross2012, richards2015}, as those confirmed by our follow-up programs. Indeed, their previous classification based on SDSS photometry and morphology would identify most of them just as compact objects (namely stars, see Table \ref{tab:GTC_obs_summary}). We suggest that this mis-classification might originate from the SDSS target selection, based on BB-colors, which might miss the presence of emission lines. On the contrary, our selection targets photometric excesses with respect to a continuum estimate, hence it can efficiently select high-z line emitters.
\begin{figure*}[t]
\centering
\includegraphics[width=0.49\textwidth]{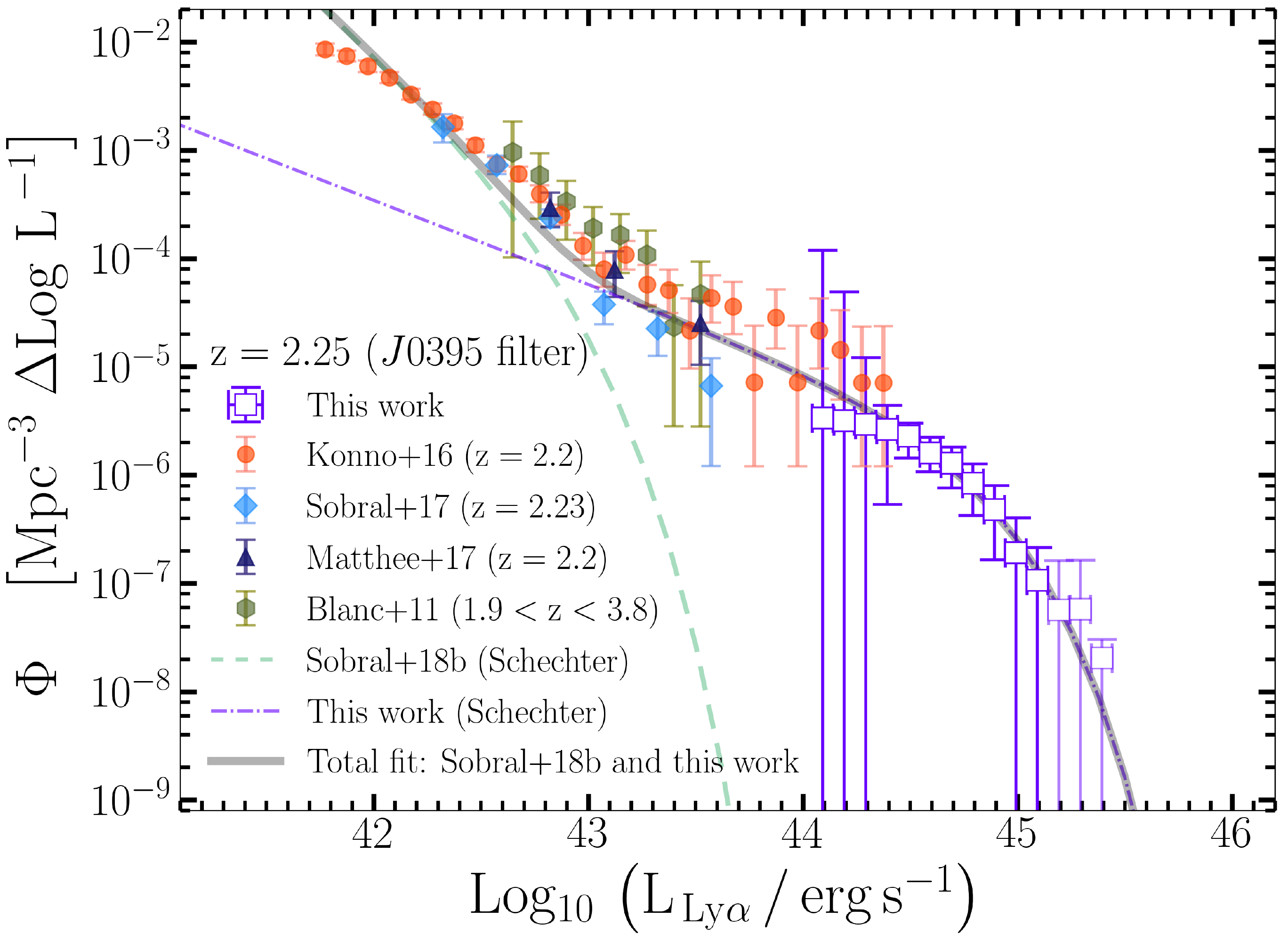}
\includegraphics[width=0.49\textwidth]{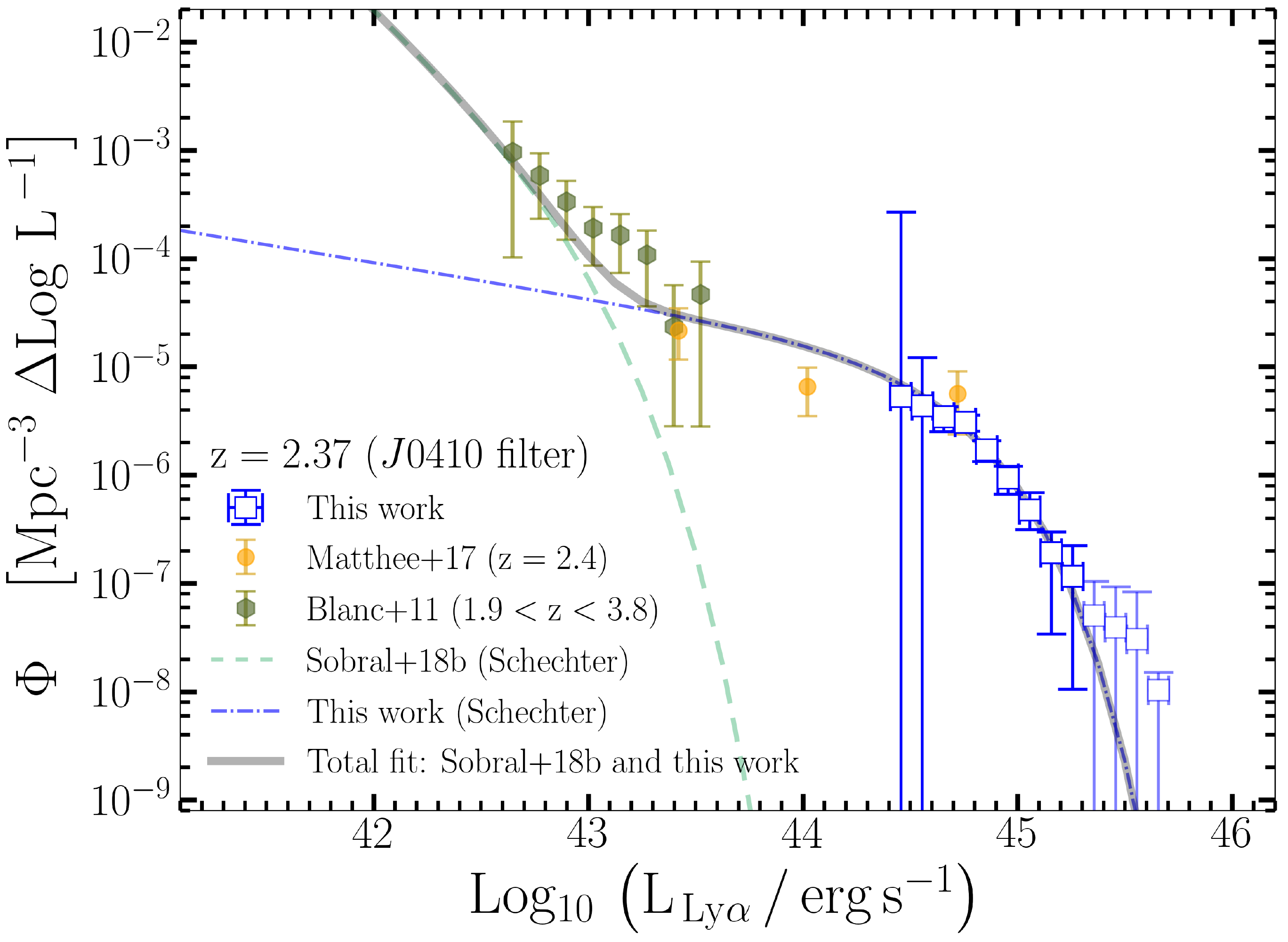}\\
\includegraphics[width=0.49\textwidth]{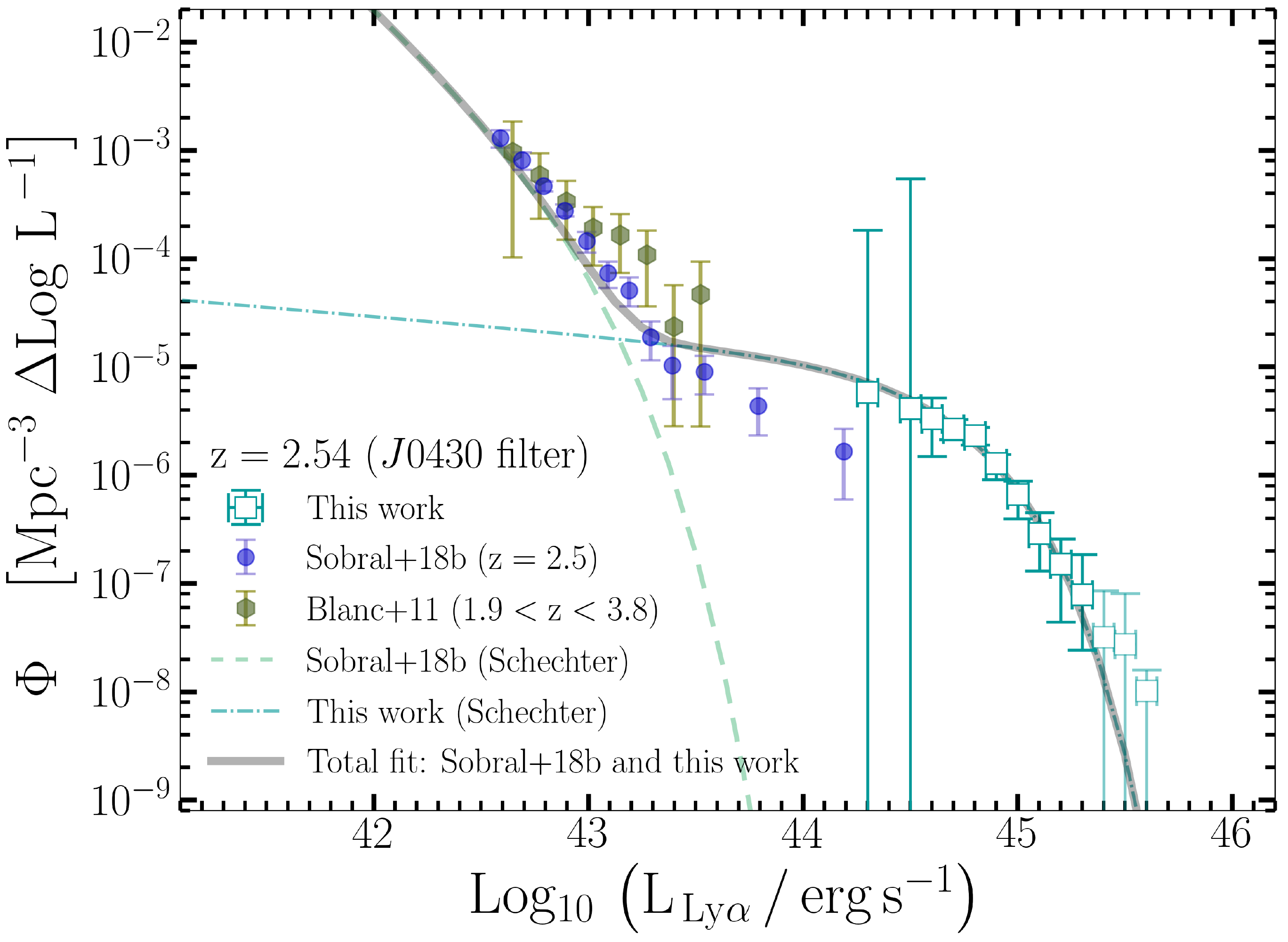}
\includegraphics[width=0.49\textwidth]{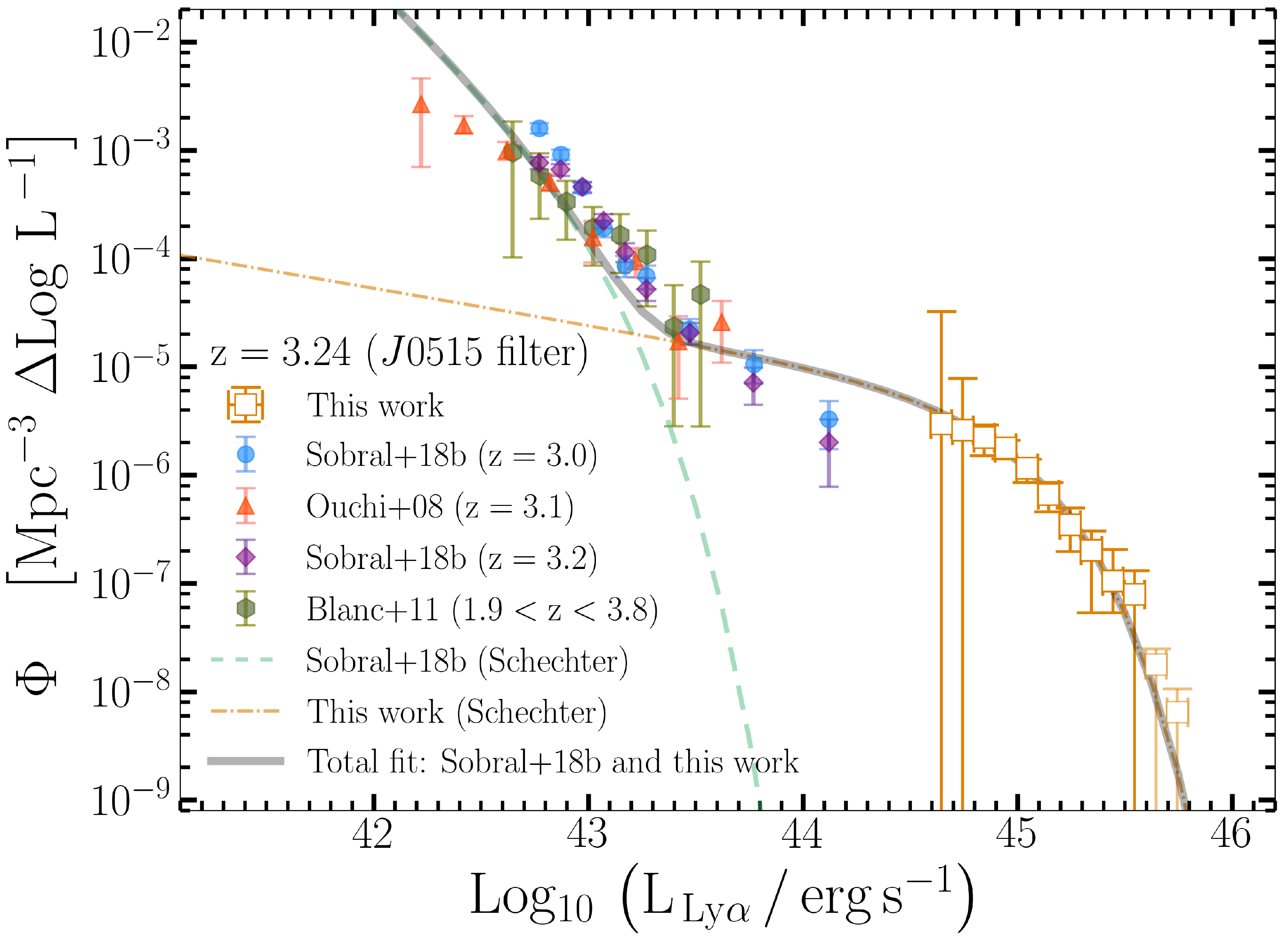}
\caption{\footnotesize Joint fit of our \lya luminosity functions and literature data with a double-Schechter model (grey solid lines in each panel). This is obtained by joining the best Schechter fit from \cite{sobral2018} at each redshift (green dashed lines in each panel) and a second Schechter function (coloured dashed-dotted lines). We jointly fit this double-Schechter model to both our data and the literature ones leaving free the parameters of the second Schechter, in order to constrain its faint-end slope $\alpha$ at each redshift.}
\label{fig:lf_doubleSchechter}
\end{figure*}

\subsection{$\rm Ly\alpha$ LF parameters}
\label{sec:schechter_fits}
\subsubsection{The faint-end slope: power-law or double-Schechter?}
\label{sec:faint_end_slope}
As suggested by e.g. \cite{konno2016, matthee2017b, sobral2018, sobral2018b} and \cite{calhau2020}, the population of bright \lya$\!\!$-emitting sources at $\rm Log(L_{Ly\alpha}/erg\,s^{-1})\!>\!43$ is likely to be composed by a mixture of SF LAEs and AGN/QSOs. In particular, \cite{matthee2017b} and \cite{sobral2018} suggest that the two source classes might be described by substantially different distributions in terms of typical number density and \lya luminosity. Interestingly, the power-law component of their studies can be explained as the faint-end of a Schechter function \citep[][see also Eq. \ref{eq:log_form_schechter}]{schechter1976} 
describing the QSOs luminosity distribution. Our data can effectively support this hypothesis by providing the bright-end complement to the AGN/QSOs Schechter distribution. At the same time, our analysis limited by the J-PLUS depth which prevents us to constrain its the faind-end slope at $\rm Log(L_{Ly\alpha}/erg\,s^{-1})\!\lesssim\!44$. This might significantly influence the determination of our Schechter paramters given their mutual correlation.
Instead of fixing the faint-end slope to a fiducial value \citep[as in e.g.,][]{gunawardhana2015, sobral2018}, we compute it by jointly exploiting our data and previous \lya LF determinations, over the whole interval $\rm41.5\!\lesssim\!Log(L_{Ly\alpha}/erg\,s^{-1})\!\lesssim\!44$. More in detail, we make use of the Schechter component from \cite{sobral2018} at each redshift to describe the \lya LF at $\rm Log(L_{Ly\alpha}/erg\,s^{-1})\!\lesssim\!43.3$, and combine it to a second Schechter function to account for $\rm Log(L_{Ly\alpha}/erg\,s^{-1})\!\gtrsim\!44.$. We then vary the faint-end slope of the latter and, for each $\alpha$, we jointly fit the complete double-Schechter model to both our data and all the literature determinations (see Fig. \ref{fig:lf_doubleSchechter}). Finally, for each NB we obtain $\alpha$ and its errors from the reduced $\chi^2$ distribution of the double-Schechter fits, namely: $\alpha_{\,J0395}\!=\!-1.77^{+0.09}_{-0.07}$, $\alpha_{\,J0410}\!=\!-1.33_{-0.22}^{+0.50}$, $\alpha_{\,J0410}\!=\!-1.17_{-0.13}^{+0.19}$ and $\alpha_{\,J0515}\!=\!-1.34_{-0.09}^{+0.12}$.
\begin{figure*}[t]
\centering
\includegraphics[width=0.49\textwidth]{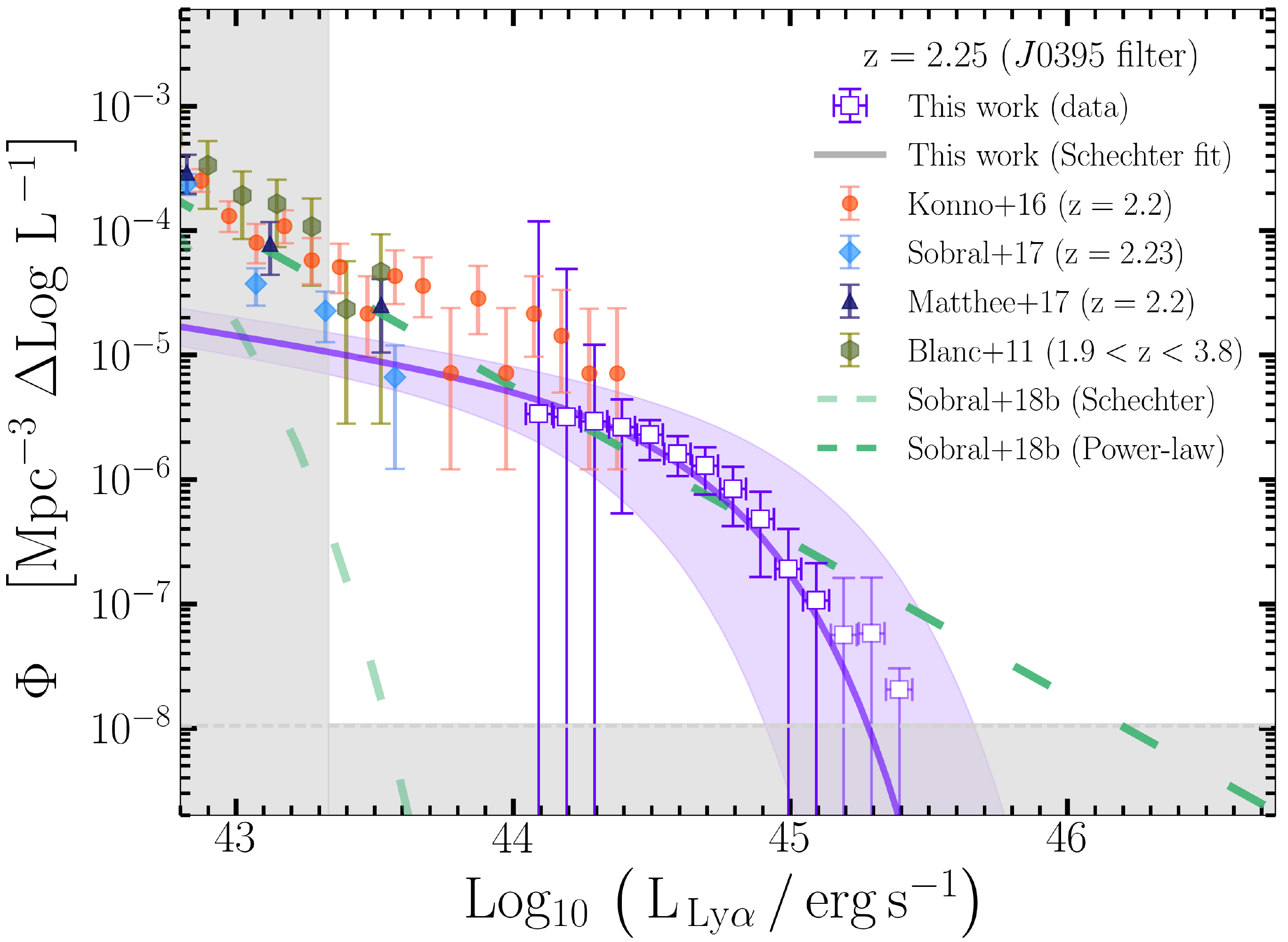}
\includegraphics[width=0.49\textwidth]{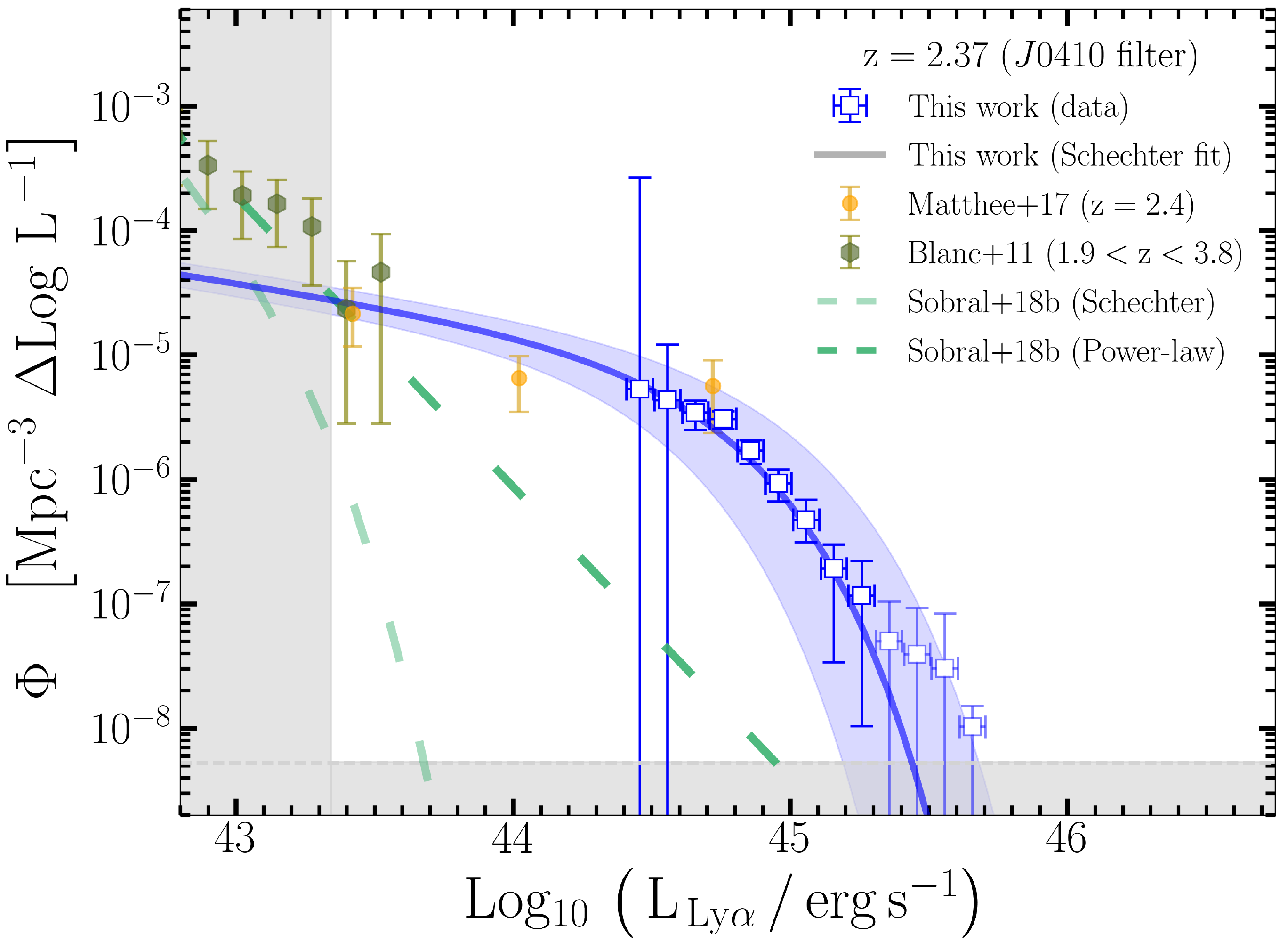}\\
\includegraphics[width=0.49\textwidth]{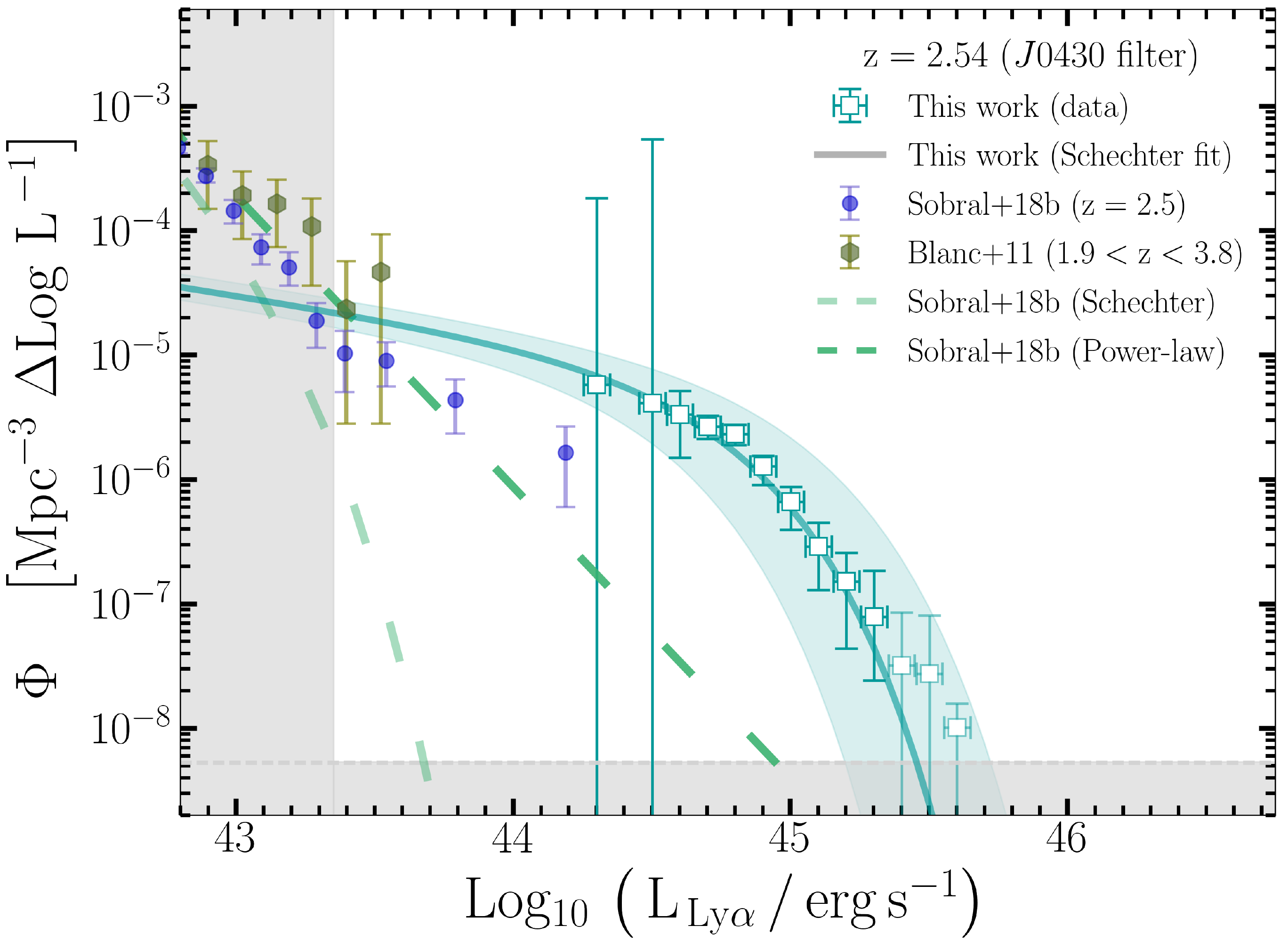}
\includegraphics[width=0.49\textwidth]{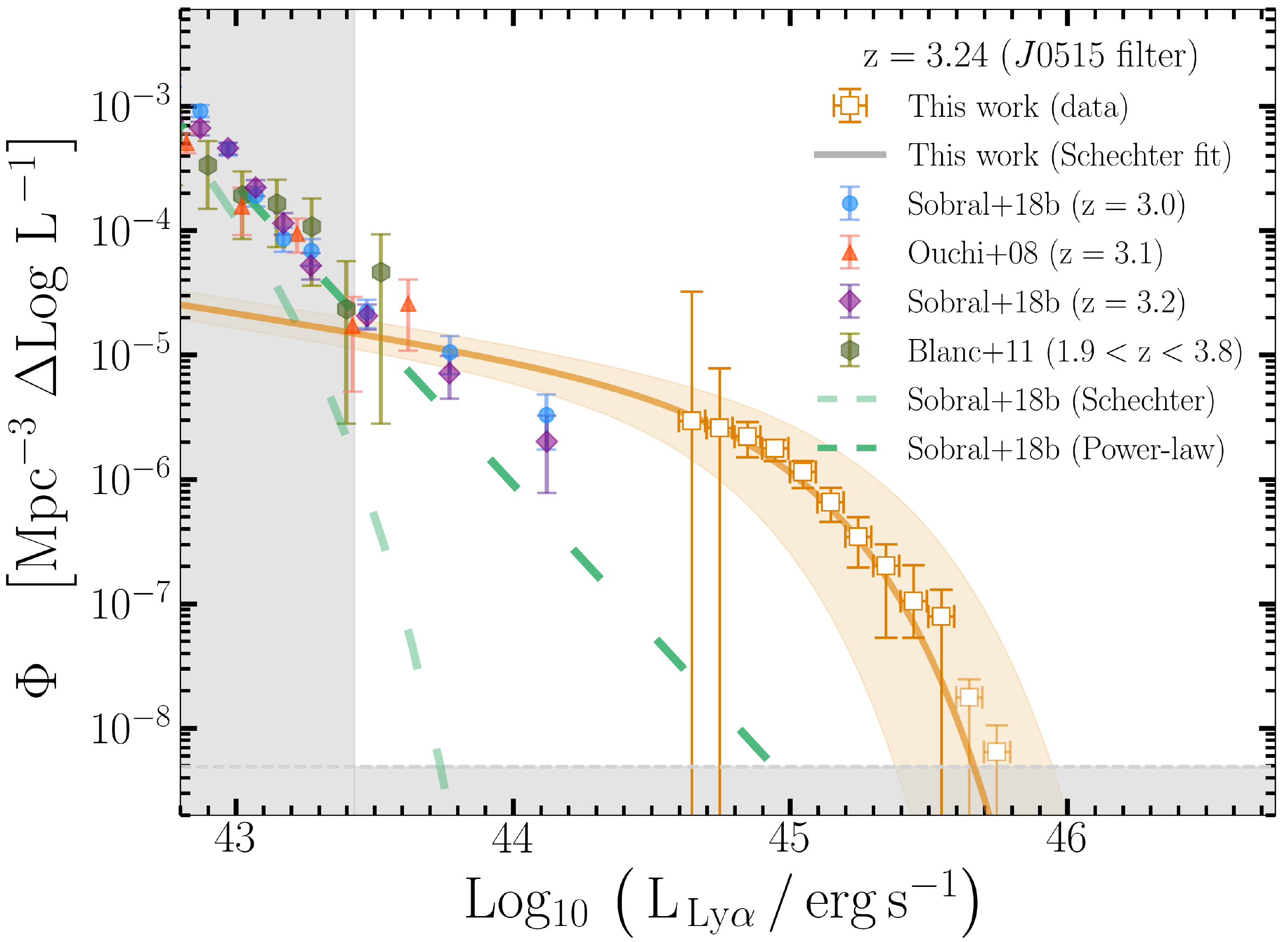}
\caption{\footnotesize Final Schechter fits of our \lya LFs (colored solid lines in each panel) performed by keeping the faint-end slope fixed to $\alpha=-1.35\pm0.84$. The colored shaded regions in each panel mark the $1\sigma$ confidence regions for the $\Phi^*$ and $\rm L^*$ parameters obtained by sampling their associated errors, obtained via monte-carlo simulations (see Sect. \ref{sec:phistar_Lstar_parameters}). The literature data shown in each panel are the same as in Fig. \ref{fig:lf}.}
\label{fig:lf_singleSchechter}
\end{figure*}

We further assume no evolution of $\alpha$ with respect to redshift since neither our data nor previous works would allow to constrain it. Under this assumption, we obtain our final $\alpha$ as the weighted average of the above values: $\rm\alpha=-1.35\pm0.84$. This high uncertainty is expected, given the limited amount of data populating the transition-regime between the two Schechter functions at $\rm Log(L_{Ly\alpha}/erg\,s^{-1})\!\sim\!43.5$ (see Fig. \ref{fig:lf_doubleSchechter}). Nevertheless, our procedure consistently accounts for available data over $\rm \sim\!3\,dex$ in luminosity, providing one of the first estimates of $\alpha$ for the Schechter LF of \lya$\!\!$-emitting sources at $\rm Log(L_{Ly\alpha}/erg\,s^{-1})\!\gtrsim\!44$. Few works have currently estimated the LF shape at these very bright regimes by usually performing a power-law fit \citep[e.g.,][]{matthee2017b,sobral2018}. Interestingly, these works respectively determined values of $(\alpha+1)=-0.75^{+0.17}_{-0.17}$ and  $(\alpha+1)=-0.74^{+0.17}_{-0.17}$ at $\rm z\!\sim\!2.2$, which are both consistent with our faint-end slopes determinations at $\rm z\!<\!2.5$ within $1\sigma$. This suggests that the power-law component observed at the bright end by previous works might be explained as the faint-end of a Schechter function describing the distribution of extremely luminous \lya$\!\!$-emitting sources (i.e. AGN/QSOs). In other words, the full \lya luminosity function at $\rm41.5\!\lesssim\!Log(L_{Ly\alpha}/erg\,s^{-1})\!\lesssim\!44$ could be effectively described by a double-Schechter model.

\subsubsection{Constraints on $\Phi^*$ and $\rm L^*$}
\label{sec:phistar_Lstar_parameters}
\begin{figure*}[t]
\centering
\includegraphics[width=0.94\textwidth]{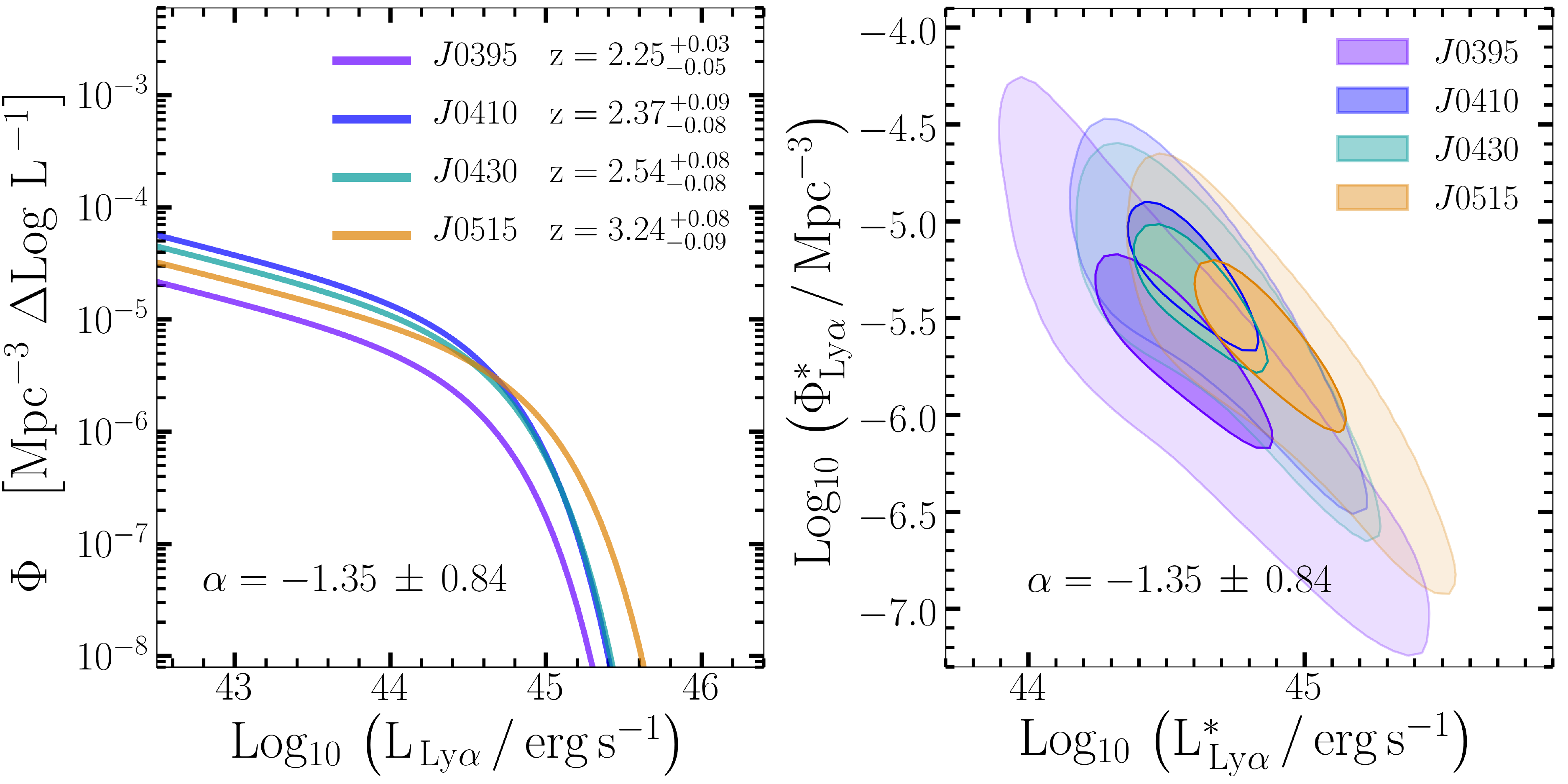}
\caption{\footnotesize Left panel: single-Schechter fits to our data computed with the fixed faint-end slope $\alpha\!=\!-1.35\pm0.84$ obtained as in Sect. \ref{sec:faint_end_slope}. We note that the difference among the four determinations (factor of $\sim\!2$ both in luminosity and normalization) are absorbed by the errors on the Schechter parameters (right panel). Right panel: distribution of $\rm\Phi^*$ and $\rm L^*$ obtained from the monte-carlo sampling of $\alpha$ errors. The contours mark the levels including $86\%$ and $39\%$ of the monte-carlo realizations (respectively faint and dark contours). This analysis shows that the parameters of the four determinations are statistically consistent, hence we do not observe hints for an evolution of the $\rm2\!\lesssim\!z\!\lesssim\!3.3$ \lya LF at $\rm Log(L_{Ly\alpha}/erg\,s^{-1})\gtrsim43.5$.}
\label{fig:lf_banana_plot}
\end{figure*}

We employ the fixed $\alpha$ computed with the above procedure to fit our data with a single-Schechter model and constrain $\rm\Phi^*$ and $\rm L^*$ at $\rm Log(L_{Ly\alpha}/erg\,s^{-1})\!>\!44$. We stress that for this step we explicitly use only our data points. The results of this procedure are compared to literature data in Fig. \ref{fig:lf_singleSchechter}, while the left panel of Fig \ref{fig:lf_banana_plot} directly compares our four redshift bins. We account for correlations between $\alpha$ and the remaining parameters by sampling the error of $\alpha$ (assumed to be Gaussian) with 50,000 monte-carlo realizations oof the single-Schechter fits, from which we extracting our final values and errors for $\rm\Phi^*$ and $\rm L^*$. Our results are listed in Table \ref{tab:schechter_parameters} and shown in the right panel of Fig. \ref{fig:lf_banana_plot}.
\begin{table}[t]
    \centering
    \resizebox{9cm}{!}{
    \begin{tabular}{c|c|c|c|c}
    \hline
    Filters  &  z  &  $\alpha$  &  $\rm\Phi^*\ [10^{-6}\,Mpc^{-3}]$  &  $\rm Log(L^*/erg\,s^{-1})$\\
    \hline
    \hlx{vvv}
    $J$0395    &  $2.25^{\,+0.03}_{\,-0.05}$  & $-1.35\pm0.84$ & $1.86^{\,+4.14}_{\,-1.60}$ &  $44.54^{\,+0.43}_{\,-0.35}$  \\
    \hlx{vvv}
    $J$0410    &  $2.37^{\,+0.09}_{\,-0.08}$  & $-1.35\pm0.84$ & $4.66^{\,+6.03}_{\,-3.25}$ &  $44.60^{\,+0.29}_{\,-0.21}$ \\
    \hlx{vvv}
    $J$0430    &  $2.53^{\,+0.09}_{\,-0.07}$  & $-1.35\pm0.84$ & $3.61^{\,+4.40}_{\,-2.57}$ &  $44.63^{\,+0.30}_{\,-0.22}$ \\
    \hlx{vvv}
    $J$0515    &  $3.24^{\,+0.07}_{\,-0.10}$  & $-1.35\pm0.84$ & $2.12^{\,+3.56}_{\,-1.55}$ &  $44.87^{\,+0.32}_{\,-0.26}$ \\
    \hlx{vvv}
    \hline
    \end{tabular}}
    \caption{\footnotesize Schechter parameters computed on our data by fixing the faint-end slope to $\alpha\!=\!-1.35\pm0.84$. The latter value was obtained as described in Sect. \ref{sec:faint_end_slope}. Errors on $\rm\Phi^*$ and $\rm L^*$ are obtained from their corresponding 1D distributions computed via monte-carlo sampling of $\alpha$ errors.}
    \label{tab:schechter_parameters}
\end{table}

Under the hypothesis that our samples are greatly dominated by AGN/QSOs, our results show that their LF is described by a clearly distinct distribution with respect to SF LAEs \citep[see also][]{matthee2017b}. In particular, by comparing our $\rm\Phi^*$ and $\rm L^*$ to previous determinations at $\rm Log(L_{Ly\alpha}/erg\,s^{-1})\!<\!43$ \citep{gronwall2007,ouchi2008,konno2016}, we measure a typical density and luminosity of AGN/QSOs respectively $\rm\sim\!\!3\,dex$ lower and  $\rm\sim\!\!2\,dex$ higher, as already suggested by e.g. \cite{matthee2017b} and \cite{sobral2018}. In turn, this would suggest that the transition between the regime dominated respectively by SF LAEs and AGN/QSOs would fall at $\rm Log(L_{Ly\alpha}/erg\,s^{-1})\!\sim\!43.5$, as also highlighted by \cite{sobral2018b} and \cite{calhau2020}.

Finally, our data do not allow to constrain the evolution of our \lya LFs determinations. Indeed the $\rm\Phi^*$ and $\rm L^*$ we obtain are statistically consistent (at $\sim2\sigma$) among the four filters, with average values $\rm\Phi^*\!=\!(3.33\pm0.19)\times10^{-6}\,Mpc^{-3}$ and $\rm L^*\!=\!44.65\pm0.65\,erg\,s^{-1}$. This is shown in the right panel of Fig. \ref{fig:lf_banana_plot}, where the faint and dark contours for each filter respectively mark the 2-$\sigma$ and 1-$\sigma$ levels (i.e. the $\!86\%$ and $\!39\%$ iso-contours) of the parameters distributions obtained from monte-carlo realizations. The wide overlap between the four filters shows the low constraining power of our data towards the evolution of the LF parameters with redshift. This was anticipated by the significant variation among the distributions of $\rm L_{\,Ly\alpha}$ and EW at each z shown in Fig. \ref{fig:ew_and_lyalum_distributions}, which ultimately hinders the possibility to disentangle the intrinsic variations of our sample properties from systematic effects. We note that 

\subsection{The AGN fraction of $\rm z\!\gtrsim\!2$ LAEs}
\label{sec:AGN_fractions_from_LFs}
By assuming that our \lya LF describes the distribution of only AGN/QSOs, we can build a simple toy model to estimate the relative fraction AGN/QSOs and SF LAEs as a function of \lya luminosity. We define the latter as:
\begin{equation}
    \rm q_{\,AGN}=\frac{LF^{\,SF\ LAEs}}{LF^{\,SF\ LAEs} + LF^{\,AGN/QSOs}}\ ,
    \label{eq:agn_fraction}
\end{equation}
where $\rm LF^{\,AGN/QSOs}$ is one of the four determinations of the Schechter function computed from our data, while $\rm LF^{\,SF\ LAEs}$ is the best fit of \cite{sobral2018} at the corresponding redshift. We use the latter since it is obtained by excluding LAE candidates with X-ray counterparts from the determination of the Schechter fit. Consequently, we assume it provides a fair estimate for the luminosity distribution of only SF LAEs. We underline that our estimate of $q_{\,AGN}$ is an illustrative application of our results rather than a rigorous measurement, given the strong assumptions on which it is based.
\begin{figure}[h]
    \centering
    \includegraphics[width=0.49\textwidth]{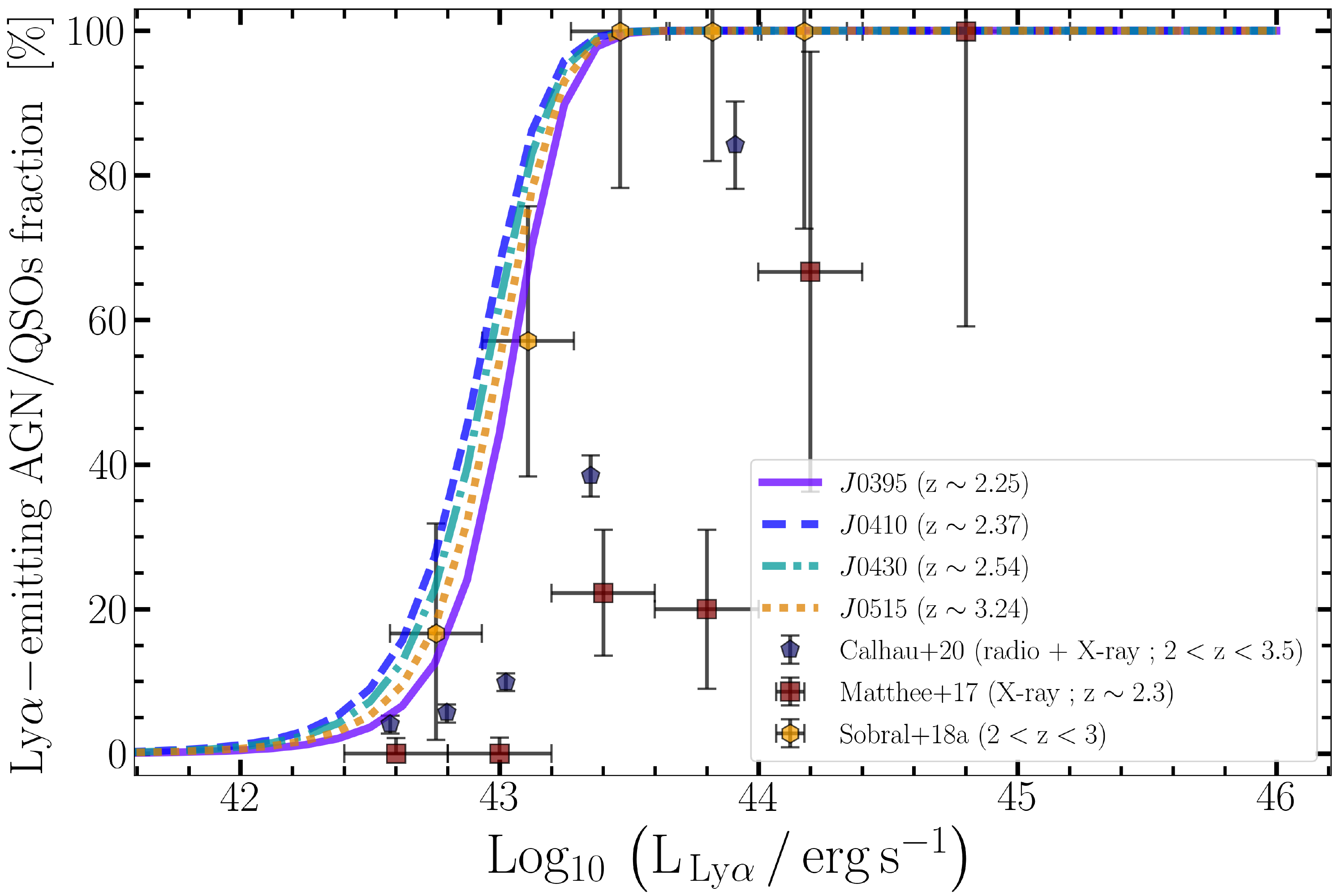}
    \caption{\footnotesize The AGN/QSO fraction as a function of luminosity for each NB. We estimated this quantity by assuming that our results are entirely dominated by AGN/QSOs and that the best Schechter fit of \cite{sobral2018} describes the distribution of SF LAEs (see Eq. \ref{eq:agn_fraction}). Our results are in agreement with the spectroscopic determination of \cite{sobral2018b}, which only employs \lya emission pre-selection for their targets. On the other hand, the estimates of \cite{matthee2017b} and \cite{calhau2020} are based on the detection of either X-ray or radio counterparts for their \lya$\!\!$-emitting candidates.}
    \label{fig:AGN_fractions}
\end{figure}

The AGN/QSOs fractions for all the redshifts we probe are shown in figure \ref{fig:AGN_fractions}. Despite our simplifying assumptions, we find a good agreement (within $1\,\sigma$) with the measurements of \cite{sobral2018b}, which are obtained from a spectroscopic follow-up of \lya$\!\!$-selected targets. On the contrary, the works of \cite{matthee2017b} and \cite{calhau2020} (also shown in Fig. \ref{fig:AGN_fractions} for comparison) are based on photometric selections which identify AGN/QSOs candidates on the basis of their X-ray and/or radio-loudness. The latter are likely to be significant only for a sub-sample of AGN/QSOs \citep[as suggested by e.g.,][and \citealt{calhau2020}]{sobral2018}, hence the discrepancy with our estimates might also be explained in terms of this incompleteness effect. 

To conclude, the good agreement between our AGN/QSOs fraction estimates and the data of \cite{sobral2018b}  supports the scenario by which our samples are strongly dominated by \lya$\!\!$-emitting AGN/QSOs. Furthermore, the discrepancy with respect to X-ray/Radio selected AGN candidates suggests that the latter are likely a sub-sample of the whole high-z AGN/QSOs population. Our selection, on the contrary, is only based on \lya emission, hence it is likely to detect previously-unidentified high-z AGN/QSOs. This is also in line with the results of our spectroscopic follow-up program (section \ref{sec:GTC_program_results}).

\section{Conclusions}
\label{sec:discussion_and_conclusions}
This work presents the determination of the bright-end of the \lya luminosity function at four redshifts in the interval $\rm 2\!\lesssim z\lesssim\!3.3$, namely $\rm z\!=\!2.25^{\,+0.03}_{\,-0.05}$, $\rm z\!=\!2.37^{\,+0.09}_{\,-0.08}$, $\rm z\!=\!2.54^{\,+0.08}_{\,-0.08}$ and $\rm z\!=\!3.24^{\,+0.08}_{\,-0.09}$. We obtain the LFs by employing four lists of \lya$\!\!$-emitting candidates selected in DR1 catalog of the J-PLUS survey, according to the significance of their photometric excess in the $J$0395, $J$0410, $J$0430 and $J$0515 narrow-bands.

We select 2547, 5556, 4994 and 1467 bright candidates ($\rm L_{Ly\alpha}>2\times10^{\,43}\,erg\,s^{-1}$), which jointly represent the largest sample of photometric \lya$\!\!$-emitting candidates at $\rm2\!\lesssim z\lesssim\!3.3$ to date. We expect our lists to include both bright star-forming LAEs (SF LAEs) and \lya$\!\!$-emitting AGN/QSOs. To identify either of these source classes in our samples, we follow-up spectroscopically a random sub-sample of our candidates (section \ref{sec:gtcProgram}). The spectroscopir data confirmed 40 out of 45 targets as genuine high-z line-emitters (with 29 out of 45 being $\rm z\!>\!2$ \lya$\!\!$-emitting QSOs) and found no star-forming LAE. In addition, we look for bi-modalities in the photometric properties of our candidates, such as \lya luminosity and EW (section \ref{sec:EW_LLya_distributions}) or colors (section \ref{sec:LAE_QSOfractions}). Overall, the properties of our candidates are consistent with those of spectroscopically-confirmed QSOs (Fig. \ref{fig:color-color_candidates_and_QSOs}) and high-z QSO templates (Fig. \ref{fig:wise-color_vs_redshift}), suggesting that the fraction of SF LAEs in our samples is negligible.

We use our candidates samples to compute the \lya LF at extremely-bright luminosity regimes for the first time, namely at $\rm44\!\lesssim\!Log(L_{Ly\alpha}/erg\,s^{-1})\!\lesssim\!45.5$, and extend by $\rm\gtrsim\!1.5\,dex$ the intervals covered by previous determinations. The extensive area observed by J-PLUS DR1 allows to access  wide cosmological volumes ($\rm\gtrsim\!1\,Gpc^3$), hence to probe number densities as low as $\rm\sim\!\!10^{-8}\,Mpc^{-3}$. This parameters-space region is unprecedented for surveys focused on bright photometrically-selected \lya$\!\!$-emitting sources. Interestingly, our \lya LFs are in line with previous results at $\rm Log(L_{Ly\alpha}/erg\,s^{-1})\!\gtrsim\!43.5$, prolonging their power-law end into a full-developed Schechter function. We derive the redshift-averaged parameters $\rm\Phi^*\!=\!(3.33\pm0.19)\times10^{-6}\,Mpc^{-3}$, $\rm L^*\!=\!44.65\pm0.65\,erg\,s^{-1}$ and $\rm\alpha\!=\!-1.35\pm0.84$ for our Schechter best-fits. This shows that the whole \lya LF, i.e. from $\rm Log(L_{Ly\alpha}/erg\,s^{-1})\!<\!42$ up to $\rm Log(L_{Ly\alpha}/erg\,s^{-1})\!>\!45$, can be effectively described by a composite model of two Schechter functions, respectively accounting for the distribution of SF LAEs and bright AGN/QSOs. These two distributions appear to be structurally different, with $\rm L^*_{QSOs}\!\sim\!100\,L^*_{SF\ LAEs}$, $\rm\Phi^*_{QSOs}\sim10^{-3}\,\Phi^*_{SF\ LAEs}$ and a transition-regime centered at $\rm Log(L_{Ly\alpha}/erg\,s^{-1})\!\sim\!43.5$ \citep[in line with e.g.,][]{konno2016,matthee2017b,sobral2018b,calhau2020}. On the whole, our results support the scenario suggested by e.g. \cite{konno2016, matthee2017b} and \cite{sobral2018}, according to which the excess of bright LAEs measured at $\rm Log\,\,L_{Ly\alpha}\gtrsim43$ with respect to a Schechter distribution is due to a population of AGN/QSOs \citep[see also][]{calhau2020}. Our findings characterize for the first time this population as being $\rm\sim\!100$ times more luminous and $\rm\sim\!1000$ times less dense than that of SF LAEs at comparable redshifts.

In addition, $\rm\sim\!70\%$ of our \lya$\!\!$-emitting candidates lacks any spectroscopic confirmation by current surveys. Based on our spectroscopic follow-up results, we suggest that our samples are dominated by high-z QSOs which are not yet identified as such, but rather mis-classified as stars by current archival data, due to their photometric colors. Indeed, even accounting for a conservative residual contamination of $\sim35\%$ in our final samples, the number of genuine $\rm z\!\gtrsim\!2$ QSOs identified for the first time by our methodology would be approximately 1300, 3200, 2900 and 900, respectively for $J$0395, $J$0410, $J$0430 and $J$0515 J-PLUS NBs. 
We ascribe this possibility to the narrow-band excess detection of our methodology, which can be particularly effective in targeting and selecting the strong line-emission features of $\rm z\!>\!2$ AGN/QSOs. Indeed, these might be missed by spectroscopic target selection based only on broad-band colors \citep[e.g.,][]{richards2009,ivezic2014,richards2015}. We stress that the confirmation of this speculative hypothesis must rely on a systematic and extensive confirmation of our candidates. The latter might be obtained via either spectroscopic analysis or by exploiting the very efficient source identification provided by multi-NB imaging. Indeed, the upcoming J-PAS survey can provide a natural setting to extend our work.

Finally, our data do not show significant evolution of the LF over the probed redshifts. Despite X-ray studies suggest little evolution of the $\rm 2\!<\!z\!<\!3.3$ AGN/QSOs population \citep[e.g.,][]{hasinger2007}, our findings might also be affected by J-PLUS detection limits. This factor could be mitigated by deeper photometric imaging, which is hardly attainable by future J-PLUS data releases. Indeed, the technical features of the T80 (80cm) telescope hinder the possibility of reaching higher depth than the nominal J-PLUS one over very wide sky areas. On the contrary, future multi-NB wide-area photometric surveys can provide a valid tools to test the LF evolution at $\rm Log\,\,L_{Ly\alpha}\gtrsim43.5$.

\begin{acknowledgements}
Based on observations made with the JAST/T80 telescope for J-PLUS project at the Observatorio Astrof\'isico de Javalambre in Teruel, a Spanish Infraestructura Cientifico-T\'ecnica Singular (ICTS) owned, managed and operated by the Centro de Estudios de F\'isica del Cosmos de Arag\'on (CEFCA). Data has been processed and provided by CEFCA's Unit of Processing and Archiving Data (UPAD). Funding for the J-PLUS Project has been provided by the Governments of Spain and Arag\'on through the Fondo de Inversiones de Teruel; the Arag\'on Government through the Research Groups E96, E103, and E16\_17R; the Spanish Ministry of Science, Innovation and Universities (MCIU/AEI/FEDER, UE) with grants PGC2018-097585-B-C21 and PGC2018-097585-B-C22; the Spanish Ministry of Economy and Competitiveness (MINECO) under AYA2015-66211-C2-1-P, AYA2015-66211-C2-2, AYA2012-30789, and ICTS-2009-14; and European FEDER funding (FCDD10-4E-867, FCDD13-4E-2685). The Brazilian agencies FINEP, FAPESP and the National Observatory of Brazil have also contributed to this project. The spectroscopic programs in this work are based on observations made with the Gran Telescopio Canarias (GTC), installed in the Spanish Observatorio del Roque de los Muchachos of the Instituto de Astrof\'isica de Canarias, in the island of La Palma. R.A.D. acknowledges support from the Conselho Nacional de Desenvolvimento Científico e Tecnológico - CNPq through BP grant 308105/2018-4, and the Financiadora de Estudos e Projetos - FINEP grants REF. 1217/13 - 01.13.0279.00 and REF 0859/10 - 01.10.0663.00 for hardware funding support for the J-PLUS project through the National Observatory of Brazil.
\end{acknowledgements}

\bibliographystyle{aa}
\bibliography{references}

\appendix
\section{Equations of the three-filters method}
\label{sec:append:3FM_equations}
Here we derive the main equations we use to extract the integrated \lya flux from J-PLUS photometry. These were originally detailed in \cite{vilellarojo2015} and similar methods are described in e.g. \cite{pascual2007} and \cite{guaita2010}. We start by defining the \textit{monochromatic} flux density of an astrophysical source (or, simply, its \textit{intrinsic} spectrum) as the emitted flux per unit frequency or wavelength: $\rm f_\nu$ or $\rm f_\lambda$, respectively in $\nu$-units ($\rm erg\,cm^{-2}\,s^{-1}\,Hz^{-1}$) and $\lambda$-units ($\rm erg\,cm^{-2}\,s^{-1}\,\text{\AA}\,^{-1}$). The two are connected by:
\begin{equation}
\rm f_\nu\,d\nu = f_\lambda\,d\lambda\quad;\quad \frac{d\nu}{d\lambda}=-\frac{c}{\lambda^2}\ ,
\end{equation}
where c is the speed of light. In this work, we only use the $\lambda$-units formalism, although magnitudes are usually defined in terms of $\rm f_\nu$. Photometric measurements are usually performed through \textit{filters} who probe $\rm f_\lambda$ over specific wavelength intervals or \textit{pass-bands}. Consequently, photometric filters are defined by their \textit{transmission curves} $\rm T_\lambda^{\,x} = T^{\,x}(\lambda)$, who describe their response\footnote{We define $\rm T_\lambda$ as the \textit{measured} transmission curve of a filter, i.e. including the quantum efficiency of the measuring device, the atmospheric transmission and the effect of telescope optics.} as a function of wavelength. All photons received within a given pass-band during the measuring process get integrated, hence the details of $\rm f_\lambda$ are lost. For this reason, the flux of a source measured in a given filter ``x'' is effectively defined as the \textit{average} flux in the pass-band weighted by the filter $\rm T_\lambda^{\,x}$, i.e. $\rm\left\langle f_\lambda^{\,x}\right\rangle$. For photon-counting devices (CCD), the latter quantity is given by \citep[e.g.,][]{tokunaga_vacca2005}:
\begin{equation}
    \rm \left\langle f_\lambda^{\,x} \right\rangle = \frac{\int f_\lambda\,T_\lambda^{\,x}\,\lambda\,d\lambda}{\int T_\lambda^{\,x}\,\lambda\,d\lambda} = \frac{\int (f_\lambda^{\,cont} + f_\lambda^{\,EL})\,T_\lambda^{\,x}\,\lambda\,d\lambda}{\int T_\lambda^{\,x}\,\lambda\,d\lambda}\ ,
    \label{eq:app:avg_flambda}
\end{equation}

\noindent
where we assume that $\rm f_\lambda$ can be written as the combination of line and continuum emission (respectively $\rm f_\lambda^{\,EL}$ and $\rm f_\lambda^{\,cont}$). In order to extract the line flux from the $\rm\left\langle f_\lambda^{\,x}\right\rangle$ measurement, we need to disentangle $\rm f_\lambda^{\,EL}$ from $\rm f_\lambda^{\,cont}$. The \textit{equivalent width} of a line measures the relative contribution of line and continuum to $\rm\left\langle f_\lambda^{\,x}\right\rangle$:
\begin{equation}
    \rm EW \equiv \int_{\lambda_{min}}^{\lambda_{max}}\left| 1 - \frac{f_\lambda^{\,tot}}{f_\lambda^{\,cont}}\right|\,d\lambda = \int_{\lambda_{min}}^{\lambda_{max}}\left| 1 - \frac{f_\lambda^{\,cont}+f_\lambda^{\,EL}}{f_\lambda^{\,cont}}\right|\,d\lambda\ ,
    \label{eq:app:EW_definition}
\end{equation}
where $\rm\lambda_{min}$ and $\rm\lambda_{max}$ encompass the whole line profile. By assuming that $\rm f_\lambda^{\,cont}$ is constant between $\rm\lambda_{min}$ and $\rm\lambda_{max}$, and denoting the wavelength of the line-profile peak as $\rm\lambda_{EL}$, we have:
\begin{equation}
    \rm EW = \frac{1}{f_{\lambda_{EL}}^{\,cont}}\,\left(\int_{\lambda_{min}}^{\lambda_{max}}f_\lambda^{\,EL}\,d\lambda\right)\, =\, \frac{F^{\,EL}}{f_{\lambda_{EL}}^{\,cont}}\ ,
    \label{eq:app:EW_definition_approx}
\end{equation}
where $\rm f_{\lambda_{EL}}^{\,cont} = f_\lambda^{\,cont}(\lambda_{EL})$. This also shows the definition of the \textit{continuum-subtracted, integrated line flux} $\rm F^{\,EL}$ ($\rm erg\,cm^{-2}\,s^{-1}$).

The above definitions allow to derive the basic equations of our methodology. We stress that this is designed to extract $\rm F^{\,EL}$ by using three photometric measurements (two BBs and one NB) and it is based on two main hypothesis, i) the emission-line profile can be approximated by a Dirac-delta, and ii) the source continuum is well-traced by a linear function of wavelength over the whole interval covered by the three filters (see also Sect. \ref{sec:3filters}):
\begin{align}
&\rm f_\lambda^{\,EL} = F^{\,EL}\cdot\delta(\lambda-\lambda_{EL})\label{eq:app:first_hypo_3fm}\ ,\\
&\rm f_\lambda^{\,cont} = A\,\lambda + B \label{eq:app:second_hypo_3fm}\ ,
\end{align}
where $\rm\delta(\lambda-\lambda_{EL})$ is centered at $\rm\lambda_{EL}$, while A and B are two scalar coefficients. Equation \ref{eq:app:first_hypo_3fm} implicitly assumes that $\rm F^{\,EL}$ is \textit{entirely} included within the NB pass-band. This might be false when part of the emission-line profile lies outside the NB pass-band, e.g. when the line-profile is wider than the NB pass-band (as for broad QSOs lines) or its peak lies close to the NB pass-band edge. The implications of this bias on our results are discussed in Sect. \ref{sec:line_flux_retrieval}. By using \ref{eq:app:second_hypo_3fm} into \ref{eq:app:avg_flambda} we get:
\begin{align}
 &\rm \left\langle f_\lambda^{\,x} \right\rangle = \frac{\int (A\,\lambda + B + f_\lambda^{\,EL})\cdot T_\lambda^{\,x}\,\lambda\,d\lambda}{\int T_\lambda^{\,x}\,\lambda\,d\lambda}\nonumber\\
 &\quad\ \ \ \, = \rm \frac{\left[ A\int \lambda^2\,T_\lambda^{\,x}\,d\lambda\ +\ B\int T_\lambda^{\,x}\,\lambda\,d\lambda\ +\ \int f_\lambda^{\,EL}\, T_\lambda^{\,x}\,\lambda\,d\lambda\right]}{\int T_\lambda^{\,x}\,\lambda\,d\lambda}\nonumber\\
 &\quad\ \ \ \, = \rm \frac{\left[ A\int \lambda^2\,T_\lambda^{\,x}\,d\lambda\ +\ B\int T_\lambda^{\,x}\,\lambda\,d\lambda\ +\ F^{\,EL}\, T_{\lambda_{\,EL}}^{\,x}\,\lambda_{EL}\right]}{\int T_\lambda^{\,x}\,\lambda\,d\lambda},\label{eq:app:almost_final_avg_flambda_3fm}
\end{align}
where $\rm T_{\lambda_{\,EL}}^{\,x} =  T_\lambda^{\,x}(\lambda_{EL})$, while the last step makes use of Eq. \ref{eq:app:first_hypo_3fm} in the last term at the numerator and the properties of the Dirac-delta. To simplify the notation we introduce:
\begin{equation}
    \rm \alpha_x = \frac{\int\lambda^2\,T_\lambda^{\,x}\,d\lambda}{\int T_\lambda^{\,x}\,\lambda\,d\lambda} \quad;\quad\quad \beta_x = \frac{T_{\lambda_{EL}}^{\,x}\,\lambda_{EL}}{\int T_\lambda^{\,x}\,\lambda\,d\lambda}\ ,
    \label{eq:app:alpha_beta_3fm}
\end{equation}
which depend \textit{only} on $\rm T^{\,x}_\lambda$ and $\rm\lambda_{EL}$. The latter is determined by each source redshift and cannot be measured without a spectroscopic observation, hence we must assume $\rm\lambda_{EL}$ a-priori. For each NB, we choose the value which maximizes the product $\rm T_\lambda\cdot [(dn/dz)\cdot(dz/d\lambda)]$, where $\rm dn/dz$ is the redshift distribution of SDSS QSOs \citep{paris2018}. This reflects that we expect most of our candidates to be $\rm z\!\gtrsim\!2$ QSOs, as discussed in Sect. \ref{sec:LAEs_classes_definition} and \ref{sec:LAE_samples}. We can now re-write \ref{eq:app:almost_final_avg_flambda_3fm} using \ref{eq:app:alpha_beta_3fm}:
\begin{equation}
    \rm \left\langle f_\lambda^{\,x} \right\rangle = A\cdot\alpha_x\ +\ B\ +\ F^{\,EL}\cdot\beta_x\ ,
    \label{eq:app:generic-filter_flux_density_3fm}
\end{equation}
which is valid for a generic filter. Note that if the targeted emission-line lies outside the pass-band we just have $\rm T_{\lambda_{EL}}^{\,x}=0$, implying $\rm\beta_x=0$. To determine A, B and $\rm F^{\,EL}$, we apply \ref{eq:app:generic-filter_flux_density_3fm} to a set of three filters: a NB, a \textit{line-contaminated} BB (denoted here by LC) and a \textit{line-uncontaminated} BB (denoted by LU):
\begin{align}
&\rm \left\langle f_\lambda^{\,NB} \right\rangle = A\cdot\alpha_{NB}\ +\ B\ +\ F^{\,EL}\cdot\beta_{NB}\label{eq:app:first_eq_in_3fm_system}\ ,\\
&\rm \left\langle f_\lambda^{\,LC} \right\rangle = A\cdot\alpha_{LC}\ +\ B\ +\ F^{\,EL}\cdot\beta_{LC}\label{eq:app:second_eq_in_3fm_system}\ ,\\
&\rm \left\langle f_\lambda^{\,LU} \right\rangle = A\cdot\alpha_{LU}\ +\ B.\label{eq:app:third_eq_in_3fm_system}
\end{align}
By solving this linear system we finally obtain $\rm F^{\,EL}$, A and B:
\begin{align}
    &\rm F^{\,EL} = \frac{ \left\langle f_\lambda^{\,LC} \right\rangle\ -\ \left\langle f_\lambda^{\,LU} \right\rangle\ +\ \frac{\alpha_{LU} - \alpha_{LC}}{\alpha_{NB} - \alpha_{LU}}\cdot\left[  \left\langle f_\lambda^{\,NB} \right\rangle - \left\langle f_\lambda^{\,LU} \right\rangle  \right]}{\beta_{LC}\ +\ \frac{\alpha_{LU} - \alpha_{LC}}{\alpha_{NB} - \alpha_{LU}}\cdot\beta_{NB}}\ ,
    \label{eq:app:fline_3fm}\\
    &\rm A = \frac{ \left\langle f_\lambda^{\,NB} \right\rangle\ -\ \left\langle f_\lambda^{\,LU} \right\rangle\ -\ \frac{\beta_{NB}}{\beta_{LC}}\cdot\left[  \left\langle f_\lambda^{\,LC} \right\rangle - \left\langle f_\lambda^{\,LU} \right\rangle  \right]}{\alpha_{NB}\ -\ \alpha_{LU}\ -\ \frac{\beta_{NB}}{\beta_{LC}}\cdot\left(\alpha_{LC}\ -\ \alpha_{LU}\right)}\ ,\\
    &\rm B = \left\langle f_\lambda^{\,LU} \right\rangle - \alpha_{LU}\cdot A\ .
\end{align}
The coefficients A and B can be used to evaluate \ref{eq:app:second_hypo_3fm} at the NB $\lambda$-pivot \citep[see][]{tokunaga_vacca2005} and get an estimate of the line-uncontaminated linear continuum in the NB $\rm\left\langle f_{\lambda\,;\,cont}^{\,NB}\right\rangle$. Equation \ref{eq:deltaNB} in Sect. \ref{sec:3filters} details how we use this continuum estimate to compute the NB excess of each J-PLUS source. Finally, we use Eq. \ref{eq:app:fline_3fm} to estimate the total line flux of our LAE candidates and construct their LFs as explained in Sect. \ref{sec:lumifunc}.

\section{Measurement of GTC spectra and follow-up results}
\label{sec:append:spectraMeasurement}
We measure the redshift of all the 37 sources identified as QSOs (either at $\rm z\!\sim\!1.5$ or at $\rm z\!\sim\!2.2$) in both spectroscopic programs. We do not aim at reaching a higher precision than $\rm \sigma_z=10^{-2}$, since the main goal of our follow-up programs is the spectroscopic confirmation of our targets. We additionally extract the \lya EW and integrated line flux $\rm F_{Ly\alpha}$ for the 29 $\rm z\!\sim\!2.2$ QSOs, from which we compute the sources \lya luminosity.

Following well-established procedures \citep[see e.g.,][]{paris2011}, we first identify the main spectral lines in our QSOs spectra, such as \texttt{CIV} and \texttt{CIII]}. We then use their profile-peaks to compute our redshift estimate. We discard the \lya profile for this analysis, since it provides a systematically biased z measure, due to the complex radiative transfer of \lya photons in the source rest frame and IGM \citep[see e.g.,][]{gronke2016, dijkstra2017, gurung2018}. We fit a double gaussian profile to both \texttt{CIV} and \texttt{CIII]} profiles, in order to trace at the same time its broad and narrow components. We use the $\lambda$ position of the narrow-component peaks to obtain two z estimates, whose average provides the final spectroscopic z of our sources.
\begin{figure}[t]
    \centering  
    \includegraphics[width=0.49\textwidth]{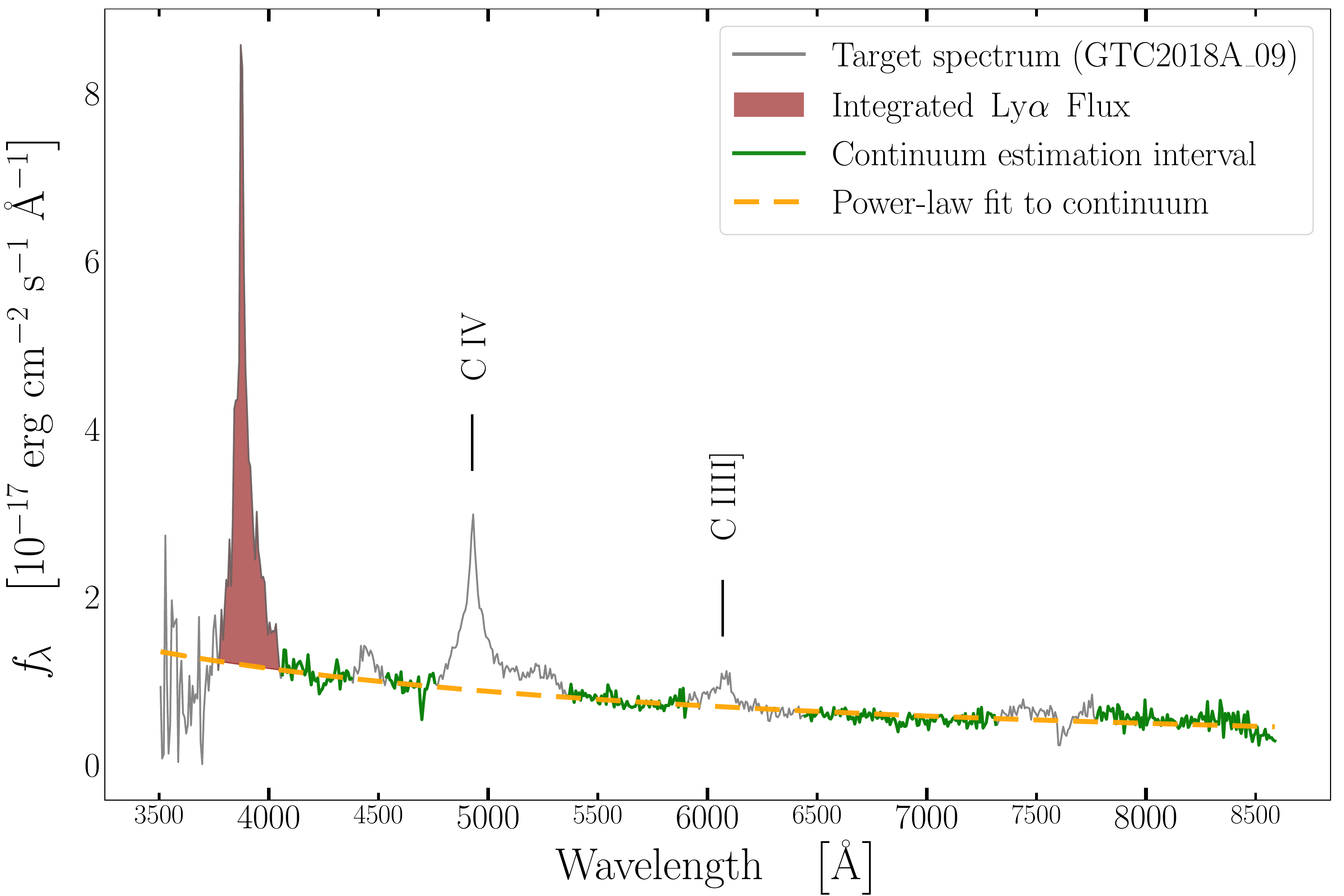}
    \caption{\footnotesize Calibrated spectrum (grey solid line) of the GTC2018A\_09 target, confirmed as $\rm z\!\sim\!2.2$ QSO. The four green regions highlight the intervals used for computing the power-law fit to the continuum (dashed yellow line). Finally, the \lya integrated flux is highlighted in red.}
    \label{fig:gtcSpectrum}
\end{figure}
The \lya line flux can only be obtained after estimating the sources continua. We then fit a power law to the wavelength regions of each spectrum which are not affected by any line feature, as shown by the yellow sections of the spectrum displayed in Fig. \ref{fig:gtcSpectrum}. We use the following simple functional form:
\begin{equation}
    \rm f_{\lambda}^{\,c}(\lambda) = k\,\lambda_{obs}^{\,\alpha}\ ,
    \label{eq:powlaw_cont}
\end{equation}
where $\rm f_{\lambda}^{\,c}(\lambda)$ is the spectrum \textit{monochromatic flux density} (in units of $\rm erg\,cm^{-2}\,s^{-1}\,\text{\AA}^{-1})$) while $\rm k$ and $\alpha$ are fit parameters. Finally, we measure the total \lya line flux by integrating the excess above the estimated continuum in the wavelength range affected by the \lya line, which is shown in Fig. \ref{fig:gtcSpectrum} as the spectral region highlighted in dark-red. As a last step, we estimate the observed \lya EW as:
\begin{equation}
    \rm EW_{obs}^{Ly\alpha} = \frac{F_{Ly\alpha}}{\rm f_{\lambda}^{\,c}(\lambda_{Ly\alpha})}\ ,
    \label{eq:ewobs}
\end{equation}
in which $\rm f_{\lambda}^{\,c}(\lambda_{Ly\alpha})$ is the value of the power-law fit to the continuum at the wavelength of the \lya line-profile peak. Figure \ref{fig:gtcSpectrum} shows the spectrum of target \texttt{GTC2018A\_09} as a visual example of our measuring procedure. The results of these measurements are shown in Table \ref{tab:GTC_obs_summary} together with a summary of the spectroscopic follow-up thechnical requrements and additional properties of the observed targets.

\begin{table*}[p!]
    \centering
    \begin{tabular}{l|c|c|c|c|c|c|c|c}
    \hline
    \textbf{Instrument} & \multicolumn{8}{ l }{\texttt{OSIRIS} spectrograph at Gran Telescopio Canarias (GTC)}\\
    \hline
    \textbf{Grism} & \multicolumn{8}{ l }{\texttt{R500B}}\\
    \hline
    \textbf{Seeing} & \multicolumn{8}{ l }{requested: 1.4 (maximum) --- effective: 1.03 (averaged on all observations)}\\
    \hline
    \textbf{Moon} & \multicolumn{8}{ l }{requested: any --- effective: $>\!90$\% Dark}\\
    \hline
    \textbf{Air mass} & \multicolumn{8}{ l }{requested: 1.5 (maximum) --- effective: 1.266 (averaged on all observations)}\\
    \hline
    \textbf{SNR} & \multicolumn{8}{ l }{$\rm\geqslant3$ at $\rm\lambda_{obs}\sim4000\text{\AA}$}\\
    \hline
    \hline
    \textbf{ID} & \textbf{Ra [hh:mm:ss]} & \textbf{Dec [hh:mm:ss]} & \textbf{Time [s]} & \textbf{SDSS class} & \textbf{GTC class} & $\rm\mathbf{z_{\,spec}}$ & $\rm\mathbf{L_{Ly\alpha}}$ & $\rm\mathbf{EW_{obs}}$\\
    \hline
    GTC2018A\_01  &  22:50:40.27  &  34:23:43.6  & 2185 &  ---     &  QSO/AGN  &  2.21  &  5.18e+44  &   513.98 \\
    \hline				         	                                          	           
    GTC2018A\_02  &  00:43:38.56  &  05:41:35.6  & 2185 &  GALAXY  &  QSO/AGN  &  2.23  &  1.70e+44  &   417.21 \\
    \hline				         	                                          	           
    GTC2018A\_03  &  23:01:08.05  &  33:44:20.0  & 2185 &  STAR    &  STAR     &  ---   &  ---	 &   ---    \\
    \hline				         	                                          	           
    GTC2018A\_04  &  16:17:15.12  &  50:25:59.2  & 2185 &  STAR    &  QSO/AGN  &  1.52  &  ---   &   --- \\
    \hline				         	                                          	           
    GTC2018A\_05  &  22:22:06.30  &  11:07:47.7  & 2335 &  STAR    &  QSO/AGN  &  2.23  &  3.87e+43  &   268.21 \\
    \hline				         	                                          	           
    GTC2018A\_06  &  18:10:22.34  &  41:49:25.3  & 2185 &  STAR    &  QSO/AGN  &  2.22  &  2.32e+44  &   741.21 \\
    \hline				         	                                          	           
    GTC2018A\_07  &  17:35:17.03  &  31:44:42.8  & 2185 &  GALAXY  &  QSO/AGN  &  2.23  &  7.46e+43  &   665.04 \\
    \hline				         	                                          	           
    GTC2018A\_08  &  01:31:29.69  &  33:55:14.9  & 2185 &  STAR    &  QSO/AGN  &  2.21  &  1.26e+44  &   637.59 \\
    \hline				         	                                          	           
    GTC2018A\_09  &  14:59:37.24  &  47:15:26.3  & 2185 &  STAR    &  QSO/AGN  &  2.18  &  2.13e+44  &   502.01 \\
    \hline				         	                                          	           
    GTC2018A\_10  &  16:11:57.72  &  46:00:45.8  & 2185 &  STAR    &  QSO/AGN  &  2.21  &  1.56e+44  &   263.97 \\
    \hline				         	                                          	           
    GTC2018A\_11  &  18:32:04.26  &  39:54:08.8  & 2185 &  ---     &  QSO/AGN  &  1.54  &  ---   &   --- \\
    \hline				         	                                          	           
    GTC2018A\_12  &  02:16:13.21  &  34:28:37.3  & 2185 &  ---     &  QSO/AGN  &  2.21  &  1.33e+44  &   272.11 \\
    \hline				         	                                          	           
    GTC2018A\_13  &  14:32:51.07  &  52:36:46.7  & 1954 &  GALAXY  &  GALAXY   &  0.51  &  ---	 &   --- \\
    \hline				         	                                          	           
    GTC2018A\_14  &  22:01:43.81  &  28:23:36.5  & 2335 &  STAR    &  QSO/AGN  &  1.53  &  ---   &   --- \\
    \hline				         	                                          	           
    GTC2018A\_15  &  16:09:37.67  &  45:29:53.6  & 2245 &  STAR    &  STAR     &  ---   &  ---	 &   --- \\
    \hline				         	                                          	           
    GTC2018A\_16  &  22:43:00.76  &  34:10:26.4  & 2335 &  STAR    &  QSO/AGN  &  2.25  &  ---	 &   --- \\
    \hline				         	                                          	           
    GTC2018A\_17  &  15:59:27.15  &  57:05:04.6  & 2245 &  STAR    &  QSO/AGN  &  2.22  &  1.08e+44  &   268.69 \\
    \hline				         	                                          	           
    GTC2018A\_18  &  23:03:24.69  &  33:20:25.5  & 2245 &  STAR    &  STAR     &  ---   &  ---	 &   --- \\
    \hline				         	                                          	           
    GTC2018A\_19  &  16:03:33.14  &  46:11:53.2  & 2245 &  ---     &  QSO/AGN  &  2.25  &  6.19e+43  &   351.25 \\
    \hline				         	                                          	           
    GTC2018A\_20  &  15:19:49.02  &  53:16:18.4  & 2335 &  STAR    &  QSO/AGN  &  2.19  &  2.04e+44  &   828.73 \\
    \hline				         	                                          	           
    GTC2018A\_21  &  14:53:19.08  &  53:02:42.3  & 2365 &  STAR    &  QSO/AGN  &  2.19  &  6.15e+43  &   217.99 \\
    \hline				         	                                          	           
    GTC2018A\_22  &  17:41:33.43  &  57:05:11.4  & 2335 &  STAR    &  QSO/AGN  &  2.27  &  8.20e+43  &   481.13 \\
    \hline				         	                                          	           
    GTC2018A\_23  &  15:38:49.50  &  48:58:13.1  & 2014 &  GALAXY  &  STAR     &  ---   &  ---	 &   --- \\
    \hline					                                          	           
    GTC2018A\_24  &  14:53:32.94  &  54:09:44.8  & 2335 &  GALAXY  &  QSO/AGN  &  1.53  &  ---   &   --- \\
    \hline					                                          	           
    \hline
    GTC2019A\_01 &  07:18:49.01  &  40:50:42.7  & 1165 &  STAR  &  STAR  &  ---  &   ---  &  ---  \\
    \hline				         	                                          	           
    GTC2019A\_02 &  22:37:58.84  &  11:41:01.4  & 1285 &  STAR  &  QSO/AGN  &  2.204  &  4.56e+44  & 706.71 \\
    \hline
    GTC2019A\_03 &  02:27:21.78  &  29:56:23.7  & 1366 & STAR   &  QSO/AGN  &  2.202  &  3.41e+44  & 859.82 \\
    \hline
    GTC2019A\_04 &  16:14:11.38  &  53:11:16.6  & 1426 &  STAR  &  BAL QSO  &  2.174  &  2.79e+44  & 290.75 \\
    \hline
    GTC2019A\_05 &  15:31:12.21  &  48:08:31.4  & 1576 &  STAR  &  QSO/AGN  &  2.209  &  2.91e+44  & 570.59 \\
    \hline
    GTC2019A\_06 &  15:34:37.75  &  46:42:36.3  & 1816 &  STAR  &  QSO/AGN  &  2.231  &  3.02e+44  & 414.59 \\
    \hline
    GTC2019A\_07 &  12:38:36.94  &  56:10:39.8  & 2086 &  STAR  &  QSO/AGN  &  2.177  &  2.02e+44  & 183.25 \\
    \hline
    GTC2019A\_08 &  12:29:07.02  &  56:03:49.2  & 2086 &  STAR  &  QSO/AGN  &  2.191  &  1.35e+44  & 230.15 \\
    \hline
    GTC2019A\_09 &  08:03:53.42  &  30:46:36.2  & 2206 & STAR   &  QSO/AGN  &  2.256  &  2.14e+44  & 380.86 \\
    \hline
    GTC2019A\_10 &  07:15:23.13  &  39:50:57.3  & 2356 & STAR   &  QSO/AGN  &  2.189  &  1.29e+44  & 625.73 \\
    \hline
    GTC2019A\_11 &  09:00:47.61  &  32:06:54.9  & 2806 & GALAXY &  QSO/AGN  &  2.253  &  9.54e+43  & 364.01 \\
    \hline
    GTC2019A\_12 &  00:51:10.29  &  03:08:25.4  & 3406 & STAR   &  QSO/AGN  &  1.543  &  8.34e+43  & 104.21 \\
    \hline
    GTC2019A\_13 &  17:25:01.20  &  33:46:50.4  & 3706 &  STAR  &  QSO/AGN  & 1.602   &  1.27e+44  & 180.52 \\
    \hline
    GTC2019A\_14 &  10:28:20.14  &  39:52:42.3  & 3706 & STAR   &  QSO/AGN  &  2.271  &  1.83e+44  & 121.36 \\
    \hline
    GTC2019A\_15 &  00:32:08.19  &  39:47:23.6  & 4476 &  STAR  &  QSO/AGN  & 1.512   &  1.33e+44  & 118.74 \\
    \hline
    GTC2019A\_16 &  16:14:30.92  &  50:12:24.2  & 4686 & GALAXY &  QSO/AGN  &  2.211  &  1.00e+44  & 705.82 \\
    \hline
    GTC2019A\_17 &  00:41:06.77  &  08:02:56.5  & 4761 &  STAR  &  GALAXY   &  ---  &  ---  &  ---  \\
    \hline
    GTC2019A\_18 &  16:10:19.83  &  45:31:49.4  & 5136 & GALAXY &  QSO/AGN  &  2.271  &  1.06e+44  & 644.06 \\
    \hline
    GTC2019A\_19 &  22:56:29.07  &  09:36:45.5  & 5136 & STAR   &  QSO/AGN  &  2.762  &  8.96e+43  & 390.88 \\
    \hline
    GTC2019A\_20 &  09:10:21.90  &  38:39:30.6  & 6246 & GALAXY &  QSO/AGN  &  1.528  &  1.61e+44  & 561.05 \\
    \hline
    GTC2019A\_21 &  13:20:29.68  &  56:31:49.6  & 6795 & GALAXY &  QSO/AGN  & 2.197   &  8.61e+43  & 154.68 \\
    \hline
    \end{tabular}
\caption{Properties retrieved from the follow-up of our 45 spectroscopic targets. These results confirm $29/45$ sources ($64.4\%$) as genuine \lya$\!$-emitting QSOs at $\rm z\!\sim\!2$. The most numerous interlopers are \texttt{CIV}-emitting QSOs at $\rm z\!\sim\!1.52$, namely $8/45$ targets \citep[$\sim18\%$, see][]{stroe2017a,stroe2017b}, and $5/45$ blue stars ($\sim11\%$). The latter are selected by our pipeline due to their strong color-gradients which mimic a NB photometric excess in the $J$0395 filter. We also note one $\rm Ly\beta$-emitting QSO contaminant at $\rm z\!\sim\!2.76$. We report the measured $\rm L_{Ly\alpha}$ and $\rm EW_{obs}$ only for the confirmed QSOs at $\rm z\!\sim\!2.2$. Among these, the spectrum of \texttt{GTC 16} could not be calibrated and measured.Finally, all the $\rm z\!\sim\!2.2$ confirmed sources in our sample are QSOs with $\rm L_{Ly\alpha}\!>\!6\times10^{43}\,erg\,s^{-1}$. This supports the results of \cite{konno2016, matthee2017b,sobral2018b,sobral2018} and \cite{calhau2020} about the strong contribution of AGN/QSOs to the \lya LF at $\rm Log(Ly\alpha)\!\gtrsim\!43.3$.}
    \label{tab:GTC_obs_summary}
\end{table*}

\clearpage
\section{Retrieval of the total line flux}
\label{sec:app:lya_flux_definitions}
\begin{figure}[t]
    \centering
    \includegraphics[width=0.48\textwidth]{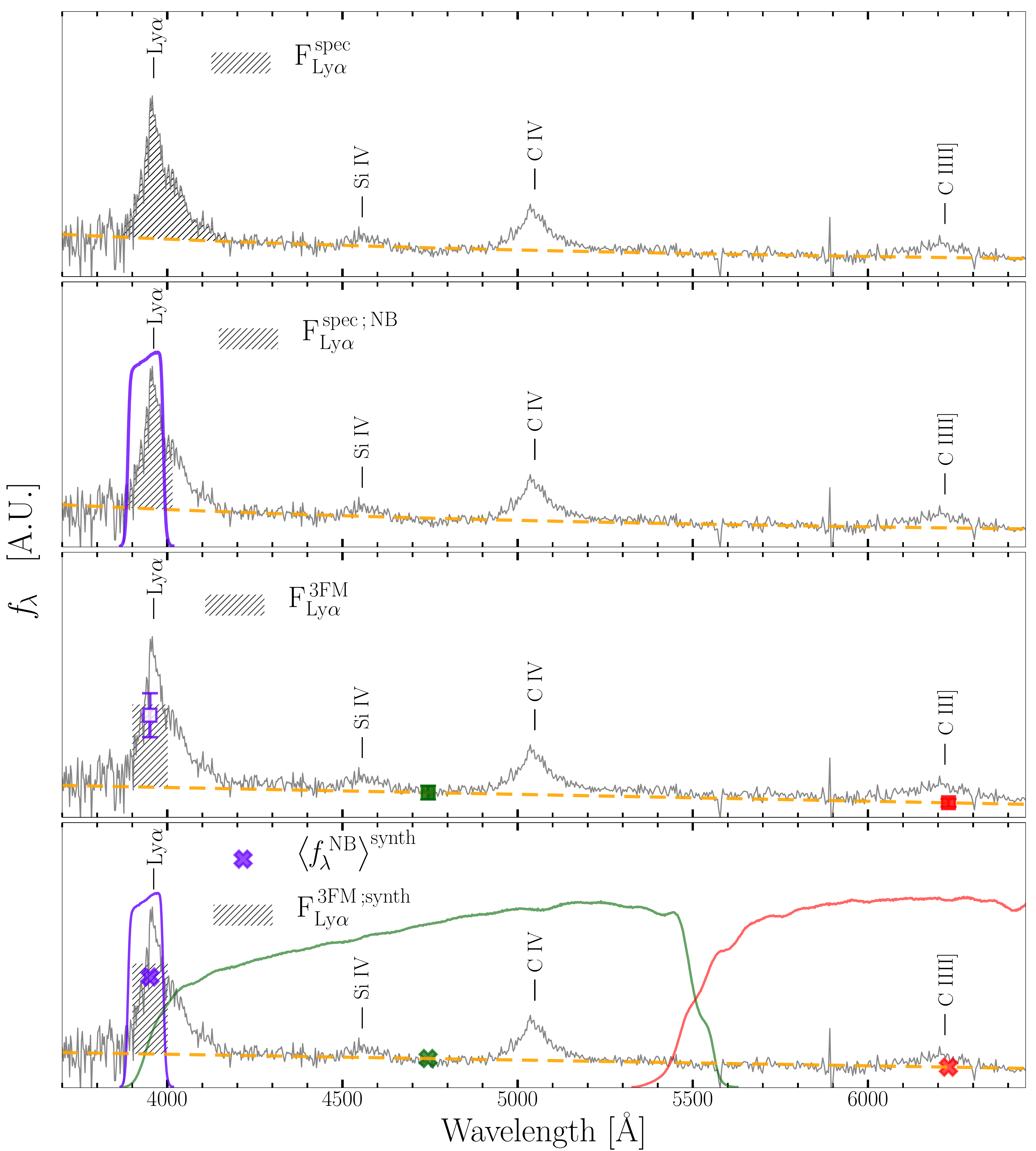}
    \caption{\footnotesize Graphic definition of the quantities we use for comparing \lya flux. The spectrum of a $\rm z\!\sim\!2.2$ QSO from \texttt{SDSS DR14} is used as example in all panels (grey lines). Each \lya flux definition is outlined by a grey shaded area (see text for details). Yellow lines in each panel show the reconstruction of the source continuum (power-law and linear approximation respectively in the first two and last two panels from above). Finally, colored squares and crosses (respectively third and last panel from the top) show respectively J-PLUS measurements and synthetic photometry performed on the SDSS spectrum with J-PLUS transmission curves (Eq. \ref{eq:app:avg_flambda}).}
    \label{fig:flux_definitions}
\end{figure}

Here we describe how we characterize and correct the bias on our measurements of \lya flux (namely $\rm F^{\,3FM}_{Ly\alpha}$). We compute a correction for each NB by exploiting the comparison between J-PLUS sources and their counterparts in the SDSS QSOs catalog \citep{paris2018}, within the redshift windows sampled by the NB. For the sake of brevity, we only show the results for either $J$0410 or $J$0430 filters. As a first step, we introduce the \lya flux quantities obtained from SDSS spectra in order to perform our comparison (see Fig. \ref{fig:flux_definitions}):\vspace{-1mm}
\begin{itemize}
    \item $\rm \langle \mathit{f}^{\,x}_{\lambda}\rangle^{synth}\,$: synthetic flux-density, in $\rm erg\,cm^{-2}\,s^{-1}\,\text{\AA}^{-1}$, measured by convolving SDSS spectra with the transmission curve of a given J-PLUS filter ``x'', as in eq. \ref{eq:app:avg_flambda}. Coloured crosses in the bottom panel of Fig. \ref{fig:flux_definitions} mark the synthetic photometry of the SDSS QSO taken as example.
    \item $\rm F^{\,spec}_{Ly\alpha}$: spectroscopic measurement of the wavelength-integrated \lya flux, in $\rm erg\,cm^{-2}\,s^{-1}$. This is obtained by fitting the QSOs continuum and integrating the spectra above it, on the wavelength range affected by the \textit{whole} \lya line profile (see appendix \ref{sec:append:spectraMeasurement} for details). This measurement does not involve the convolution of SDSS spectra with the filters transmission curves.
    \item $\rm F^{\,spec\,;\,NB}_{Ly\alpha}$: a version of $\rm F^{\,spec}_{Ly\alpha}$ obtained by integrating the spectra above the continuum-fit exclusively over the wavelength range covered by a J-PLUS NBs. If the \lya line of a given QSO is wider than the NB, this measurement is just a fraction of $\rm F^{\,spec}_{Ly\alpha}$, since the \lya flux lying outside the transmission curve would not be accounted for (see Fig. \ref{fig:flux_definitions}).
    \item $\rm F^{\,3FM\,;\,synth}_{Ly\alpha}$: this photometric quantity is analogous to $\rm F^{\,3FM}_{Ly\alpha}$, with the only difference of being computed on the synthetic photometry of SDSS QSOs (coloured crosses in the bottom panel of Fig. \ref{fig:flux_definitions}).
\end{itemize}
The photometric quantity $\rm F^{\,3FM}_{Ly\alpha}$ is directly comparable to the spectroscopic $\rm F^{\,spec;\,NB}_{Ly\alpha}$, since the method of \cite{vilellarojo2015} is designed to remove the effect of the filter transmission curve (see also appendix \ref{sec:append:3FM_equations}). Moreover, any photometric measurement is only sensitive to the flux received within a given band, hence to $\rm F^{\,spec;NB}_{Ly\alpha}$ and not to $\rm F^{\,spec}_{Ly\alpha}$, in our case.

\subsection{$r$-band \texttt{auto}-aperture flux}
\label{sec:app:aperture_correction}
\begin{figure}[h]
    \centering
    \includegraphics[width=0.46\textwidth]{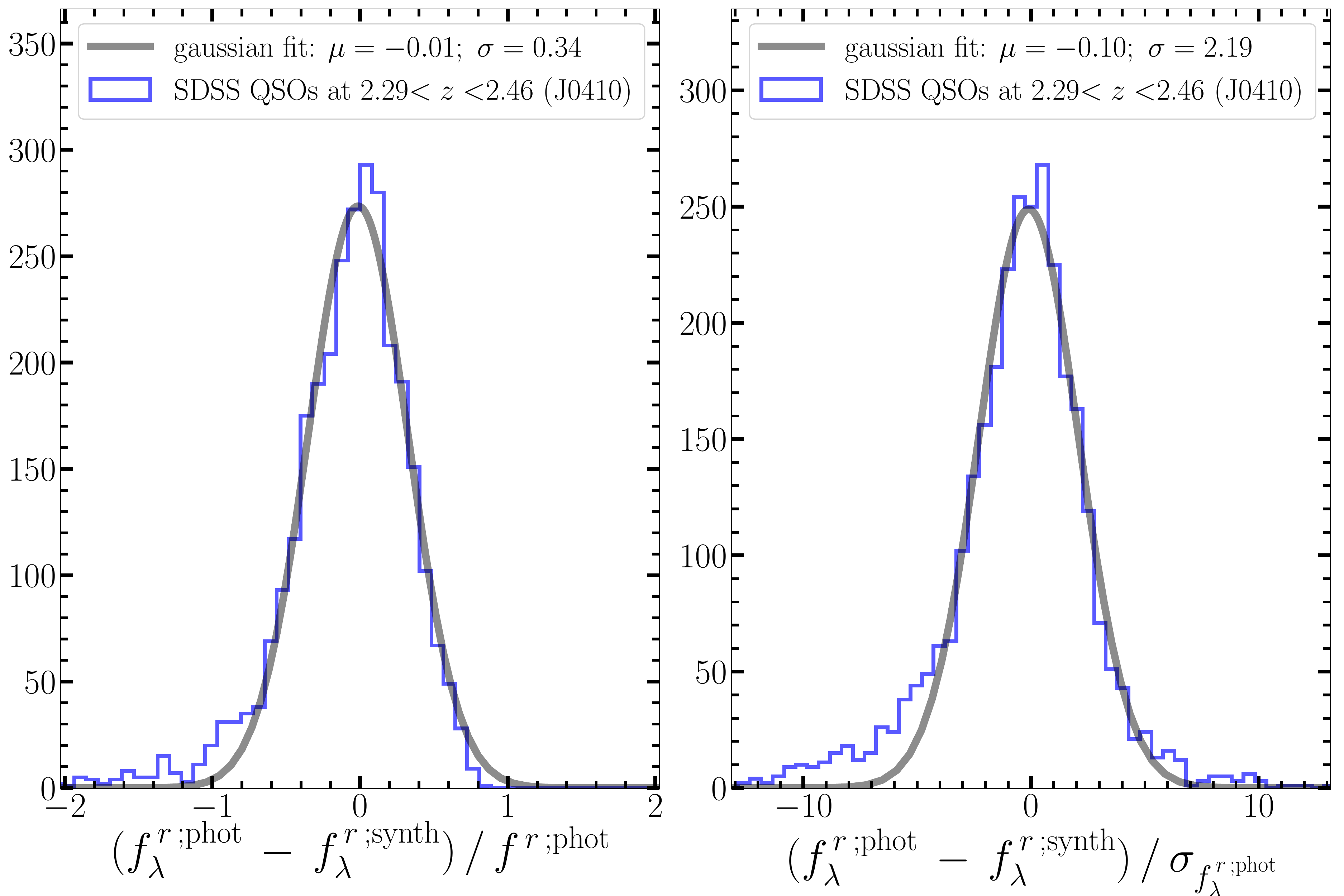}
    \caption{\footnotesize Blue solid lines in both panels show the histograms of the difference between J-PLUS $r$ flux (i.e. $\rm \mathit{f}_\lambda^{\,\mathit{r}}$) and $\rm\mathit{f}^{\,\mathit{r}\,;\,synth}_\lambda$ obtained from SDSS QSOs spectra (see Fig. \ref{fig:flux_definitions}). The distributions are normalized by respectively $\rm \mathit{f}_\lambda^{\,\mathit{r}}$ (left panel) and its photometric error $\rm\sigma_{\mathit{f}_\lambda^{\,\mathit{r}}}$ (right panel). Both distributions are centered in zero (see plot legends), meaning that $\rm \mathit{f}_\lambda^{\,\mathit{r}}$ and $\rm\mathit{f}^{\,\mathit{r}\,;\,synth}_\lambda$ values are statistically equivalent. On the other hand, the distribution spread in the right panel is significantly bigger than one, hence photometric errors do not fully account for the flux difference. Consequently, we re-scale $\rm\sigma_{\mathit{f}_\lambda^{\,\mathit{r}}}$ to the value obtained by the Gaussian fit (right-panel legend).}
    \label{fig:aperture_flux_comparison}
\end{figure}

Figure \ref{fig:aperture_flux_comparison} shows the comparison between $r$-band flux and $\rm \langle \mathit{f}^{\,\,\mathit{r}}_{\lambda}\,\rangle^{\,synth}\,$ for J-PLUS sources and their QSOs counterparts with \lya in the $J$0410 filter. The left panel displays no systematic shift, hence no strong aperture bias ($\lesssim0.2\,\sigma_r$) affects the \texttt{auto}-aperture flux of point-like sources. Since $\rm z\!\gtrsim\!2$ sources appear point-like in J-PLUS (see Sect. \ref{sec:morpho}), we conclude that \texttt{auto}-aperture photometry collects the total light of our \lya$\!$-emitting candidates. On the other hand, the spread of the distribution in the right panel is significantly greater than one, hence we need to account for this additional statistic uncertainty on top of J-PLUS photometric errors when computing $\rm F^{\,3FM}_{Ly\alpha}$. We re-scale the photometric errors of $r$ band photometry and propagate the resulting $\sigma_r$ on $\rm F^{\,3FM}_{Ly\alpha}$. The latter is accounted for in the errors of our LFs as discussed in Sect. \ref{sec:errors_on_the_LF}.
\begin{figure}[h!]
    \centering
    \includegraphics[width=0.48\textwidth]{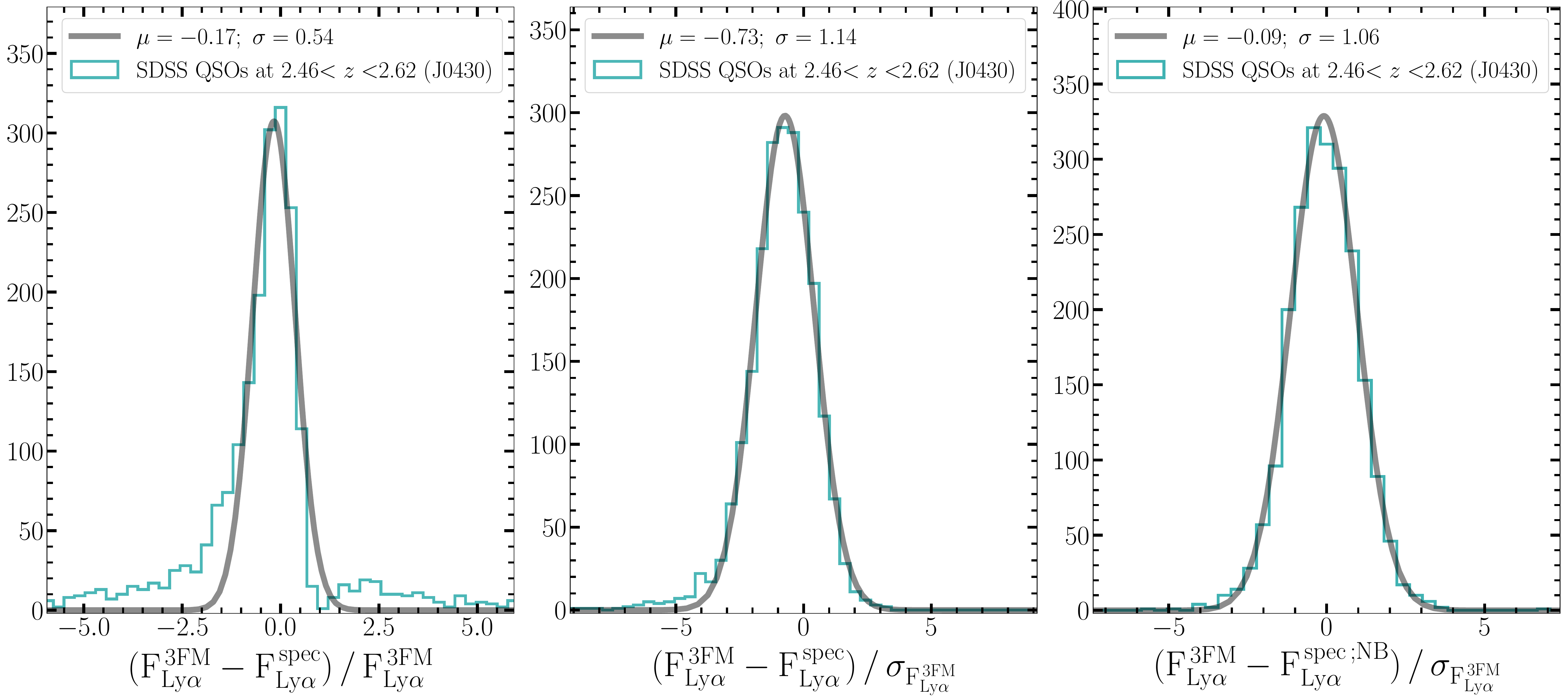}
    \caption{\footnotesize Cyan solid lines in all panels show the histograms of differences between line flux measurements performed on J-PLUS photometry (section \ref{sec:3filters}) and on SDSS spectroscopy (Fig. \ref{fig:flux_definitions}), for the case of $J$0430. The difference between $\rm F^{\,3FM}_{Ly\alpha}$ and $\rm F^{\,spec}_{Ly\alpha}$ is normalized respectively by $\rm F^{\,3FM}_{Ly\alpha}$ and by $\rm\sigma_{F^{\,3FM}_{Ly\alpha}}$ in the left and middle panels. We use the systematic shift of the distributions in the left panel to statistically correct $\rm F^{\,3FM}_{Ly\alpha}$, while the distribution spread in the middle panel allows to account for residual statistical errors not included in $\rm\sigma_{F^{\,3FM}_{Ly\alpha}}$. Section \ref{sec:errors_on_the_LF} details how the systematic offset (left panel) and distribution spread (middle panel) concur to the errors on our final LFs. Finally, right panel shows that $\rm F^{\,3FM}_{Ly\alpha}$ well compares to the spectroscopic measure $\rm F^{\,spec\,;\,NB}_{Ly\alpha}$.}
    \label{fig:filter-width_flux_correction}
\end{figure}

\subsection{Filter-width effect on the line flux}
Since the observed \lya line profile of QSOs is generally wider than the FWHM of J-PLUS NBs, we need to account for line-flux losses. For this, we obtain two corrections, respectively for $\rm F^{\,3FM}_{Ly\alpha}$ and its errors\footnote{We stress that $\rm\sigma_{F^{\,3FM}_{Ly\alpha}}$ is the error on $\rm F^{\,3FM}_{Ly\alpha}$ computed by propagating the photometric errors in Eq. \ref{eq:line_flux}, as detailed in appendix \ref{sec:append:3FM_equations}.} $\rm\sigma_{F^{\,3FM}_{Ly\alpha}}$. These affect the \lya luminosity of our candidates and its errors, hence we account for them on our final LFs (as discussed in Sect. \ref{sec:errors_on_the_LF}). Figure \ref{fig:filter-width_flux_correction} shows the comparison between  $\rm F^{\,3FM}_{Ly\alpha}$ and $\rm F^{\,spec}_{Ly\alpha}$ for $J$0430 filter. Their flux-difference is presented in the left and middle panel, respectively normalized by $\rm F^{\,3FM}_{Ly\alpha}$ and by $\rm\sigma_{F^{\,3FM}_{Ly\alpha}}$. The left panel shows clear evidences of a systematic offset $\rm \Delta F$ between the two flux quantities. We measure $\rm \Delta F$ with a gaussian fit (see Fig. \ref{fig:filter-width_flux_correction}) and use it to correct $\rm F^{\,3FM}_{Ly\alpha}$ as follows:
\begin{equation}
    \rm F^{\,3FM\,;\,corr}_{Ly\alpha} = (1-\Delta F)\cdot F^{\,3FM}_{Ly\alpha}\ .
\end{equation}
We stress that by directly comparing $\rm F^{\,3FM}_{Ly\alpha}$ to $\rm F^{\,spec}_{Ly\alpha}$, our statistical correction accounts for any systematic bias of our measurements, such as the linear-continuum approximation or the line-peak position of \lya within the NB. The spread of the distribution in the middle panel of Fig. \ref{fig:filter-width_flux_correction} is significantly bigger than unity. This shows that $\rm\sigma_{F^{\,3FM}_{Ly\alpha}}$ cannot fully account for the difference between $\rm F^{\,3FM}_{Ly\alpha}$ and $\rm F^{\,spec}_{Ly\alpha}$. Consequently, we re-scale $\rm\sigma_{F^{\,3FM}_{Ly\alpha}}$ according to the measured spread of the distribution in the middle panel of Fig. \ref{fig:filter-width_flux_correction}. Finally, we account for these errors on our final LFs (see Sect. \ref{sec:errors_on_the_LF}).
To conclude, the comparison between middle and right panels of Fig. \ref{fig:filter-width_flux_correction} clearly shows how $\rm F^{\,3FM}_{Ly\alpha}$ better compares to the fraction of spectroscopic line-flux measured \textit{only} on the wavelength range covered by the J-PLUS NB (i.e. $\rm F^{\,spec\,;\,NB}_{Ly\alpha}$, see Fig. \ref{fig:flux_definitions} and Sect. \ref{sec:app:lya_flux_definitions}). This is a direct effect of the filter-width bias, because our methodology is only sensitive to the flux captured by J-PLUS photometry within the NB wavelength range, as discussed in Sect. \ref{sec:app:lya_flux_definitions}.

\section{Multi-variate completeness computation}
\label{sec:app:completeness_computation}
Here we describe the computation of the corrections accounting for incompleteness due to our selection methodology (section \ref{sec:app:selection_completeness}) and the use of $r$-band detected catalogs (section \ref{sec:app:bivariate_completeness}).

\subsection{Selection completeness}
\label{sec:app:selection_completeness}
\begin{figure}[h]
    \centering
    \includegraphics[width=0.49\textwidth]{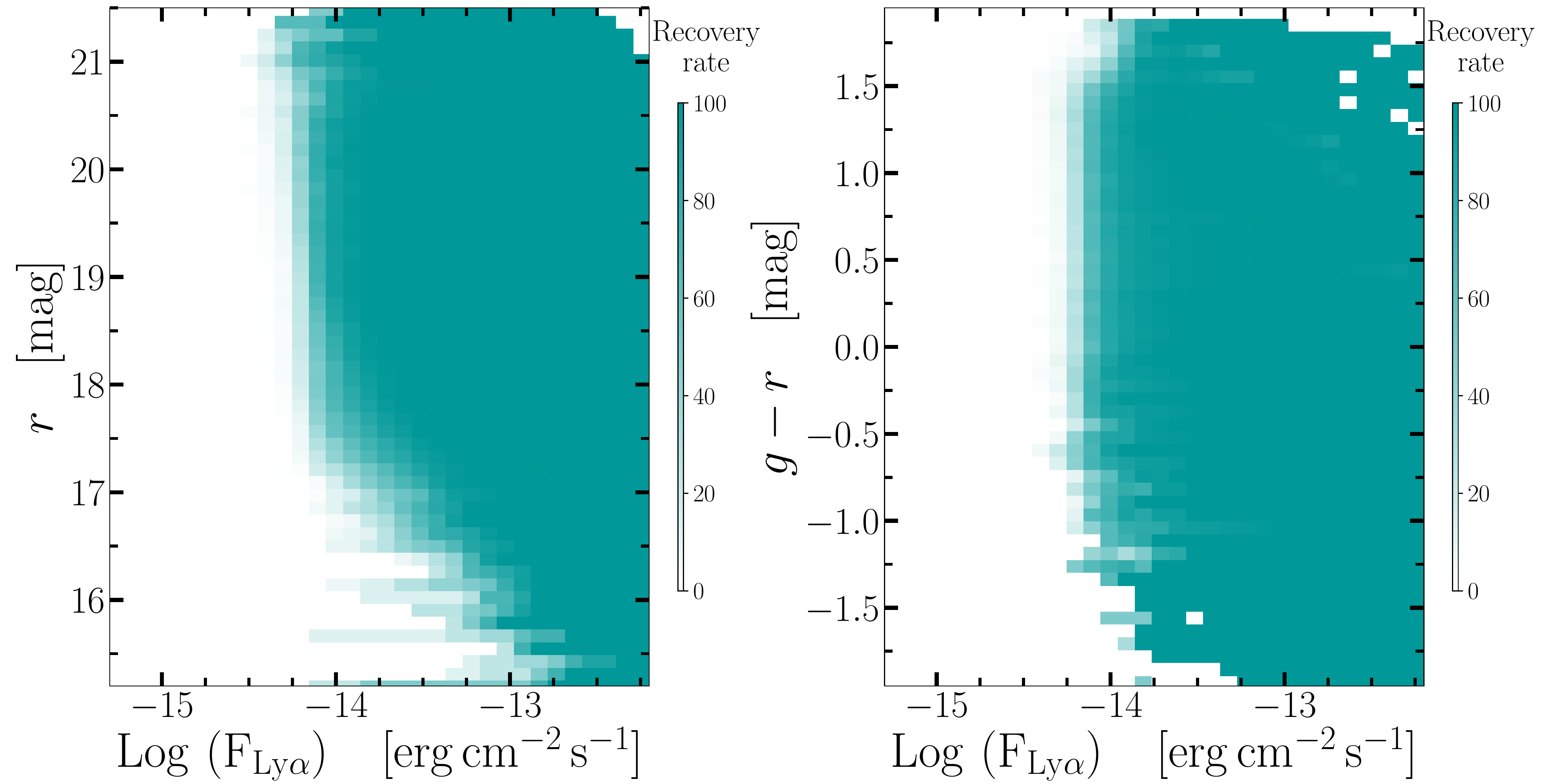}
    \caption{\footnotesize{3D grid of recovery rates for the $J$0430 filter, taken as example. Left and right panels respectively show the projections of recovery rates in the \lya flux vs. $r$ plane and \lya flux vs. $g\!-\!r$ plane. We note that the recovery rates show noisy values at $r\!<\!17$ and $g\!-\!r\!<\!-1.25$ due to the low number of sources in these magnitude and color bins. Nevertheless, these regions of the 3D parameter space are excluded from the LF computation by the purity weight (section \ref{sec:purity_of_samples}).}}
    \label{fig:3D_completeness}
\end{figure}

To simplify, our selection depends on i) the linear approximation of the sources continuum, ii) their \lya flux and iii) their \lya EW. To account for these dependencies, we measure the recovery rate of our methodology as a function of i) $g\!-\!r$ color, ii) \lya flux and iii) $r$-band magnitude. Indeed, $g\!-\!r$ color can be thought as a proxy of our linear continuum approximation (see e.g. Fig. \ref{fig:LAEspectra_3FM}), hence the EW dependence of our selection is accounted for by independently varying the \lya flux with respect to $g\!-\!r$ and $r$ magnitude. More in detail:
\begin{enumerate}
    \item we first subtract the measured $\rm F^{\,3FM}_{Ly\alpha}$ from the sources photometry of our candidates, in order to get a list of \textit{non-emitters}, i.e. sources without a significant NB excess according to our measuring method,
    \item then we artificially re-add increasing values of line flux to the sources photometry (i.e. to the NB and the $g$ band, since both are affected by the emission line),
    \item for each value of the re-added flux, we apply our complete set of selection rules (section \ref{sec:selection_rules}) and we store the number of re-selected sources as a function of their $r$ magnitude and $g-r$ color,
    \item we finally compute the sources-recovery rate of our selection as $\rm C=N_{selected}/N_{total}$ in each bin of $r$ magnitude, $g-r$ color and artificially-injected \lya flux.
\end{enumerate}
\noindent
With this method we obtain a 3D grid of recovery rates which can be interpolated in order to compute a selection weight $\rm C^{\,s}_i$ for each source. The latter accounts for the loss of candidates due to our selection within the total completeness correction we apply to the final LFs. The computation of $\rm C^{\,s}_i$ depends on \lya$\!\!$-flux since this is the observable tackled by our measuring method (see e.g. Sect. \ref{sec:3filters} and \ref{sec:append:3FM_equations}). Nevertheless, this dependence can be directly converted into a $\rm L_{Ly\alpha}$ dependence by assuming a redshift for every source (see Sect. \ref{sec:lyalum_and_volume_computation}). Figure \ref{fig:3D_completeness} shows the recovery-rates grid for $J$0430 filter as an examples. The values are projected in the planes $\rm F_{Ly\alpha}$ vs. $r$ (left panel) and $\rm F_{Ly\alpha}$ vs. $g\!-\!r$ (right panel).

\subsection{Bivariate completeness model}
\label{sec:app:bivariate_completeness}
Since we use $r$-band detected catalogs, we need to take into account the loss of undetected continuum-faint \lya$\!\!$-emitting sources. In other words, we need to estimate the distribution of sources in regions of the $\rm L_{Lya}$ vs. $r$ plane which lie outside the thresholds on $r$ and NB-excess significance imposed by J-PLUS detection limits. To carry out this analysis we closely follow the methods of \cite{gunawardhana2015}, who tackled a very similar issue with a multi-variate approach. More in detail, we build the \textit{bivariate} luminosity function of our candidates, defined as the number density of sources in each bin of $\rm Log\,(L_{Ly\alpha})$ and $r$, weighted by the purity and completeness corrections:
\begin{equation}
    \rm \Phi[\,Log\,(L_{Ly\alpha}),\,\mathit{r}\,]_{\,j\,k}\,=\,\frac{\sum_i\ [P_i\,/\,(C^{\,d}_i\cdot C^{\,s}_i)]}{V\ \cdot\,\Delta_j\, Log(L_{Ly\alpha})\ \cdot\,\Delta_k\mathit{r}}\ ,
    \label{eq:bivariate_LF}
\end{equation}
where j and k indexes identify the 2D bins of $\rm Log(L_{Ly\alpha})$ and $r$, while the index i runs over the total number of sources in each 2D bin. $\rm P_i$ is the purity weight of each candidate (see Sect. \ref{sec:purity_of_samples}), while $\rm C^{\,d}_i$ and $\rm C^{\,s}_i$ are respectively its detection-completeness (section \ref{sec:detection_completeness}) and selection-completeness (section \ref{sec:flux_completeness} and appendix \ref{sec:app:selection_completeness}) weights. Finally, $\rm V$ is the survey effective volume for a given NB filter (see Sect. \ref{sec:lyalum_and_volume_computation} and Table \ref{tab:QSOcontamin}). By following \cite{gunawardhana2015}, we assume that the 2D LF can be modelled by the product of two functions, describing respectively the $r$ and $\rm Log(L_{Ly\alpha})$ distributions \citep[see also][]{corbelli_salepeter_dickey1991}. We choose to employ the combination of a Schechter function (in logaritmic form) for $r$ and a Gaussian in $\rm Log\,L_{Ly\alpha}$ \citep[as in][]{gunawardhana2015}:
\begin{equation}
\begin{array}{l}
 \rm\Phi(\mathit{r}) = 0.4\ ln(10)\ \Phi_{\,r}^*\ 10^{\,0.4(\,r^*\,-\,r\,)(\alpha_{\,r}+1)}\ exp[-10^{0.4(\,r^*\,-\,r\,)}]\vspace{2mm}\\
 \rm\Phi(LogL) = \frac{\Phi_L^*}{\sigma_{L}\,\sqrt{2\pi}}\,exp\left[-\frac{1}{2\,\sigma_{L}^2}\,\left(LogL\,-\,LogL^*\right)^2\right]\ ,
\end{array}
\label{eq:log_form_schechter}
\end{equation}
where $\rm\Phi_r^*$, $\rm r^*$, $\rm\alpha_r$ are the ordinary Schechter parameters \citep[see][]{schechter1976}, while $\rm\Phi_L^*$, $\rm L^*$, $\rm\sigma_{L}$ describe respectively the number-density normalization, the average luminosity and the spread of the $\rm L_{Ly\alpha}$ distribution in each $r$ bin. In order to obtain the bivariate model, we follow \cite{gunawardhana2015} and join the two univariate distributions presented in Eq. \ref{eq:log_form_schechter} with an equation between their structural parameters:
\begin{equation}
    \rm L^*(\mathit{r})=10^{\,A\,(r\,-\,r_0^*)\,+\,B}\ ,
    \label{eq:lstar_to_magnitude}
\end{equation}
where A and B are free parameters to be determined by fitting the 2D model to our data, while $\rm r_0^*=19.5$ \citep[as in][]{gunawardhana2015}. By substituting Eq. \ref{eq:lstar_to_magnitude} into \ref{eq:log_form_schechter} and multiplying the two univariate distributions, we obtain the full bivariate model:
\begin{equation}
\begin{array}{l}
 \rm \Phi[\,Log\,(L_{Ly\alpha}),\,\mathit{r}\,] = \rm\Phi(\mathit{r})\,\times\,\Phi(LogL_{Ly\alpha}\,;\,\mathit{r}) = \vspace{2mm} \\
 \rm 0.4\ \,ln(10)\ \,\Phi_{\,r}^*\ \,10^{\,0.4(\,r^*\,-\,r\,)(\alpha_{\,r}+1)}\ \,\rm exp\left[-10^{0.4(\,r^*\,-\,r\,)}\right]\ \,\times\vspace{2mm}\\
 \rm \frac{\Phi_L^*}{\sigma\,\sqrt{2\pi}}\,exp\left\{-\frac{1}{2\,\sigma^2}\,\left[LogL\,-\,\left(A\,(r\,-\,r_0^*)\,+\,B\right)\right]^2\right\}\ .
 \end{array}
 \label{eq:bivariate_model}
\end{equation}
We fit the seven free parameters of this function using our measured 2D luminosity function. In particular, we only use the candidates with a completeness weight $\rm(C^{\,d}_i\cdot C^{\,s}_i)\!>\!0.85$, in order to avoid biasing our fit with regions affected by the incompleteness of our selection. Finally, the ratio between the measured 2D LF of our candidates and the fitted model (over the whole $r$ - $\rm L_{Ly\alpha}$ plane) provides an estimate of the incompleteness of our selection in each $\rm [Log(L_{Ly\alpha}),\mathit{r}]$ bin. We use this ratio to compute a \textit{bivariate weight} $\rm C^{\,b}_i$ for each of our candidates. These weights are then combined to $\rm C^{\,d}_i$ and $\rm C^{\,s}_i$ in order to obtain our total completeness correction (see Sect. \ref{sec:completeness}).

\end{document}